\documentclass[a4paper,font=10pt,UKenglish,cleveref,autoref,sigplan,screen]{acmart}
\usepackage{hyperref}
\usepackage{graphicx}
\usepackage[strings]{underscore}

\usepackage{xcolor}
\usepackage{amsmath}
\usepackage[shortlabels]{enumitem}
\usepackage{booktabs}
\usepackage{listings}
\usepackage{pifont}
\usepackage{cancel}
\usepackage{multirow}
\usepackage{bm}
\usepackage{grace}
\usepackage[scaled=.8]{helvet}

\usepackage{thmtools}
\usepackage{thm-restate}

\newcommand{\cmark}{\ding{51}}%
\newcommand{\xmark}{\ding{55}}%

\usepackage[capitalise]{cleveref}
\Crefname{section}{\S$\!$}{\S\S$\!$}

\lstdefinestyle{grace}
{
  language=java,
  basicstyle=\fontsize{9}{9}\selectfont\tt,
  keywordstyle=\fontsize{9}{9}\selectfont\tt\bfseries,
  commentstyle=\fontfamily{ptm}\selectfont\itshape\color{magenta},
  numbers=none,
  numberstyle=\fontsize{6}{9}\fontfamily{ptm}\selectfont,
  literate={~~>}{$\leadsto\ $}{2} {lambda}{$\lambda$}{0} {<<}{$\llbracket$}{1}
               {>>}{$\rrbracket$}{1} {->}{$\to\ $}{1} {<-}{$\leftarrow\ $}{1},
  aboveskip=1ex,
  belowskip=1ex,
  tabsize=1,
  columns=fullflexible,
  xleftmargin=0ex,
  resetmargins=false,
  showstringspaces=false,
  escapeinside={(*@}{@*)},
  morecomment=[l]{--},
  backgroundcolor=\color{white},%
morekeywords={trait,async,spawn,get,this,string,uint,int,real,bool,var,val,then,fun,match,
                          lin,imm,local,method,local,self,consume,iso,obj,object,chan,copy,let,in,unsafe,unknown,use,isolate,def}
}

\lstdefinestyle{minigrace}
{
  language=java,
  basicstyle=\fontsize{9}{9}\selectfont\tt,
  keywordstyle=\fontsize{9}{9}\selectfont\tt\bfseries,
  commentstyle=\fontfamily{ptm}\selectfont\itshape\color{magenta},
  numbers=none,
  numberstyle=\fontsize{6}{9}\fontfamily{ptm}\selectfont,
  literate={~~>}{$\leadsto\ $}{2} {lambda}{$\lambda$}{0} {<<}{$\llbracket$}{1}
               {>>}{$\rrbracket$}{1} {->}{$\to$}{1} {<-}{$\leftarrow$}{1},
  aboveskip=1ex,
  belowskip=1ex,
  tabsize=1,
  columns=fullflexible,
  xleftmargin=0ex,
  resetmargins=false,
  showstringspaces=false,
  escapeinside={(*@}{@*)},
  morecomment=[l]{--},
  backgroundcolor=\color{white},%
morekeywords={trait,async,spawn,get,this,string,uint,obj,copy,imm,iso,unsafe,int,real,bool,var,val,then,fun,match,let,in,actor,method,self,consume,object,use,def,local}
}

\lstdefinestyle{micrograce}
{
  language=java,
  basicstyle=\fontsize{9}{9}\selectfont\tt,
  keywordstyle=\fontsize{9}{9}\selectfont\tt\bfseries,
  commentstyle=\fontfamily{ptm}\selectfont\itshape\color{magenta},
  numbers=none,
  numberstyle=\fontsize{6}{9}\fontfamily{ptm}\selectfont,
  literate={~~>}{$\leadsto\ $}{2} {lambda}{$\lambda$}{0} {<<}{$\llbracket$}{1}
               {>>}{$\rrbracket$}{1} {->}{$\to$}{1} {<-}{$\leftarrow$}{1},
  aboveskip=1ex,
  belowskip=1ex,
  tabsize=1,
  columns=fullflexible,
  xleftmargin=0ex,
  resetmargins=false,
  showstringspaces=false,
  escapeinside={(*@}{@*)},
  morecomment=[l]{--},
  backgroundcolor=\color{white},%
morekeywords={trait,async,spawn,this,string,uint,int,real,bool,var,then,fun,match,let,in,actor,method,self,consume,object,use,def,local}
}

\newcommand{\Hd}[1]{\textsf{#1}}
\newcommand{\ie}{\textit{i.e.,}}
\newcommand{\wrt}{\textit{w.r.t.}}
\newcommand{\eg}{\textit{e.g.,}}

\newcommand{\CK}[2]{\ensuremath{Cap}(#1, #2)}

\newcommand{\erase}[1]{\ensuremath{(#1)^e}}

\newcommand{\ProblemOne}{Safe One-Size-Fits-All Concurrency}
\newcommand{\ProblemTwo}{Balancing Safety, Complexity and Performance}

\newcommand{\RaceUnsafe}{race unsafe}
\newcommand{\Safe}{safe}

\newcommand{\LangName}{Dala}
\newcommand{\LangNameImpl}{Daddala}
\newcommand{\LangMoth}{Daddala}
\newcommand{\LangFormal}{Dalarna}
\newcommand{\etal}{\textit{et al}}

\newcommand{\ec}[1]{\lstinline[style=minigrace,breaklines=true]@#1@}
\renewcommand{\c}[1]{\lstinline[style=micrograce]@#1@}
\newcommand{\kw}[1]{\text{\ec{#1}}}
\lstnewenvironment{code}{\lstset{style=grace}}{}
\lstnewenvironment{codep}[1][numbers=left]{\lstset{style=grace,#1}}{}

\newcommand{\isIso}[1]{\Helper{isIso}{#1}}
\newcommand{\By}[2]{By \RN{#1} the configuration steps to $#2$.}

\newcommand{\ByWithWhere}[4]{By \RN{#1} with (\ref{#2}) the configuration steps to $#3$.
}

\newcommand{\K}{\ensuremath{K}}
\newcommand{\self}{\ensuremath{\kw{self}}}
\newcommand{\move}[2]{\ensuremath{#1 \leftarrow #2}}
\newcommand{\consume}[1]{\ensuremath{\kw{consume}\ #1}}
\newcommand{\recv}[1]{\ensuremath{\ \leftarrow #1}}
\newcommand{\kopy}[2]{\ensuremath{#1\ \kw{copy}\ #2}}
\newcommand{\loc}{\ensuremath{\iota}}
\newcommand{\CErr}{\ensuremath{\textmd{Err}_A}}
\newcommand{\CastErr}{\ensuremath{\textmd{Err}_C}}
\newcommand{\Absent}{\ensuremath{\top}}
\newcommand{\PErr}{\ensuremath{\textmd{Err}_P}}
\newcommand{\err}{\ensuremath{\textmd{Err}_N}}

\newcommand{\OkField}[2]{#1 \leq #2}

\newcommand{\OkRef}[2]{\Helper{OkRef}{H, #1, #2}}
\newcommand{\OkRefEnv}[2]{\Helper{OkRefEnv}{\Gamma, #1, #2}}
\newcommand{\OkDup}[3]{\Helper{OkDup}{#1, #2, #3}}
\newcommand{\OR}{\ensuremath{\hspace*{1ex}|\hspace*{1ex}}}
\newcommand{\block}[1]{\ensuremath{\{ #1 \} }}
\newcommand{\ObjOrChan}[3]{\ensuremath{#1\,#2\,\{ #3 \} }}
\newcommand{\Obj}[3]{\ObjOrChan{#1}{\kw{obj}}{#2\,#3}}
\newcommand{\Chan}[1]{\ObjOrChan{}{\!\kw{chan}}{#1}}
\newcommand{\Many}[1]{\ensuremath{\overline{#1}}}
\newcommand{\Ms}{\Many{M}}
\newcommand{\Fs}{\Many{F}}
\newcommand{\ReducesTo}{\rightsquigarrow}
\newcommand{\Replays}{\rightsquigarrow}

\newcommand{\Blocked}[2]{\ensuremath{\blacksquare_{#1}\,#2}}
\newcommand{\Method}[3]{\ensuremath{\kw{method}~#1(#2)~\block{#3}}}

\newcommand{\Spawn}[2]{\ensuremath{\kw{spawn}~(#1)~\block{#2}}}
\newcommand{\LetIn}[3]{\ensuremath{\kw{let}~#1~=~#2~\kw{in}~#3}}

\newcommand{\CC}[2]{\ensuremath{(#1)~#2}}

\newcommand{\RN}[1]{\textnormal{\textsc{\fontsize{6pt}{12pt}#1}}}
\newcommand{\PN}[1]{\textsf{#1}}
\newcommand{\Helper}[2]{\PN{#1}\ensuremath{(#2)}}

\SetTracking[spacing={25*,166,}]{encoding=*,shape=sc}{25}

\newcommand{\dom}[1]{\ensuremath{\textit{dom}}(#1)}

\newcommand{\isokw}{\ensuremath{\kw{iso}}}
\newcommand{\unkw}{\ensuremath{\kw{unsafe}}}
\newcommand{\immkw}{\ensuremath{\kw{imm}}}
\newcommand{\localkw}{\ensuremath{\kw{local}}}

\newcommand{\Remark}[1]{\hfill\textup{\small#1}}

\newcommand{\ntyperule}[3]{
\begingroup
   \fontsize{8pt}{12pt}\selectfont
  \begin{array}{c}
    \textsc{(\RN{#1})} \\
    #2 \\
    \hline
    \raisebox{-1pt}{$#3$}
  \end{array}
\endgroup
}

\begin{document}

\title{Dala: A Simple Capability-Based Dynamic Language Design For Data Race-Freedom}
\keywords{concurrency, capability, permission, isolation, immutability}



\author[K. Fernandez-Reyes]{Kiko Fernandez-Reyes}
\affiliation{
  \institution{Uppsala University}
  \country{Sweden}
}
\email{kiko.fernandez@it.uu.se}

\author[J. Noble]{James Noble}
\affiliation{
  \institution{Victoria University of Wellington}
  \country{New Zealand}
}
\email{kjx@ecs.vuw.ac.nz}

\author[Isaac O. G.]{Isaac Oscar Gariano}
\affiliation{
  \institution{Victoria University of Wellington}
  \country{New Zealand}
}
\email{isaac@ecs.vuw.ac.nz}

\author[E. Greenwood-Thessman]{Erin Greenwood-Thessman}
\affiliation{
  \institution{Victoria University of Wellington}
  \country{New Zealand}
}
\email{erin.greenwood-thessman@ecs.vuw.ac.nz}

\author[M. Homer]{Michael Homer}
\affiliation{
  \institution{Victoria University of Wellington}
  \country{New Zealand}
}
\email{michael.homer@vuw.ac.nz}

\author[T. Wrigstad]{Tobias Wrigstad}
\affiliation{
  \institution{Uppsala University}
  \country{Sweden}
}
\email{tobias.wrigstad@it.uu.se}

\begin{abstract}
  Dynamic languages like Erlang, Clojure, JavaScript, and E
  adopted data-race freedom by design. To enforce data-race
  freedom, these languages either deep copy objects during actor
  (thread) communication or proxy back to their owning thread. We
  present Dala, a simple programming model that ensures data-race
  freedom while supporting efficient inter-thread communication.
  Dala is a dynamic, concurrent, capability-based language that
  relies on three core capabilities: immutable values can be
  shared freely; isolated mutable objects can be transferred
  between threads but not aliased; local objects can be aliased
  within their owning thread but not dereferenced by other
  threads. Objects with capabilities can co-exist with unsafe
  objects, that are unchecked and may suffer data races, without
  compromising the safety of safe objects. We present a formal
  model of Dala, prove data race-freedom and state and prove a
  dynamic gradual guarantee. These theorems guarantee data
  race-freedom when using safe capabilities and show that the
  addition of capabilities is semantics preserving modulo
  permission and cast errors.
\end{abstract}

\maketitle              
\renewcommand\footnotetextcopyrightpermission[1]{}
\pagestyle{plain}

\section{Introduction}

Most mainstream object-oriented languages do not rule out
\emph{data races} -- read-write or write-write accesses to a
variable by different threads without any interleaving
synchronisation. This makes it hard to reason about the
correctness, or in programming languages like C and C++ even the
meaning of programs.
Static languages such as Java or Go have higher level
constructs to control concurrency but, ultimately, nothing strictly prevents data
races from happening.  Dynamic languages such as Ruby or Python repeat the same
story: nothing prevents data races, not even a global interpreter lock.

Many object-oriented languages added concurrency constructs as an
afterthought, and objects may suffer data races. Data race free languages have implicit concurrent
properties as part of the ``object model'' and guarantee data race-freedom. Examples of
data race free languages are:
E which uses object capabilities and far references to forbid access to (global) and
un-owned resources \cite{E}, capability- or ownership-based languages such as
Pony~\cite{PonyLang,PonyTS} or Rust~\cite{Rust},
or languages without mutable state,
\eg{} Erlang~\cite{Erlang}.

Freedom from data races simplifies avoidance of \emph{race
  conditions}, which happen when behaviour is controlled by
factors outside of the program's control, such as the scheduling
of two threads. Data race free languages thus have a leg up on
``racy'' languages in this respect, but data race freedom always
comes at a cost: languages either deliver ``efficient
concurrency'' or ``simple concurrency'' but not both.

\begin{table*}[t]
  \centering
  \caption{\label{tbl:static capability}Summary of features of capability-based static languages}
  \begin{tabular}{r|c@{~}c@{~}c@{~}c|c@{~}c@{~}c@{~}c|c}
    \toprule
    \Hd{Complexity \textbackslash{}  Languages} & \Hd{Pony} & \Hd{Rust} & \Hd{Encore} & \Hd{RefImm} & \Hd{E}   & \Hd{Newspeak} & \Hd{AmbientTalk} & \Hd{Erlang} & \Hd{Dala} \\
    \midrule
    Capabilities                                & 6         & 5         & 7+          & 4
                                                & \cmark    & \cmark    & \cmark      & \xmark      & 3                                                                     \\
    Capability Subtyping                        & \cmark    & \xmark
                                                & \cmark    & \cmark
                                                & \xmark    & \xmark    & \xmark      & \xmark      & \xmark{}                                                              \\
    Promotion, Recovery, Borrowing              & \cmark    & \cmark
                                                & \cmark    & \cmark
                                                & \xmark    & \xmark    & \xmark      & \xmark      & \xmark{}                                                              \\
    Compositional Capabilities                  & \cmark    & \cmark
                                                & \cmark    & \xmark
                                                & \xmark    & \xmark    & \xmark      & \xmark      & \xmark{}                                                              \\ 
    Deep copying                                & \xmark{}  & \xmark{}  & \xmark{}    & \xmark{}    & \cmark{} & \cmark{}      & \cmark           & \cmark      & \xmark{}
                                                                                                                                                                            \\
    Far References                              & \xmark{}  & \xmark{}  & \xmark{}    & \xmark{}    & \cmark   & \cmark        & \cmark           & \xmark      & \xmark{}  \\
    Data-Race Freedom                           & \cmark{}  & \cmark{}  & \cmark{}    & \cmark{}    & \cmark   & \cmark        & \cmark           & \cmark      & \cmark{}  \\

    \bottomrule
  \end{tabular}
  \vspace*{-1em}
\end{table*}

\cref{tbl:static capability} shows the features of eight
data race free languages. (\LangName{} is our proposal, and we
discuss its features later in the paper.) The first four languages
are statically typed. To maintain data race-freedom, these
introduce new concepts which permeate a system to: ownership,
capabilities, capability composition, capability subtyping,
capability promotion and recovery, and viewpoint
adaption~\cite{ViewpointAdaptation, EncoreTS, PonyTS, Rust,
  DBLP:conf/oopsla/ClarkeD02} (though not necessarily all at
once). This may have lead to a steep learning curve~\cite{RustOS}.
In return, these languages deliver efficient concurrency:
they allow large object graphs to be shared or passed around
safely by pointer, or allow multiple threads to access different
places of a single data structure at once.

The next four languages are dynamically typed. These
maintain data race-freedom implicitly and requires the programmer
to do little or nothing: objects are either copied between
``threads''~\cite{destructive-read} (which, while simplifying
garbage collection, may be expensive \cite{CopyExpensive} and
loses object identity), or proxied back to the ``thread'' that
owns them~\cite{JSproxies, AmbientTalk, RobustComposition,
  CASTEGREN2018130}, which adds latency and makes performance
hard to reason about unless it is clear what operations are
asynchronous.\footnote{E, Erlang, and AmbientTalk are also distributed languages
which motivates their copying and proxied reference approaches.}

The goal of this work is to deliver a design that provides both
simplicity and performance, and that may work both in dynamically
and statically typed programming languages. To this end we present
\LangName{},\footnote{A ``Dala Horse'' 
  is a small 
  carved wooden toy horse --- a handmade little pony \cite{PonyTS,
    MyLittlePony, PonyCulture}).} a capability-based and dynamic
approach to data race-freedom without mandating deep copying and
without the typical complexity of capability systems. The
\LangName{} model allows mixing objects which are guaranteed to be
safe from data-races with objects that are not and thereby
supports the gradual migration of programs to using only ``safe
objects,'' by converting unsafe objects to safe objects one at a time.
\LangName{} uses three object capabilities to maintain data race-freedom:
\underline{imm}utable values that can be shared freely;
\underline{iso}lated mutable objects that cannot be aliased but can be
transferred between threads, and thread-\underline{local} objects that
can be aliased across threads but only dereferenced by the thread that created them.
In this paper, we study our capabilities in a very simple setting:
an untyped object-based language, leaving optional static typing for
future work.

To support the design of \LangName{} we contribute \LangFormal{},
a formal model of the core of \LangName{}. We use this model
to prove that objects with capabilities cannot be subject to data races,
nor can they observe a data race except via explicitly unsafe parameters
passed to their methods. We also show that \LangName{} supports a
form of gradual guarantee~\cite{GradualTyping,RefinedGradualTyping}:
the addition of capabilities preserves the
dynamic semantics, modulo run-time errors that check that the program
behaves according to the programmer's explicitly stated intentions, \eg{}
disallowing a write to an immutable object.
 The \LangName{} model can be
embedded into a wide range of garbage-collected
object-oriented languages, such as Java, TypeScript, Ruby, OCAML, Swift, Scala,
Go etc. To demonstrate that \LangName{} has the potential to be practicable, for
both statically and dynamically typed programs, we have a proof-of-concept
implementation, \LangMoth, which embeds \LangName{} in
Grace~\cite{grace,GraceDialects,GracesInheritance}, built on top of
Moth VM \cite{TransientChecks}.

\subsubsection*{Contributions and Outline}

\begin{enumerate}\raggedright
\item We overview three inherent problems in \RaceUnsafe{} and \Safe{}
  programming languages, and discuss the current approaches
  in ownership- and capability-based systems~(\cref{sec:motivation}).
\item We introduce the \LangName{} capabilities that
  allow safe interaction between programs containing data races
  from parts that should remain data race-free~(\cref{sec:dala}).
\item We show how \LangName{} tackles the three inherent problems in~\cref{sec:motivation}
  (\cref{sec:solutions}).
\item We provide a formal description of \LangName{}
  and its core properties (\cref{sec:formalism,sec:properties}).
\item We provide \LangMoth{}, a proof-of-concept implementation that
  (anecdotally) shows the relative ease of embedding \LangName{} in an existing
  system (\cref{sec:implementation}).
\end{enumerate}

\noindent
\cref{sec:related} places \LangName{} in the
context of related work,
and \cref{sec:conclusion} concludes.

\section{Background: Perils of Concurrent Programming}
\label{sec:motivation}

To set the scene for this paper we discuss common problems in \RaceUnsafe{} (\eg{}
Java) and \Safe{} concurrent languages (\eg{} Rust). 
%
First we discuss balancing complexity and performance
(\cref{sec:complex-simple dichotomy});
then how tying safety to particular concurrency abstractions leads to a
one-size-fits-all model which leads to problems with compositionality of
concurrent abstractions (\cref{sec:unsafe concurrency});
last we discuss the problem of providing escape hatches to permit behaviour
that is not supported by the programming language
(\cref{sec:undefined}).




\subsection{\ProblemTwo{}}
\label{sec:complex-simple dichotomy}

Safe languages have (implicit) concurrency mechanisms to
prevent data races. From \cref{tbl:static capability}, we argue
that the constructs or systems fall into three categories:
\emph{complex and efficient}, \emph{simple and inefficient}, and
\emph{complex and inefficient}.

\subparagraph{Complex and Efficient}

Pony, Rust, Encore as well as Gordon's work on reference
immutability are race-safe by controlling access to shared data,
rather than banning its
existence~\cite{PonyTS,PonyLang,Rust,SFMEncore,Gordon}. This is
achieved by providing concepts that are not common to every day
developers, such as capability and ownership type systems. These
allow both \emph{efficient} and \emph{data race-safe} sharing and
transfer of ownership with reference semantics. This allows \eg{}
passing a large data structure across actors by reference, which
subsequently allows fast synchronous access by the receiving
actor. Implementing a concurrent hashmap in these systems requires
up-front thinking about how keys and values may be accessed across
different threads, and mapping the intended semantics onto the
types/capabilities that the languages provide. For example, in
Pony, a concurrent hashmap will be an actor; to allow multiple
threads to know the existence of the same keys, the keys must be
immutable; etc.

The cost of safe efficient data sharing is complexity: capability
type systems introduce complex semantics such as capability
promotion, capability subtyping, capability recoverability (\eg{}
getting back a linear reference after it was shared),
compositional capability reasoning (\ie{} combining capabilities
to produce new capabilities), and viewpoint adaption (\ie{} how to
view an object from another object's perspective, \eg{} Pony uses
viewpoint adaption to write parametric
polymorphism~\cite{PonyGenerics}). Understanding these concepts is
key to write code that is efficient and free from data
races.

Some form of unique/affine/linear reference is often the
cornerstone of many of the static capability systems. Such
references are extremely powerful: they provide reasoning power,
they can often be converted into other capabilities (\eg{} to
create cyclic immutable data structures) or to transfer ownership
of objects across threads. However, polymorphic behaviour is
typically a source of great pain for these systems. For example,
consider a simple hashmap -- concurrent or not. If values can be
unique, a lookup must remove the value out of the data structure
to preserve uniqueness (and the associated entry to reflect this
in the hashmap). If values are not unique, this behaviour is
counter-productive. Behaviours like this force duplication of
code.





\subparagraph{Simple and Inefficient}

ASP~\cite{ASP}, ProActive~\cite{ProActive}, Erlang~\cite{Erlang}, E~\cite{E}, AmbientTalk~\cite{AmbientTalk1,
  AmbientTalk2}, Newspeak~\cite{Newspeak2010, bracha2017newspeak},
functional objects in ABS~\cite{ABS}, among others, avoid data
races by deep copying objects in messages (for some languages
modulo far references, see \emph{Complex and Inefficient}). This
approach is relatively simple and the price for data race-freedom
is copying overhead paid \emph{on every message send}.

In addition to consuming CPU cycles and increasing the memory
pressure, deep copying also loses object identity, and requires
traversal of the objects transferred, a $O(\#\mathit{objects})$
operation, possibly requiring auxilliary data structures (adding
overhead) to preserve internal aliasing of the copied structure.

Notably, Erlang requires two such traversals: the first
calculates the size of all objects to enlarge the receiving
process' heap and the second copies them across. Erlang's
deep copying is key to keeping process' heaps disjoint, which
simplifies concurrent garbage collection and reduce overall system
latency. In languages that compile to Java, like ABS, ASP,
ProActive, etc., the run-time is unable to see or leverage such
isolation.

A concurrent hashmap in Erlang has no choice: its keys and values
must be immutable, and are therefore safe to share across multiple
processes. The internal hashmap data structures must be immutable too,
which might be less efficient.\footnote{Erlang Term Storage
  provides a way to escape this design of Erlang, but at a cost of
  dropping to a much lower-level of programming and manual memory
  management.}

\subparagraph{Complex and Inefficient}

To support safe sharing without losing object identity,
E~\cite{E}, AmbientTalk~\cite{AmbientTalk1, AmbientTalk2},
and Newspeak~\cite{Newspeak2010, bracha2017newspeak} support far
references. A far reference is a proxy that allows an object owned by an actor/process
to directly reference another's innards, but all interaction with
the proxy is lifted into an asynchronous message and sent back to
the owner, to be executed there. Thus, despite the fact that many
actors can point directly to an object, only its owning process
will ever read or write the object. Thus, with proxies, there is a
cost \emph{per access} which can be expensive~\cite{CopyExpensive}.

%
In E and AmbientTalk, proxied (\emph{far}) references must be
operated on asynchronously and only non-proxied (\emph{local})
references allow synchronous access. Code that needs to be
``proxy-agnostic'' must use asynchronous access. In E and
AmbientTalk, promises are also implicit proxied references.%
\footnote{E calls them promises, AmbientTalk calls them
  futures. We will refer to them as
  promises~\cite{Promise-LiskovS88}.} This means that asynchronous sends (\c{x<-})
may be delayed indefinitely if the promise is never fulfilled,
meaning the \c{x}'s value is never materialised so there is no
recipient of the message.
%


Implementing a concurrent hashmap in AmbientTalk (E and Newspeak
are similar) does not need any capability annotations to track how
values stored in the map may be shared across threads. Promises
remove the need for callbacks, but indirections make the
implementation more complex or convoluted. As a blocking synchronisation on the
result of a promise is not possible, we must use the promise
chaining operator to access the (possible) value in the promise.




While syntactically simple, the inability to access
resources directly makes reasoning about performance hard. The main implication of not allowing
direct access is that we neither know if far references are ever going to be
fulfilled nor if the promise chaining combinator (on far references) makes
the owning actor (of the promise) the bottleneck of the system.

\subsection{\ProblemOne}
\label{sec:unsafe concurrency}



In \RaceUnsafe{} languages, abstractions are broken by the addition of concurrency
constructs. For example, inconsistencies in an object's internal state \emph{during a method's execution}, which are hidden in a purely sequential system, may be \emph{observed} if the object can be accessed concurrently.  This problem of object-oriented languages and their ``unsafe''
concurrency features was first studied by M. Papathomas~\cite{OO-Conc-Issues-papathomas89}.
Yet, there are plenty of abstractions and programming models that guarantee data-race
freedom, such as implementations of the actor
model~\cite{ActorSurvey, Akka, CAF}. In languages like
\textit{Akka}~\cite{Akka}, the model is data race-free as long as all code in a system
adheres to a set of guidelines~\cite{AkkaDocs}.

These guidelines tie concurrency safety properties to their concurrency model,
suffering from ``one-size-fits-all'' problem. For example, Akka can only
guarantee data race-freedom when the program (follows the guidelines and) stays
within the actor model. Spawning threads within an actor can easily break its
concurrency safety properties, \eg{} data race-freedom.

Thus, to implement a concurrent hashmap, an Akka program might
simply wrap the standard Java hashmap in an actor. As long as the
hashmap itself is never leaked from the actor, and the actor does
not create additional threads, all updates the hashmap will be
sequential and therefore free from data races. Whether the keys
and values are safe from data races is beyond the control of the
hashmap, and is ultimately up to the diligence of the programmers
(\eg{} to maintain uniqueness or immutability).

In E, AmbientTalk, and Newspeak, (as well as many languages that did not fit in \cref{tbl:static capability} such as ABS, Pony, Proactive and others), the mechanisms that guarantee data race-freedom are inherently linked to the
languages' chosen concurrency models. In this case, the near
and far references span (affect) other concurrency models. Using the example
from before, global objects can be safely protected by locks 
but this cannot be easily
accommodated\footnote{This can be partially mitigated by adding futures, which imposes again
other concurrent semantics to maintain the safety properties~\cite{AmbientTalkJavaSymbiosis-CutsemMM09}}
and the run-time may throw an exception when multiple threads have access
to a shared object~\cite{AmbientTalkJavaSymbiosis-CutsemMM09}.

\subsection{Safety May Beget Unsafety}
\label{sec:undefined}

In \Safe{} languages based on capabilities or ownership
types (\eg{}~\cite{SFMEncore, PonyLang, PonyTS, Rust}) it is
sometimes necessary to side-step the type checker, to write
low-level code that interacts with hardware, or when the type
system is not ``clever enough'' to allow a correct behaviour. We
exemplify these cases in turn.

\subparagraph{Hardware Meets Software.}
Graphic cards constantly read from video memory and developers can write directly
on the frame buffer that points to the video memory to update the image in the
next refreshing cycle. Implicitly, this means that there is a data race between
the graphics card and the main thread; such racing behaviour may show
flickering of the image on screen.
A double buffering technique removes this flickering,
using an on-screen buffer that the graphics card reads, and
one off-screen where developers write the next scene. A swap operation
swaps the buffer pointers from off- to on-screen. This common -- and racy --
approach prevents the flickering.

\subparagraph{Trust Me, I Know What I Am Doing.}
As all statically typed languages, capability- and ownership-based
programming languages are engaged in a balancing act of
expressivity and complexity. Simplifications made to the systems
to reduce the programmer's overhead will invariably lead to
exclusion of valid programs, simply because the system is not
powerful enough to express its behaviour within its model. 

We can illustrate this using the concurrent hashmap example in the
context of the Encore programming language. Encore suffers from
the problem mentioned in \cref{sec:complex-simple dichotomy} where
a hashmap must choose between supporting unique references (and
therefore always moving values in an out), or not (allowing them
to be in the hashmap at the same time as they are referenced
elsewhere). This can be solved in Encore by dropping to the
underlying language to which Encore compiles where no such checks
are made, and implement, for example, parallel put and get methods
that accepts linear values even though this is unsound in the type
system, and transfers them (hopefully) correctly. Other languages
(\eg{} Rust) may have an unsafe block, or provide reflective
constructs that are unsafe from a capability perspective. Naturally,
all such code gives rise to technical debt.



In order to circumvent shortcomings of the capability systems, programmers
may resort to escape hatches that void the guarantees of data-race freedom.
If data races happen inside an escape hatch, the
behaviour is undefined~\cite{klabnik2019rust, Orca}. In Pony (and Encore), if an unsafe block
introduces a data race the run-time (garbage collector) might eventually
crash~\cite{Orca}. There is no safe \emph{interoperability} between unsafe and
safe code!

\subsection{Summary}
The data race-freedom guarantee of safe languages
comes at a cost of complexity or inefficiency, or both.
Furthermore, most or all languages' safety is tied to a specific concurrency model,
and may not compose with others.
Finally, data races are sometimes desired, or a systems' notion of safe is too safe
to express correct code. Escape hatches overcome these problems, but at a cost of
losing data race-freedom.

In the next section, we describe our simple model that is
efficient, concurrency-model agnostic, and provides safe
interoperability between unsafe and safe code.

\section{An Overview of The \LangName{} Model}
\label{sec:dala}


\newcommand{\allow}{$\bullet$}
\newcommand{\deny}{$\times$}

\begin{figure*}[t]
  \centering\footnotesize
  \begin{minipage}{0.295\linewidth}
    \includegraphics[width=\linewidth]{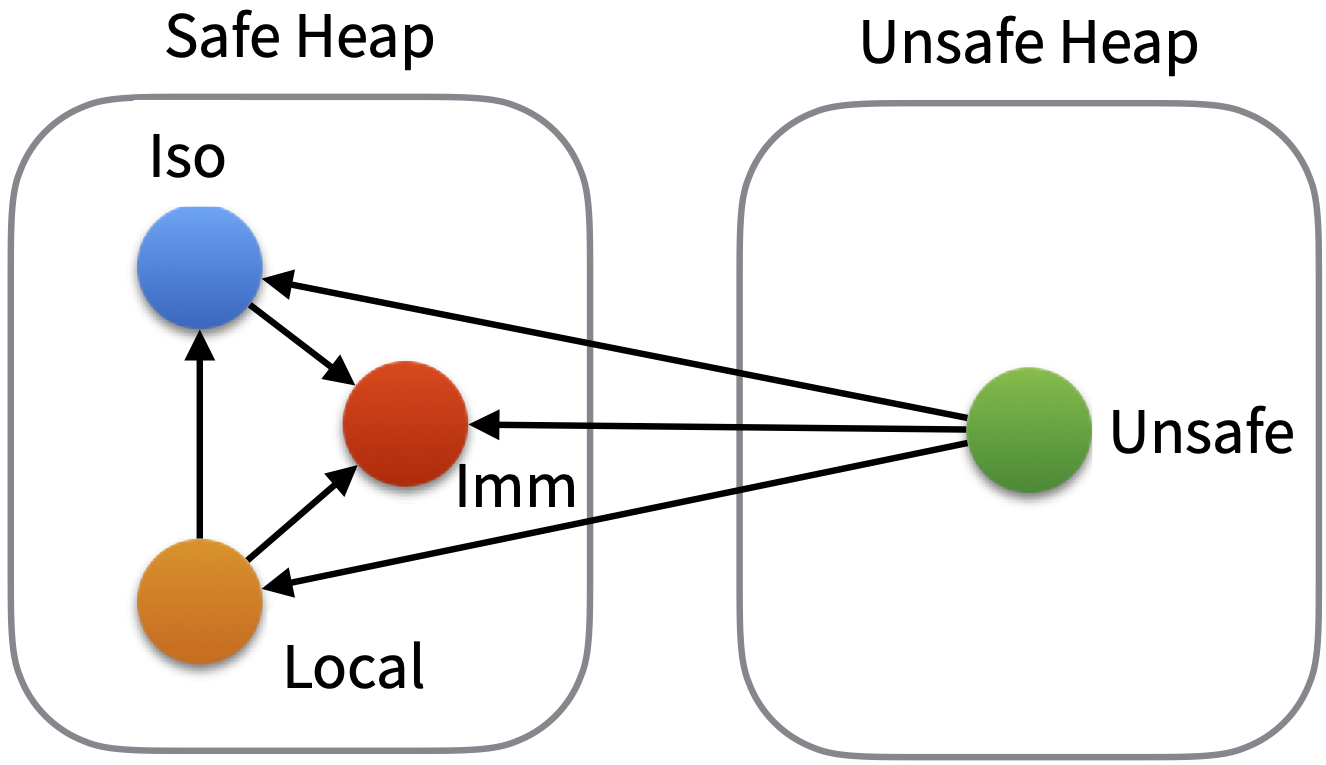}
    \caption{Dala Heap.}
    \label{fig:structural-restriction}
  \end{minipage}
  \hspace{.1cm}
  \begin{minipage}{0.34\linewidth}
  \begin{tabular}{@{}r@{\quad}|@{\quad}c@{~}c@{~}c@{~}c@{}}
    \toprule
                & \multicolumn{4}{c}{\Hd{Field contents}}            \\
    \Hd{Object} & \Hd{Imm} & \Hd{Iso}  & \Hd{Local}   & \Hd{Unsafe}  \\
    \midrule
    Immutable   & \allow   & \deny     & \deny        & \deny        \\
    Isolated    & \allow   & \allow    & \deny        & \deny        \\
    Local       & \allow   & \allow    & \allow       & \deny        \\
    Unsafe      & \allow   & \allow    & \allow       & \allow       \\
    \bottomrule
  \end{tabular}
  \caption{Structural restrictions.}
  \label{tab:deep-capabilities}
  \end{minipage}
  \begin{minipage}{0.34\linewidth}
  \begin{tabular}{@{}r@{\quad}|@{\quad}c@{~}c@{~}c@{~}c@{}}
    \toprule
                    & \multicolumn{4}{c}{\Hd{Effects}}                \\
    \Hd{Object}   & \Hd{Read}    & \Hd{Write} & \Hd{Alias}  & \Hd{Transfer}   \\
    \midrule
    Immutable     & \allow       & \deny      & \allow      & \allow          \\
    Isolated      & \allow       & \allow     & \deny       & \allow          \\
    Local         & \allow       & \allow     & \allow      & \deny           \\
    Unsafe        & \allow       & \allow     & \allow      & \allow          \\
    \bottomrule
  \end{tabular}
  \caption{Capabilities and effects.}
  \label{tab:permissions-and-capabilities}
\end{minipage}
\vspace*{-1em}
  \end{figure*}

Objects in \LangName{} are associated with capabilities that describe
how they can interact with other objects. The association
happens at creation time and is fixed for life.
We refer to \emph{\underline{imm}utable}, \emph{\underline{iso}lated},
and \emph{thread \underline{local}} as the ``safe capabilities''.
Programs which only operate on a safe heap are guaranteed to be
data-race free.
When it is not important to distinguish between an object and its
capability, we will say \eg{} ``a local object'' to mean an object
with a local capability or a ``safe object'' etc.

The following code creates an immutable object with a field $f$
with value $v$ and a unary method $m$ with body $t$: \c{object \{
  use imm;} \c{def f = v; method m(x)\{t\} \}}.
\cref{fig:structural-restriction} shows the interaction of
capabilities: arrows show all legal references from an object with
one capability to another, modulo reflexivity.

The \LangName{} capabilities form a hierarchy;
imm $<$ iso $<$ local $<$ unsafe. Objects can only refer
to other objects with the same or lesser capabilities: an immutable
object can only refer to other immutable objects; an isolate
object can refer to other isos or imms, and a local object
can refer to imms, isos, and other local objects.
The model treats objects outside \LangName{}
as having a fourth unrestricted \emph{\underline{unsafe}} capability for uniformity.
Using
unsafe objects in a \LangName{} program makes it susceptible to data
races. Because it is the top capability, objects in the safe heap (imms,
isos, locals) cannot refer to objects in the unsafe heap, while
unsafe objects can refer to anything (see
\cref{tab:deep-capabilities}). The implication of this model is
that once an object is safe, its entire reachable object graph is
safe as well. \emph{Thus: data races can only occur in unsafe objects.}
By including unsafe objects in the model, we can describe the
semantics of a program partially annotated with \LangName{} capabilities.

Avoiding data-races by ensuring that two threads do not
concurrently execute a code block that accesses the field $f$ of
some object $o$ at the same time focuses on \emph{code}. If there
is another place in the program that also accesses $o$'s $f$ field
which could be run at the same time, a data race could still
happen. Making the \emph{object} safe means that all \emph{code}
that interacts with $o$ must follow the rules that make $o$ safe
from data races. If $o$ is local, it cannot be shared across
threads, so all accesses to its $f$ field will come from the same
thread. If $o$ is immutable, all accesses to $f$ will be read
accesses which are benign. If $o$ is isolated, two accesses to its
$f$ field by different threads require an explicit transfer from
the first thread to the second thread (possibly via additional
``stop-overs'').


\LangName{} capabilities are self-protecting in the sense that
safety stems from a capability's own internal restrictions, not
from restrictions elsewhere in the system. Writing to a field of
an imm throws an error; so does aliasing an iso\footnote{This property can be implemented by ensuring that all code external to an iso accesses it via a proxy object with movement semantics, \eg{} by overloading the \c{=} operator. A simple implementation is possibly using linear proxies and allows unsafe objects to race on iso stack variables (as they already can on fields in Dala).}, or accessing a
local object from outside of the thread it is local to. Imagine that objects' fields
are private, and that field accesses implicitly go through a getter/setter indirection -- in this
case \emph{all the checks necessary for an object to maintain its invariants are in its own
internal code}. This is an important part of the design and key to adding unsafe
capabilities in cases where these cannot be expected to be cooperative in
avoiding data races. \cref{tab:permissions-and-capabilities} shows
how capabilities restrict certain effects in the system. Due to the absence of static types, these
restrictions are enforced at run-time, meaning the cost of data-race freedom is
(roughly) \emph{per access}.


\LangName{}'s isolated objects are the key to efficient transfer of mutable state
between different threads.  (Immutable objects can be shared directly without causing
problems, and thread local objects are permanently confined within their owning
thread.)  Isolates have only one unique incoming reference, and some extra care
must be taken to preserve this uniqueness.  \LangName{} incorporates
an explicit \emph{consume} operation that destructively reads \cite{Islands} the
contents of a variable, and prevents those contents from being used again, similar to
C++'s ``move semantics''.  The contents of any mutable variable may be consumed
but variables containing isolates \emph{must} be consumed---otherwise the
attempt to read the isolate will fail. Experiences by Gordon et al.
\cite{Gordon} suggest explicit consume is to be favoured over implicit

The consume operator succeeds on all variables except the special
\c{self} variable.  This makes an isolated self variable effectively
\emph{borrowed}~\cite{Boyland01}, meaning its value is tied to the current stack frame.
Because the value of self cannot escape, expressions such as \c{x.m(y)} when \c{x} is
an isolated object do not need to consume \c{x} and are guaranteed
not to introduce any alias to \c{x}, other than through \c{self} on subsequent stack frames.
With this borrowing-like behaviour, it is possible to traverse
isolated structures without consuming them (a well-known problem
when dealing with unique or linear values), but notably only using
\emph{internal} methods: \c{x.m} is allowed to borrow whereas
\c{m(x)} is not. Since overcoming this limitation is well-known, we
refrain from discussing this any further.

\LangName{} guarantees that data races can only happen in places with unsafe
objects, never in objects with safe capabilities (\cref{th: drf}: \emph{Data Race
  Freedom}).  Developers can easily add capabilities to migrate from a ``racy''
program to a data race-free program with the certainty that this migration is
semantics preserving, modulo permission and cast errors (\cref{th: dyn gradual}:
\emph{Dynamic Gradual Guarantee}).  These two properties are key in \LangName{}
and we will formally state them in \cref{sec:properties}.


\subsection{Simple Case Study: A Concurrent Hash Map}

To illustrate the simplicity of the Dala capabilities, consider
the implementation of a simple concurrent hashmap in
\cref{fig:map}. Assume that the hashmap is created inside a
lightweight process, eventually calling the tail-recursive
\c{run()} method with the channel \c{msgs}. The \c{run()}
method reacts to input sent on the channel. It dispatches on the
\c{op} field of messages received, and additionally expects the
fields \c{key} and \c{val} to be present depending on operation.

A hashmap has five moving pieces: the hashmap object itself, the
array of buckets, the entries in the buckets, and the keys and
values of the entries. We examine the possible capabilities of
these in turn to ensure thread-safety of the hashmap
implementation.

\begin{figure*}
  \begin{minipage}{0.6\linewidth}
\begin{lstlisting}[style=micrograce,numbers=left]
map = object { // hashmap 
  use local 

  method run(msgs) {
    msg <- msgs 
    if (msg.op == "done") return "done"
    k = msg.key.freeze(); c = msg.reply
    if (msg.op == "get") c <- get(k) 
    if (msg.op == "put") c <- put(k, msg.val = null)
    if (msg.op == "update") { 
      get(k); c <- put(k, msg.val = null); }
    run(msgs)
  }
  method get(key) {
    link = buckets.get(key.hash() % buckets.size)
    while (link.next && link.next.key != key) {
      link = link.next
    }
    if (link.next != null) {
      target = link.next = link.next.next // unlink
      return target.val = null // return result
    } else {
      return "Failure: No such key"
    }
  }
\end{lstlisting}
  \end{minipage}
  \begin{minipage}{0.37\linewidth}
\begin{lstlisting}[style=micrograce,numbers=left,firstnumber=26]
  method put(key, val) {
    link = buckets
      .get(key.hash() % buckets.size)
    while (link.next && 
           link.next.key != key) {
      link = link.next
    }
    if (link.next != null) {
      return link.val = consume val
    } else {
      link.next = object { // link in bucket
        use local 

        key = key
        val = consume val 
        next = null
      }
      return "Success"
    }
  }

  buckets = object { // array implementation 
    use local ...
  }
}
\end{lstlisting}
  \end{minipage}
  \caption{Simple concurrent hashmap using the Dala capabilities.
    Note that \c{x.f = new} returns the old value of \c{x.f} and
    is used to move isos in and out of the heap.}
  \label{fig:map}
\end{figure*}

For keys and values, there are two possibilities: iso and imm. In
the former case, the keys and values can be transferred between
the hashmap and its clients. In the latter case, they can be
shared but also never change. It makes sense for keys to be
immutable so that any client thread can know of their existence
and ask for their associated value. We capture this in the code on
Line 7 by calling the built-in \c{freeze()} operator that is the
identity function on immutables, or creates an immutable copy
otherwise (like in Ruby).

For values, iso and imm make sense under different circumstances.
A sensible hashmap implementation should store only a single
reference to its values, so the same code base should be reusable
in both scenarios. The code in \cref{fig:map} support values which are
both iso and imm. The key lines are 9, 11, 21, 34, and 40 which
always \emph{move} a value. Our field assignment makes use of
``swap semantics'' where the old value of the field is returned on
an update. As a result, a get operation will move the value
associated with a key out of the hashmap and remove the
corresponding entry (Lines 20--21). (But see \cref{sec:supp-both-immut}!)

From a thread-safety perspective, linked entries that constitute a
bucket could be either immutable, iso, or local. Immutable entries
would complicate the code when entries are unlinked (Line 20). Iso
entries would slightly complicate searching through linked entries
while maintaining uniqueness (lines 16--18 and 29--32). Thus, our
entries are local (Line 37).

A similar thought process applies to the array of buckets.
However, as our entries are local, we have no choice but to make
the array local as well. For simplicity, the implementation of the
array is elided above but we include Line 48 to explicitly show
this choice. (As suggested by lines 15--16 and 27--29, our
implementation assumes an empty bucket has a dummy entry for
simplicity.)

Consequently, the hashmap itself must be local (Line 2). This ties
the hashmap to the thread or lightweight process where it was
created which does not seem unreasonable.

\subsubsection{Gradual Transition to Dala Capabilities}

Notice the bottom-up thinking when assigning the capabilities in
the previous section. In a top-down approach, we might have
decided to make the hashmap an iso to support its movement across
threads or processes. This would have excluded local from the
possible capabilities for the buckets array and its entries. When
retrofitting existing code to use Dala capabilities, the bottom-up
approach is superior to the top-down approach as it will incur the
smallest possible change to the program. Assume the code in
\cref{fig:map} was written as now, but without capabilities in
mind: including no \c{use} declarations on Lines 2, 37 and 48, no
consumes on Lines 34 and 40, and no \c{freeze()} on Line 39, etc.
In this case, changing the keys to be immutable is simply adding
the \c{freeze()} call on Line 39. All other objects in the
hashmap can remain as-is. Similar, making values isos only needs
the two \c{consume} operations, and changing \c{msg.val} to
\c{msg.val = null} on lines 9 and 11.

Because of the structural constraints imposed on objects with a
capability, top-down migration is likely to require bigger
changes. For example, adding the \c{use local} on Line 2 will
require adding a capability to the bucket array on Line 47--49.
This will propagate to the entries and their keys and values.

A good regression test suite might be useful to help drive adding
annotations regardless of the approach.

\subsubsection{Constructing and Using the Hashmap}

While orthogonal to the Dala capabilities, \cref{fig:client} illustrates
how a program might construct a hashmap (lines 1--8) by spawning a
new process with an associated channel that creates the hashmap
and connects it to the channel.

When calling the constructor (Line 9), the caller gets a channel
that can be used to send messages to the hashmap. It is easy to
construct a proxy for a map that captures the channel used to
communicate with the map, and uses a dedicated channel to get the
reply (lines 11--24). The current \c{map_proxy} creates a new
channel per interaction with the concurrent hashmap. This allows
it to be immutable. If the creation of \c{map_ch} was moved outside
of the object, a local proxy per client would make more sense. This
change would be captured by changing Line 12 to \c{use local}.
(Line 18 is just defensive programming.)

\begin{figure*}[!h]
  \begin{minipage}{0.37\linewidth}
\begin{lstlisting}[style=micrograce,numbers=left]
method new_hashmap() {
  return spawn (msgs) {
    map = object {
      // Code from Fig 4
    }
    map.run(msgs)
  } // returns channel
}
\end{lstlisting}
  \end{minipage}
  \begin{minipage}{0.36\linewidth}
\begin{lstlisting}[style=micrograce,numbers=left,firstnumber=9]
map = new_hashmap()

map_proxy = object {
  use imm

  method put(key, val) {
    map_ch = ... // channel
    map <- object {
\end{lstlisting}
  \end{minipage}
  \begin{minipage}{0.26\linewidth}
\begin{lstlisting}[style=micrograce,numbers=left,firstnumber=17]
      use iso
      key = key
      val = consume val
      reply = map_ch
    }
    return <- map_ch
  }
}
\end{lstlisting}
  \end{minipage}
  \caption{Constructing and using the hashmap in \cref{fig:map}}
  \label{fig:client}
\end{figure*}

With the code in \cref{fig:client} in place, \c{map_proxy.put(k, v)} will
asynchronously communicate with the concurrent hashmap even though
it looks like a synchronous operation.

The message object on Lines 16--21 is the first example of
creating an iso. Because isos can only hold other isos or
immutables, \c{key} must be immutable. If it was local, it would
err due to the structural constraints. If if was an iso, it would
err due to the lack of a \c{consume}. As a consequence of the
\c{use iso} on Line 17, there is thus a guarantee that any call
to \c{map_proxy.put(k, v)} that would share mutable state across
threads would throw an error.

\subsubsection{Dala Properties and Concurrency Models}

Recall that the Dala capabilities are a set of rules for
constructing objects (structural constraints) that guarantee that
-- with the exception of explicitly unsafe objects -- programs are
safe from data-races. Unless a reference to an unsafe object is
passed in as an argument, a method inside a safe object
cannot see an unsafe object. Furthermore, while unsafe objects can
store references to safe objects, they cannot violate their
properties. Assume for example that an unsafe object $u$ is shared
across multiple threads and that the safe object $s$ is stored in
a field $f$ of $u$: $u.f = s$.

\begin{enumerate}
\item When $s$ is an imm, it cannot be subject to data-races
  because it cannot be updated. If all field accesses are required
  to go via setters, this can be implemented by having all setters
  throw an error on use.

\item When $s$ is an iso, it cannot be subject to data-races
  because iso's can only be dereferenced on the stack (\ie{} we
  allow $f.g$ but not $x.f.g$ when $f$ contains an iso). Thus, any
  thread wanting to do $u.f.g$ must first transfer the contents of
  $f$ to its local stack where it is unreachable from all other
  threads. This means that multiple threads can race on the
  $f$ field of the unsafe object, which is not a race on $s$. The simplest
  implementation of iso's ensure that variables containing iso's
  are destructively read. (This however requires that unsafe
  objects cooperate in preserving the properties of isos.)

\item When $s$ is a local, it cannot be subject to data-races
  because it can only be dereferenced by its creating thread. All
  other attempts to dereference throw an error. Passing the local
  around freely and using its identity is allowed, but this is not
  a data-race. Forbidding dereferences can be implemented by recording the identity
  of the creating thread and checking it against the identity of
  the current thread at the beginning of each method/getter/setter.
\end{enumerate}

(In addition to the checks mentioned above, all setters must check
that the structural constraints are satisfied, informally:
$OK(o.f = o')$ if
$\mathit{capability}(o') \leq{} \mathit{capability}(o)$. Getters
must also throw an error if used to retrieve iso's.)

In our concurrent hashmap example, we used an Erlang-esque model
with lightweight processes and channels for communication between
processes. Dala capabilities are not inherently tied to a specific
concurrency model. For example, in a language with support for
actors/active objects, the hashmap in \cref{fig:map} might have
been an actor, and \c{run()} might be replaced by different
methods called asynchronously. In this case, the process to which
the bucket array and its entries are local would be to the
implicit thread of the active
object. 

\section{How the Dala Model Addresses the Concurrent Problems in \cref{sec:motivation}}
\label{sec:solutions}

We now revisit the problems in \cref{sec:motivation} and show how \LangName{}
addresses these problems.

\subsection{\ProblemTwo{}}

First, programs with only safe objects are safe from data races.
Allowing parts of a program to be unsafe is useful for several reasons: transition to guaranteed
safety can happen incrementally without a full-blown rewrite; there may be elements in a program's
surroundings that are unsafe but that we still may need to access.

With \LangName{}, we set out to deliver both simplicity\footnote{The design is both simple and simplistic.
  Additional expressivity might be gained for example by adding a
  notion of ownership, or borrowing. How to compare complexity of
  different capability systems is not clear. For example, let us
  briefly compare the number of rules and concepts in formalisms (on purpose in a footnote).
  Dala: 32 run-time rules (omitting helper predicates); 30
  well-formed rules. Total: 62~rules.
  Encore's type system \cite{EncoreTS}:
  73 rules for well-formed declaration and configurations, environment, type equivalence and expression typing;
  40 reduction rules.
  Total: 113~rules.
  Pony \cite{PonyGenerics}:
3 Table/Matrix with viewpoint-adaptation matrix, safe-to-write and capability constraints;
7 definitions for Restricted syntaxes of types and bounds;
16 reduction rules;
10 typing rules;
13 rules for reduction of types and bounds with a partial reification;
38 rules for inheritance, nominal, structural and bound inheritance,
  capability and reified subtyping, bound compliance, sub-bound compliance, method subtyping;
23 rules for safe-to-write, sendable types, reduction of types and translation of expressions
Total: 110 rules.}  and
efficiency.\footnote{See \cref{sec:future} for a discussion on efficiency.} \LangName{} has only three
capabilities (not counting unsafe). Our design avoids complexities
like capability subtyping, promotion and recovery. As will be exemplified in \cref{sec:supp-both-immut},
dynamic checking enables flexible programs without relying on concepts like
capability subtyping. Our
capabilities are also orthogonal to concepts like deep copying,
and far references (summary \cref{tbl:static capability}) and
inter-thread communication can be efficient as \LangName{} avoids deep copying.

\subsection{\ProblemOne{}}
\label{sec: solution unsafe compositional}

Adding the \LangName{} capabilities to an unsafe language
does not change its existing semantics or the fact that data races can (still) happen (in unsafe objects).
Capabilities prevent the introduction of data races in the safe heap,
empowering the objects with \Safe{} semantics.

While our formal model in \cref{sec:formalism} uses channels, this
choice is ``unimportant'' and was driven by the desire to reduce
the complexity of the formal model. (An earlier draft of this paper
used actors, but this introduced unnecessary complexity.) The
\LangName{} model can be applied to other concurrent models, \eg{}
actor- or lock-based concurrency models, and combinations.

For example, we could allow far references to local objects. With
this design, a client of the hashmap in \cref{fig:map} could call
\c{get} and \c{put} etc. directly. This works well with our
design with keys are values being isolated or immutable. If Dala
capabilities are used in an actor-based system, the story would be
similar, and local objects would constitute an actor's private
(mutable) state, and isolated objects enable efficient transfer of
arguments in message sends. If locks are available, they could be
useful for operating on unsafe objects, which might potentially be
shared across threads.

\subsection{Safety Begetting Unsafety}

\LangName{} introduces safe interoperation between unsafe and data race-free
fragments.  From the Data Race-Free Theorem (\cref{th: drf}) and Progress and
Preservation (\cref{th: progress dyn,th: preserv dyn}), unsafe objects may be
involved in data races but do not produce undefined (untrapped) behaviour.
\emph{More importantly, unsafe objects cannot create errors that are observable
in the safe heap.}

An implementation of a racy double-buffering can use unsafe
capabilities without compromising the safety of any safe
capabilities. However, with the current rules, unsafe capabilities
cannot \emph{read} local capabilities, which may lead to the
programmer wanting to propagate the unsafe capability annotation
through the system. A slightly weaker version of our system would
allow unsafe capabilities to \emph{read} fields of local
capabilities freely. Since local objects cannot contain unsafe
objects, any data race due to this weakening is not visible to the
local object, which voids the need for unsafe propagation.

\subsubsection{Supporting both Immutable and Isolated Values in a Hashmap}
\label{sec:supp-both-immut}

Because of the inherent flexibility of dynamic checking, it is
simple to add an additional operation that behaves like \c{get},
but returns an \emph{alias} to the value in the hashmap of
\cref{fig:map}. This operation would throw an error if used on
isos, but work fine on immutable objects. We can even implement it
by extending the existing \c{get} method:

\begin{lstlisting}[style=micrograce]
method get(key, move) {
  // additonal parameter controls movement semantics
  // lines 15-19
  if (move) { /* lines 20-21 */ } 
  else { return link.next.val; /* create alias */ }
  // lines 22-24
}
\end{lstlisting}
This overcomes the problem pointed out previously forcing
developers to choose one particular semantics or duplicate code, and
does not compromise soundness.

\begin{figure}[t]
 \begin{tabular}{@{}r@{\quad}c@{\quad}c@{\quad}l}
 \textit{Programs}       & \textit{P} & $::=$ & $t$                                                          \\
 \textit{Fields}         & \textit{F} & $::=$ & $f=w$                                                       \\
 \textit{Methods}        & \textit{M} & $::=$ & $\Method{m}{x}{t}$                                           \\
 \textit{Terms}          & $t$        & $::=$ & $w \OR \LetIn{x}{e}{t}$                         \\
    \textit{Expressions} & $e$        & $::=$ & $w \OR x.f \OR x.f = w
                         \OR x.m(w)$ \\
                         &
                         $\OR$ & $::=$ &  $\move{x}{w} \OR \recv{x}
                         \OR \Spawn{x}{t} \OR \blacksquare_i\,\loc$ \\
                         &
                         $\OR$ & $::=$ &  $\kopy{K}{x} \OR \Obj{K}{\Fs}{\Ms} \OR \CC{K}{w} \OR v$  \\
 \textit{Variables}      & $w$        & $::=$ & $x \OR \consume{x}$                             \\
 \textit{Values}         & $v$        & $::=$ & $\loc \OR \varnothing \OR \Absent$ \\
 \textit{Capabilities}   & $K$     & $::=$ & $\immkw{} \OR\ \localkw{} \OR \isokw
                         \OR \unkw{}$                                     \\
  \end{tabular}
\caption{\label{fig: permission language} Syntax of \LangFormal{}. $m$, $f$, and
  $x$ are meta-variables representing method, fields, and variable names;
  metavariable $x$ includes \self{}. $\varnothing$ is a run-time value
  representing an empty channel.
}
\end{figure}

\section{Formalising the Dala Capability and Concurrency Models}
\label{sec:formalism}

\noindent
To formally study the \LangName{} model and clearly state and prove its key
properties (\textit{Data Race-Freedom} and \textit{Dynamic Gradual Guarantee}),
we formalise \LangName{} in a minimal concurrent object-based language which we
call \LangFormal{}, which we describe in this section. In \LangFormal{}, objects
have the usual fields and methods, and there are no classes and no
inheritance. Threads are created by a \c{spawn} operation which also sets up a
channel for communication. Channels support both reference semantics and value
semantics for objects depending on their capability meaning \LangFormal{} is a
shared-memory model. However, with the exception of unsafe objects, two threads
sharing a common object $o$ cannot implicitly transfer objects between each other
by reading and writing fields in $o$. Hence, with the exception of unsafe
objects, objects are effectively running in a message passing model, which may use
shared memory under the hood for efficiency, without compromising safety.

Not modelling classes or inheritance is a choice driven by the desire to keep the system minimal.
We note that dynamically typed languages are less dependent on inheritance,
because of the non-need to establish nominal subtyping relations. Our
simplifications allow us to focus on the most important aspects of
our work. Handling permissions and capabilities in the presence of
various forms of inheritance is well-known (\eg{} \cite{EncoreTS,
  OTSurvey, PermissionGradualTyping,
  DBLP:conf/tase/ZhaoB08, Krishnaswami-Aldrich:PLDI05}) in a
statically typed world including problems that may arise
  in an untyped setting. Therefore, we do not believe that these
simplifications accidentally suppress any fundamental
limitations of our approach.

For simplicity, channels themselves cannot be transferred. There
is nothing fundamental about this simplification but undoing it
requires some design thinking which is not important for the paper
at hand, such as, whether or not we allow multiple threads
connected to a single channel to race on taking the next message,
etc.

Figure~\ref{fig: permission language} shows the syntax of \LangFormal{}.
As is common, overbars (\eg{} $\overline{f}$) indicate possibly empty
sequences (\eg{} $f_1, f_2, f_3, \ldots$).
To simplify the presentation of the calculus programs are in A-normal
form~\cite{ANormalForm}: all subexpressions are named except for the \kw{consume}
expression. We further assume that programs use static single-assignment
  form~\cite{SSA}, \ie{} the \kw{let-in} term always introduces a new variable, field
reads are assigned to variables before they are bound to other variables, etc,
and that \self{} is never aliased. None of these constraints are essential for the soundness of our approach.

A program ($P$) is a term ($t$). Terms are variables and let-bound expressions
(\ie{} $x$ and $\LetIn{x}{e}{t}$).
An unusual design choice borrowed from \cite{PonyTS} is that assignment, $L = R$,
binds the left-hand side $L$ to the value of the right-hand side, $R$, and
returns the \emph{previous value} of $L$. (This is not uncommon when dealing with
\isokw{} fields, and previous work in this area enforce it via an explicit swap
operation~\cite{destructive-read,LaCasa,Haller10}).
Expressions are variables ($w$), destructive reads ($\consume{x}$), field reads
($x.f$), field assignments ($x.f = w$), method calls ($x.m(w)$), a deep copy
operation ($\kopy{\K}{x}$) that returns a copy of an object graph with capability
$\K$, an object literal,  a casting operation that asserts a capability ($\CC{K}{v}$), and a spawn operation that creates a new thread. At
run-time, the expression syntax also includes values.
An object consists of fields ($\overline{f = x}$) and methods
($\overline{M}$) and are instantiated with a given capability $\K$.
For simplicity, methods have a single argument
(\eg{} $x$) and more can be modelled using an object indirection using an unsafe object.

Spawning a new thread using $\Spawn{x}{t}$ introduces a new channel $x$ both
at the spawn-site, and inside the scope of the new thread whose initial term is $t$.
The $t$ is closed,
\ie{} it cannot access variables declared from an outer scope. Channels are
bidirectional unbuffered with blocking semantics on sending and receiving
operations. The send operation $\move{x}{w}$ puts $w$ on
channel $x$, if necessary blocking until the channel is ``free'' (\ie{} contains $\varnothing$).
The sender then blocks ($\blacksquare_i\,\loc$) until the message $i$ is received by the  thread on the other end of the channel $\loc$. The
receive operation $\recv{y}$ is similar to a send and blocks the current thread while
the channel is ``free''. Values are locations ($\loc$),
the ``absent value''
$\Absent$ used to populate a consumed variable or field, and
$\varnothing$, used at run time to indicate that a channel is empty.

For simplicity, we assume that programs do not attempt to consume
\c{self}, which can be enforced through a simple syntactic check,
and that all variable/method-parameter/channel names are distinct
and none is called \self{}. This is a common restriction in the literature, and
key to avoiding breaking of abstraction~\cite{ExternalUniqueness}.

\subsection{Dynamic Semantics}

\begin{figure}[t]
\begin{tabular}{rccl}
& \textit{E}   & $::=$ & $\bullet \OR \LetIn{x}{E}{t}  \OR x.f = E
                         \OR \move{E}{w}$ \\
                         &              & $\OR$ & $\move{v}{E}  \OR \recv{E} \OR \Obj{K}{\overline{f=v}\ f'=E\ \Fs}{\Ms}$  \\
                         &              & $\OR$ & $x.m(E) \OR \CC{K}{E}$ \\
& \textit{H}   & $::=$ & $\epsilon \OR H, \loc \mapsto \Obj{K^i}{\overline{f=v}}{\Ms}
\OR H, x \mapsto v$ \\
                         &              & $\OR$ & $H, \loc \mapsto \Chan{i, v}$ \\
& \textit{Cfg} & $::=$ & $H;  \overline{T} $                                                                                         \\
& \textit{T}   & $::=$ & $\epsilon \OR t^i \OR \textit{Err}$                                                                                        \\
& \textit{Err} & $::=$ & $\err \OR \CErr \OR \PErr \OR \CastErr$
\end{tabular}
\caption{\label{fig:runtime}Definitions of evaluation context, store and run-time configurations.}
\end{figure}

We formalise the dynamic semantics of \LangName{} as a small-step
operational semantics with reduction-based, contextual rules for
evaluation within threads (Fig.~\ref{fig:runtime}). The
evaluation context $E$ contains a hole $\bullet$ that denotes the
location of the next reduction step~\cite{EvContext}. We write the
reduction step relation
$H;  \overline{T}  \ReducesTo H';  \overline{T'} $ which takes a
reduction step from heap H and a collection of threads in $\overline{T}$, to a new
heap H' and a new thread state $\overline{T'}$.
A store $H$ is either empty ($\epsilon$), or contains mappings from variables to
values, and from locations to objects and channels (\cref{fig:runtime}). The
superscript $i$ in $K^i$ represents the object's thread owner and we used it to
keep track of ownership of \localkw{} objects (omitted from the rules when not relevant).

A configuration \textit{Cfg} is a heap $H$ and a collection of concurrently
executing threads $\overline{T}$. A thread is either finished ($\epsilon$), a
term $t^i$ (where $i$ represents the thread owner id, omitted
when not necessary),  or a run-time error
(\textit{Err}). There are four kinds of run-time errors: consumption errors
($\CErr$, which occur when a program accesses a consumed value); permission
errors ($\PErr$, which occur when a program violates the structural constraints
imposed by its capabilities); cast errors ($\CastErr$, which occur when a program has a different
capability than the one casted to); and normal errors ($\err$), such as accessing
a non-existing field, calling a non-existing method, etc. The execution of
threads is concurrent and non-deterministic.  The non-determinism comes from
\RN{C-Eval} and the commutativity equivalence rule. \label{fig: config
  concurrency}
\[
  \ntyperule{C-Eval}{
    H;  t
    \ReducesTo
    H';  \overline{T'}
    }{
    H;   t\ \overline{T} \
    \ReducesTo
    H';  \overline{T'} \ \overline{T}
    }
\qquad
    \begin{array}{c}
      H; v  \equiv H; \ \epsilon \\
      \overline{T}\ \textit{Err}\  \equiv \textit{Err} \\
      \overline{T}\ \overline{T'} \  \equiv \overline{T'}\ \overline{T} \\
            \overline{F}\ \ f = v \equiv f = v\ \ \overline{F} \\
      \overline{M}\ M \equiv M\ \overline{M}
    \end{array}
  \]
The reduction of a program $t$ begins in an initial configuration with an empty heap $\epsilon; t$ (\cref{def:init}, \cref{sec:properties}).
In the remainder of this section, we go through the reduction rules, ending with a discussion of the error trapping rules
that dynamically trap actions which (might) lead to data races.
For capabilities, the following relations hold:
  $\unkw{} \le{} \localkw{}$,
  $\localkw{} \le{} \isokw{}$, and
  $\isokw{} \le{} \immkw{}$. The $\le$ relation is reflexive and transitive.
  $\Helper{isImm}{H, \loc}$ (etc. for other capabilities) holds
  if $H(\loc)=\Obj{K}{\_}{\_}$ and $K = \immkw$.

The reduction of the $\kw{let}$ term
updates the heap with a stack variable $x$ pointing to the value $v$
(\RN{R-Let}). Reading a variable with a non-isolated object reduces to a location
(\RN{R-Var}); reading an isolated object involves moving semantics: 
moving the contents of a variable using \c{consume} reduces to a location, and leaves a $\Absent$ token in the variable which will cause an error if accessed before overwritten (\RN{R-Consume}).
Consuming fields is not allowed. Instead, one consumes a field when doing an assignment, \eg{}
\LetIn{x}{(y.f = z)}{\dots}, places on \kw{x} the object pointed by \kw{y.f} and places \kw{z} in \kw{y.f}.
\begin{figure*}
\[
  \ntyperule{R-Let}
  {~ \\
  x \notin \dom{H}
  }
  {
  H;  \LetIn{x}{v}{t}
  \ReducesTo
  H, x \mapsto v;  t
  }
  \qquad\!\!
  \ntyperule{R-Var}
  {
  H(x) = \loc
  \\
  \neg \isIso{H, \loc}
  }
  {
  H;  E[x]
  \ReducesTo
  H;  E[\loc]
  }
  \qquad\!\!
  \ntyperule{R-Consume}
  {
    H(x) = \loc
    \\
    H' = H[x \mapsto \Absent]
  }
  {
  H;  E[\consume{x}]
  \ReducesTo
  H';  E[\loc]
  }
  \qquad
  \ntyperule{R-Field}
{
H(x) = \loc \quad
H(\loc) = \Obj{\_}{\_ \ f = v}{\overline{M}}
\\
\neg \isIso{H, v}
\quad
\Helper{localOwner}{H, i, \loc}
}
{
H;  E[x.f]^i  \ReducesTo H;  E[v]^i
}
\]
\[
\ntyperule{R-FieldAssign}
{
H(x) = \loc \quad
  H(\loc) = \Obj{K}{\overline{f=v}\ f = v'}{\overline{M}}
  \\
  \neg \Helper{isImm}{H, \loc}
  \quad
  \OkRef{K}{v}
  \\
  \Helper{isLocal}{H, \loc} \Rightarrow
  (\Helper{isOwner}{H, i, \loc} \land
    \Helper{localOwner}{H, i, v})
}
{
H;  E[x.f = v]^i
\ReducesTo
H[\iota\mapsto\Obj{K}{\overline{f=v}\ f = v}{\overline{M}}];  E[v']^i
}
\quad
\ntyperule{R-New}
{
   ~ \\ 
  \forall f = v \in \overline{f = v}.\
  \OkRef{K}{v} \land
  (K = \localkw{} \land \Helper{isLocal}{H, v}) \Rightarrow \Helper{isOwner}{H, i, v}
  \\
  \loc\ \textit{fresh}
  \qquad
  H' = H, \loc \mapsto \Obj{K^i}{\overline{f = v}}{\overline{M}}
}
{
H;  E[\Obj{K}{\overline{f = v}}{\overline{M}}]^i
\ReducesTo
H';  E[\loc]^i
}
\]
\[
\ntyperule{R-Call}
{
   ~ \\ 
  x', y'\,\mathit{fresh} \qquad
  H(x) = \loc \qquad
  H(\loc) = \Obj{\_}{\_}{\overline{M}\ \Method{m}{y}{t}}
}
{
H;  E[x.m(v)]
\ReducesTo
H,x'\mapsto\loc, y'\mapsto v;   E[t[\self{} = x'][y = y']]
}
\quad
  \ntyperule{R-CastLoc}
  {
   ~ \\ H(\loc) = \Obj{K}{\_}{\_} }
  {H; E[\CC{K}{\loc}] \ReducesTo H; E[\loc]}
\quad
\ntyperule{R-Spawn}
{
  \loc, i, j \textit{ fresh}
  \quad
  x \notin \dom{H}
  \\
  H' = H, x \mapsto \loc, \loc \mapsto \Chan{i, \varnothing}
}
{
H;  E[\Spawn{x}{t}]
\ReducesTo
H';  E[\loc]\ t^j
}
\]
\[
\ntyperule{R-Recv}
{
  H(\loc) = \Chan{i, \loc'}
  \\
H' = H[\loc \mapsto \Chan{i, \varnothing}]
}
{H;  E[\recv{\loc}]  \ReducesTo H';  E[\loc'] }
\quad
\ntyperule{R-SendBlock}
{
  H(\loc) = \Chan{\_, \varnothing} \qquad
  \\
    i~\mathit{fresh}
    \qquad
  H'=H[\loc \mapsto \Chan{i, v}]
}
{
H;  E[\move{\loc}{v}]  \ReducesTo
H';  E[\blacksquare_i\,\loc]  }
\quad
\ntyperule{R-SendUnblock}
{
  H(\loc) = \Chan{i', v} \\
  v=\varnothing \lor i \neq i'
}
{
H;  E[\blacksquare_i\,\loc]  \ReducesTo H;  E[\loc]
}
\quad
\ntyperule{R-Copy}
{
  \isokw{} \neq K
  \quad
  H(x) = \loc'
  \quad
  \Helper{localOwner}{H, i, \loc'}
  \\
  \OkDup{H}{K}{H(x)} = (H', \loc)
}
{
H;  E[\kopy{K}{x}]^i
\ReducesTo
H';  E[\loc]^i
}
\]
\caption{\label{runtime semantics}Runtime semantics}
\end{figure*}
Reading a field ($x.f$) reduces to the value stored in the field
(\RN{R-Field}). Note that to read an isolated object's field one must update the
field and place another object in its stead --- directly reading an isolated field
would create a new alias.
The helper predicate \Helper{localOwner}{...} checks
that if the target object is \localkw{}, then the current thread is its owner.
(This prevents threads to access unowned local objects.)


Updating a field ($x.f$) with a value ($v$) reduces
to returning the previously held value in the field and updating
the field $f$ to point to value $v$.
There is a check that prevents mutating immutable objects, $\OkRef{K}{v}$ ensures
that the object $v$ can be placed under an object with capability $K$, and the
remaining helper predicates check that if the target object ($x$) is local, then
its owner is the current thread and if the source ($v$) is local then its owner
is the current thread.\footnote{This design is good for JIT compilation since
  local objects have guarantees to not point to unowned local objects (\cref{th:
    thread affinity}), thus removing some unnecessary dynamic checks from
  possible implementations.}  (Other reduction rules repeat this local ownership
check and for space reasons we omit mentions in the remaining rules.)

An object literal (\RN{R-New}) checks that values of its fields do not
violate the structural constraints imposed by its capability $K$, and if $K$
is \localkw{} then the current thread must own the local fields. \textsc{R-New} returns
the (fresh) location of the object.

Calling a method on an object referenced by $x$ and with argument $v$ reduces to
the body $t$ of the method with \c{self} substituted for a fresh variable
bound to the location of $x$
and the singular argument substituted for a fresh variable bound to $v$ (\RN{R-Call}).

%
A new thread is introduced by a \c{spawn} operator which
introduces a new channel connecting the spawned thread with its
``parent'' (\RN{R-Spawn}). Rules (\RN{R-SendBlock}), (\RN{R-SendUnblock})
and (\RN{R-Recv}) handle sending and receiving values on a channel.
Sending on a channel $\loc$ blocks until the channel is empty, and
subsequently blocks the sending thread until the value has been
received on the other side. Reading on a channel blocks until
there is a value that can be retrieved.

Casting an object (\RN{E-CastLoc}) checks that the object has the specified capability,
throwing a permission error, otherwise.
The function \textsc{R-Copy} deep copies the object pointed by $\loc$, returning
a heap that contains the copy of the object graph with capability $K$
and a fresh location that points to the object copied.\footnote{The helper function $\Helper{OkDup}{H, K, v}$
  is a standard deep-copying operation.~\cite{SheepCloning,MinimalOwnership}}
The helper functions used above are defined thus:
\[
\ntyperule{RefCheck}
{
H(\loc) = \Obj{K'}{\_}{\_} \quad
\OkField{K}{K'}
}
{\OkRef{K}{\loc}}
\qquad
  \ntyperule{Helper-LocalOwner}{
    \Helper{isLocal}{H, v} \Rightarrow \Helper{isOwner}{H, i, v}
    }{
      \Helper{localOwner}{H, i, v}
    }
\]

For simplicity, we have gathered some rules that trap capability errors at
run-time in \cref{fig:errors}. Common errors when accessing non-existent fields
and methods throw a \err{} error (\eg{} \textsc{E-NoSuchField}).  Accessing values
which are absent due to a destructive read yields a $\CErr$ (\eg{}
\RN{E-Consume}). Assigning an illegal
value to a field is not allowed (\eg{} \RN{E-AliasIso} and \RN{E-IsoField}). Casts to the wrong capability reduce to $\CastErr$. (Remaining rules in \cref{sec: expression rules errors}, \cref{fig:errors extended}.)
\begin{figure*}[t]
\vspace*{-1em}
\[
\ntyperule{E-NoSuchField}
{~\\
  H(x.f)=\bot
}
{
H;  E[x.f]
\ReducesTo
H;  \err
}
\quad
\ntyperule{E-Consume}
{
~\\
H(x) = \Absent
}
{
H;  E[\consume{x}]
\ReducesTo
H;  \CErr
}
\quad
\ntyperule{E-AliasIso}
{
  H(x)=\loc \\
  \Helper{isIso}{H, \loc}
}
{
H;  E[x]
\ReducesTo
H;  \PErr
}
\quad
\ntyperule{E-IsoField}{
H(x) = \loc' \quad   H(x.f)=\loc \\
\Helper{localOwner}{H, i, \loc'} \quad
\Helper{isIso}{H, \loc} 
}{
  H;  E[x.f]^i
  \ReducesTo
  H;  \PErr
}
\quad
\ntyperule{E-CastError}
{H(\loc) = \Obj{K'}{\_}{\_} \\
  K' \neq K}
{H; E[\CC{K}{\loc}]
\ReducesTo
H; \CastErr
}
\]
\caption{Expression rules producing errors. To reduce clutter, we
  write $H(x.f)=v$ when
  $H(x) = \loc \land H(\iota)=\Obj{\_}{F}{\_}$ and $f=v\in F$. (Remaining rules in \cref{sec: expression rules errors}, \cref{fig:errors extended}.)}
\label{fig:errors}
\vspace*{-1em}
\end{figure*}

\subsection{Well-Formedness}


We define the environment (also used as store typing~\cite{TAPL}) as $\Gamma ::= \epsilon \OR
\Gamma, x : K \OR \Gamma, \loc : K$, where $\epsilon$ represents the empty
environment, and $x : K$ and $\loc : K$ mean variable $x$ and location $\loc$
have capability $K$.

\[
\ntyperule{WF-Env-Empty}{}{\vdash \epsilon}
\qquad
\ntyperule{WF-Env-Var}{x \notin \dom{\Gamma} \quad \vdash \Gamma}{\vdash \Gamma, x : K}
\qquad
\ntyperule{WF-Env-Loc}{\loc \notin \dom{\Gamma} \quad \vdash \Gamma}{\vdash \Gamma, \loc : K}
\]
Well-formedness rules are mostly standard and straightforward. Objects'
thread-locality and proper isolation (for objects with local and isolated
capabilities respectively) fall out of well-formedness and appear in the
Appendix. A well-formed configuration $H; \Gamma \vdash \overline{T}$
(\RN{WF-Configuration}) consists of predicates establishing that \localkw{}
objects (in the heap) cannot be reachable from multiple threads and that isolated
objects (in the heap) have a single reference modulo borrowing (Appendix,
\cref{def: thread-local,def: object iso}). Essentially, the well-formed rules
guarantee that the heap is well-formed \wrt{} object capability (and its fields) and
that variables are not duplicated when introduced, but they do not statically forbid
violation of object capabilities (which will throw a permission run-time error).
\[
    \ntyperule{WF-Configuration}{
      \dom{\Gamma} = \dom{H} \quad
      \Gamma \vdash H \\
      \forall t\in \overline{T}. \Gamma \vdash H;  t
      \quad
      \Helper{Local}{H, \overline{T}}\quad \Helper{Isolated}{H, \overline{T}}
    }{
      \Gamma \vdash H;  \overline{T}
    }
\qquad
  \ntyperule{WF-Term}
  {~\\
  \Gamma \vdash H \quad \Gamma\vdash t
  }
  {\Gamma \vdash H;  t}
\]
Terms are well-formed (\RN{WF-Term})
if the store is well-formed \wrt{} an environment ($\Gamma \vdash H$) and the
term is well-typed ($\Gamma \vdash t$).
The store is well-formed \wrt{} an environment if every variable and
location in the store is defined in the environment (\RN{WF-H-Absent},
\RN{WF-H-Chan}, and \RN{WF-H-Var}), if there is a match between the object
capability and the variable's (or location's) expected capability
(\RN{WF-H-Object}), and object fields are compatible with their capability object. The environment is well-formed if there are no
duplicate locations or variables (\RN{WF-Env-$\star$} rules).
\[
\ntyperule{WF-H-Empty}{~ \\ ~\\ \vdash \Gamma}{\Gamma \vdash \epsilon}
\quad
  \ntyperule{WF-H-Absent}
  {~\\
    x \in \dom{\Gamma} \\
    \Gamma \vdash H
  }
  {
    \Gamma \vdash H, x \mapsto \Absent
  }
\quad
  \ntyperule{WF-H-Chan}
  {
  \loc \notin \dom{H} \\ 
  v \neq \Absent \quad \Gamma(\loc) = \localkw{} \\
   \Gamma(v) \neq \localkw{} \quad \Gamma \vdash H
  }
  {
  \Gamma \vdash H, \loc \mapsto \Chan{i, v}
  }
\]
\[
  \ntyperule{WF-H-Var}
  {
  x \notin \dom{H} \quad \loc \in \dom{H} \\
  \Gamma(x) = \Gamma(\loc) = K \quad
  \Gamma \vdash H
  }
  {\Gamma \vdash H, x \mapsto \loc}
\qquad
  \ntyperule{Helper-OkRefEnv}
  {
    \Gamma(\loc) = K' \quad \OkField{K}{K'}
  }
  {
    \OkRefEnv{K}{\loc}
  }
\]
\[
  \ntyperule{WF-H-Object}
  {
    \Gamma(\loc) = K \qquad \loc \notin \dom{H} \qquad
    \forall v \in \overline{v}.\ \Gamma \vdash H \land \OkRefEnv{K}{v} \land v \neq \varnothing
  }
  {
  \Gamma \vdash H, \loc \mapsto \Obj{K}{\overline{f=v}}{\_}
  }
\]
Programs are well-formed \textit{w.r.t} an environment $\Gamma$ (\cref{fig: wf
  declarations}) when \textit{let} terms always introduce new variables, and
when all accessible variables are defined in $\Gamma$, \eg{} \RN{WF-Var}.

\begin{figure*}[t]
\[
\ntyperule{WF-Program}
          {\epsilon \vdash t}
          {\vdash t}
\qquad
\ntyperule{WF-Env}{ }{\vdash \epsilon}
\qquad
\ntyperule{WF-Env-Var}
          {\vdash \Gamma\qquad
           x \notin \dom{\Gamma}}
          {\vdash \Gamma, x : K}
\qquad
\ntyperule{WF-Let}
          {x \notin \dom{\Gamma}\qquad
           \Gamma \vdash e \qquad
           \Gamma, x : K \vdash t}
          {\Gamma \vdash \LetIn{x}{e}{t}}
\qquad
\ntyperule{WF-Var}
          {x \in \dom{\Gamma} \qquad
           \vdash \Gamma}
          {\Gamma \vdash x}
\]
\[
\ntyperule{WF-Loc}
          {\loc \in \dom{\Gamma} \quad \vdash \Gamma}
          {\Gamma \vdash \loc}
\qquad
\ntyperule{WF-Absent}
          {\vdash \Gamma}
          {\Gamma \vdash \Absent}
\qquad
\ntyperule{WF-Consume}
          {\Gamma \vdash x }
          {\Gamma \vdash \consume{x}}
\qquad
\ntyperule{WF-Field}
{\Gamma \vdash x}
{\Gamma \vdash x.f}
\qquad
\ntyperule{WF-Assignment}
{
  \Gamma \vdash x \qquad
  \Gamma \vdash w
}
{\Gamma \vdash x.f = w}
\qquad
\ntyperule{WF-MethodCall}
{\Gamma \vdash x \qquad
 \Gamma \vdash w}
{\Gamma \vdash x.m(w)}
\]
\[
\ntyperule{WF-Unblock}
{\loc \in \dom{\Gamma} \qquad
 \vdash \Gamma}
{\Gamma \vdash \Blocked{i}{\loc}}
\qquad
\ntyperule{WF-Send}{\Gamma \vdash x \qquad \Gamma \vdash w}{\Gamma \vdash \move{x}{w} }
\qquad
\ntyperule{WF-Recv}{\Gamma \vdash x}{\Gamma \vdash \recv{x}}
\qquad
\ntyperule{WF-Copy}{\isokw{} \neq K \qquad  \Gamma \vdash x}{\Gamma \vdash \kopy{K}{x}}
\qquad
\ntyperule{WF-Spawn}
{
x \notin \dom{\Gamma} \quad
x : \localkw{} \vdash t \quad \Helper{FreeVars}{t} \subseteq \{ x \}
}
{\Gamma \vdash \Spawn{x}{t}}
\]
\[
\ntyperule{WF-Cast}{\Gamma \vdash w}{\Gamma \vdash \CC{K}{w}}
\qquad
\ntyperule{WF-Method}
{
\Gamma, x : K \vdash t
}
{\Gamma \vdash \Method{m}{x}{ t }}
\qquad
\ntyperule{WF-Object}
          {\forall w \in \overline{w}.\ \Gamma \vdash w \qquad
           \forall\ \Method{m}{x}{t} \in
               \overline{M}.\
               \self{} : K \vdash \Method{m}{x}{t}}
          {\Gamma \vdash \Obj{K}{\overline{f = w}}{\overline{M}}}
\]
\caption{\label{fig: wf declarations}Well-formed declarations, terms and expressions. $\Gamma ::=
  \epsilon \OR \Gamma, x : K \OR \Gamma, \loc : K $}
\end{figure*}

\subsection{Properties of Well-formed Programs}
\label{sec:properties}

We highlight the properties satisfied by well-formed programs (proofs in the Appendix):
\begin{itemize}
\item Progress and Preservation (\cref{th: preserv dyn,th: progress dyn}). This
  means that if a well-formed program is not finished (empty state), is not an
  error (normal, absent, permission, or cast error), or is not in a deadlock state
  (terminal configuration), then it can take a reduction step until it ends in a
  terminal configuration state and the result of each reduction step is well-formed.

\item Data-Race Freedom (\cref{th: drf}). Programs without unsafe objects are
  data-race free by construction.

\item Dynamic Gradual Guarantee (\cref{th: dyn gradual}). Adaptation of the
  gradual guarantee~\cite{GradualTyping,RefinedGradualTyping} stating
  capabilities do not affect the run-time semantics, modulo casts and capability
  errors.  Essentially, if an unsafe program is well-formed and takes a reduction
  step, the same program with capability annotations either reduces to the same
  run-time configuration (modulo safe erasure, \cref{sec: defs lems proofs} \cref{def:safe stripping})
  or throws an error due to a cast or capability violation.
\end{itemize}

Programs start in an initial well-formed configuration $\epsilon; t$ (Appendix, \cref{def:init}),
and reduce to new configurations. Progress (\cref{th: progress dyn}) guarantees that
a well-formed configuration reduces to a new configuration, or it is terminal (Appendix, \cref{def:terminal}).
From Preservation (\cref{th: preserv dyn}), a reduction step always leads to
a well-formed configuration. Terminal configurations are either finished programs,
errors (with $\overline{T}\ \textit{Err} \equiv \textit{Err}$ from equivalence rules on Page~\pageref{fig: config concurrency}), or a deadlock configuration (Appendix, \cref{def: deadlock}).
A deadlock configuration happens when all threads are either waiting on a receive
or on a send operation.

\begin{restatable}[Progress]{theorem}{PROGRESS}\label{th: progress dyn}
A well-formed configuration $\Gamma \vdash H;  \overline{T} $ is either
a terminal configuration or
$H; \overline{T} \ReducesTo H'; \overline{T'} $.
\end{restatable}

\begin{restatable}[Preservation]{theorem}{PRESERVATION}\label{th: preserv dyn}
If $\Gamma \vdash H;   t\ \overline{T}$ is a well-formed configuration, and $H;
 t\ \overline{T}  \ReducesTo H';  \overline{T'}\ \overline{T}$ then, there exists a
$\Gamma'$ s.t. $\Gamma' \supseteq \Gamma$ and $\Gamma' \vdash H';   \overline{T'}\ \overline{T}$
\end{restatable}

\begin{restatable}[Thread-Affinity of Thread-Local Fields]{corollary}{THREADLOCAL}\label{th: thread affinity}
  Implied by Preservation, a thread local object with owner $i$ cannot contain a thread local
  object with owner $j$, where $i \neq j$. The only way a local object can reference another local object
  of a different owner is via field assignment (\textsc{R-FieldAssign}). But \textsc{R-FieldAssign}
  checks that target and source share owners. Thus, thread local objects can only reference
  thread local objects of the same owner.
\end{restatable}

\begin{definition}[Data Race]\label{def: informal data race}
  \textup{Informally}, a data race is defined as two threads accessing
  (write-write or read-write) the same field without any
  interleaving synchronisation. In our setting, this translates to
  two accesses to the same field of an object $o$ in threads $i$
  and $j$ without an interleaving explicit transfer of $o$ from
  $i$ to $j$. Thus, a data race in Dalarna requires the ability of a mutable object to be
  referenced from two threads at the same time (formal definition in \cref{def:data race}).
\end{definition}

\noindent
\cref{th: drf} states that \LangFormal{} is data race-free modulo unsafe objects.
Objects that use safe capabilities cannot introduce a data race; unsafe objects
may introduce data races.

\begin{restatable}[\LangFormal{} is Data-Race Free Modulo Unsafe Objects]{theorem}{DRF}\label{th: drf}
  All data races directly or indirectly involve an
  unsafe object. A data race is defined as a read/write or
  write/write access to an object by different threads without
  interleaving synchronisation (In our formalism, this
  means an interleaving transfer of an object to another thread; formal proof in Appendix).
\end{restatable}

\noindent
We now proceed to sketch the proof of data-race freedom for safe
objects by showing that it is not possible for a mutable safe
object to be aliased from two different threads at the same time.
We refer to $l$, $i$, $j$, $r_1$ and $r_2$ from \cref{def: informal data race} for
clarity.

Let us examine the implications of $l$ having any of the three
safe capabilities (and ignore unsafe objects for now).

\begin{enumerate}
\item \emph{$l$ is immutable}. By \RN{E-BadFieldAssign}, attempts to
  write fields of an immutable object will err. Thus, if $l$ is
  immutable, then $R$ will contain an error, which it did not by
  assumption.

\item \emph{$l$ is local}. By \RN{R-Field} and \RN{R-FieldAssign},
  attempts to write a field of a local from outside of its
  creating thread will err. Thus, if $l$ is local, then $R$ will
  contain an error (because $i\neq j$), which it did not by
  assumption.

\item \emph{$l$ is isolated}. Fields of isolated objects can be
  read or written freely. Thus, we must show that $l$ can be
  accessible in two threads at the same time. In the initial heap,
  no objects exist that is shared across threads and our only way
  to share objects across threads is by sending them on a channel.
  An isolated $l$ can be sent on a channel. However, this requires
  that $l$ is consumed, meaning it will no longer be accessible by
  the sender. To avoid the consumption, we could store $l$ in a
  field of an object, and then transfer the object. Such objects,
  would have to be immutable or local. However, by
  \RN{R-FieldAssign}, immutable or local objects cannot contain
  isolated objects.

  Thus, we cannot create a situation where $l$ is in both $r_1$
  and $r_2$ without any interleaving send.

\end{enumerate}

A program that uses unsafe objects may use these to store local
and isolated objects. Thus, an unsafe object $u$ aliased across
threads could store a local or isolated $l$ in $u.f$ (see 3. above). This would
allow threads $i$ and $j$ to do $u.f.g$. If $l$ is local,
unless $i=j$, at least one of the accesses will err (or both if
neither $i$ nor $j$ is the creating thread of $l$, see 2. above). If $l$ is
isolated, $u.f.g$ will err by \RN{E-IsoField}. Thus, even in the
presence of unsafe objects, safe objects will not participate in
data races.

\noindent
Progress and preservation guarantee the absence of untrapped errors and \cref{th: drf}
shows that all data races can be blamed on unsafe objects. Next, we show that capability
annotations do not affect the run-time semantics, modulo cast or capability
violations which trap operations that if allowed could lead to a data race.
We show this in the Dynamic Gradual Guarantee theorem (\cref{th: dyn
  gradual}, adapted from~\cite{RefinedGradualTyping}).
We unpack this theorem before stating it formally. Let $P$ be a
well-formed program, and $S$ its ``stripped equivalent'', where
all safe capabilities have been erased (and thus replaced with unsafe).
Below, ``reduces'' denotes a single reduction step.

\begin{enumerate}
\item The well-formedness of $S$ follows from the well-formedness
  of $P$, as an unsafe object can reference any object
  (\cref{sec: defs lems proofs} \cref{def:safe stripping}).

\item If $P$ reduces to a non-error configuration, $S$ reduces to
  the same configuration and the same heap (modulo heap capability
  erasure); in case $P$ throws an error caused by absent values
  or normal errors, $S$ will throw the same error. The two
  evaluations only diverge if $P$ throws a permission error in a
  dynamic check -- this will never happen in $S$ because it
  does not have any safe capabilities. (See \cref{lem:unsafe err} \eg{}
  \textsc{E-BadFieldAssign} or \textsc{E-BadInstantiation}
  among others.)

\item If $S$ reduces to an error configuration, $P$ will reduce to
  the same error configuration.
If $S$ reduces to a non-error
  configuration, $P$ will either reduce to the same
  configuration (modulo capability stripping), or throw a
  capability\,or\,permission error.
\end{enumerate}

\begin{restatable}[Dynamic Gradual Guarantee]{theorem}{DynamicGradualGuarantee}\label{th: dyn gradual}
  Let $H; t\ \overline{T_0}$ be a configuration and $\Gamma$ a store type such that $\Gamma\vdash H; t\ \overline{T_0}$.
  Let $^e$ be a function that replaces safe capabilities with unsafe in heaps, terms, etc (\cref{def:safe stripping}). Then:

  \begin{enumerate}\raggedright
  \item $\Gamma^e \vdash (H; t\ \overline{T_0})^e$. \label{proof: drf 1}

  \item \label{th: dyn gradual 2}
    \begin{enumerate}
    \item
    If $H; t\ \overline{T_0}  \ReducesTo H'; \overline{T_1}\ \overline{T_0}$
    and $\overline{T_1} \neq \textit{Err}$ then,
    $(H; t\ \overline{T_0})^e  \ReducesTo (H'; \overline{T_1}\ \overline{T_0})^e$.

    \item
    If $H; t\ \overline{T_0}  \ReducesTo H';  \overline{T_1}\ \overline{T_0}$ and
    $\overline{T_1} = \CErr \lor \err$ then,
    $(H; t\ \overline{T})^e  \ReducesTo H'';  \overline{T_1}\ \overline{T_0}$
    \end{enumerate}

  \item \label{th: dyn gradual 3}
    \begin{enumerate}
    \item
      If $(H; t\ \overline{T_0})^e  \ReducesTo (H';  \textit{Err}\ \overline{T_0})^e$
      then, $H; t\ \overline{T_0}  \ReducesTo H'; \textit{Err}\ \overline{T_0}$
    \item
      If $(H; t\ \overline{T_0})^e \ReducesTo (H'; \overline{T}\ \overline{T_0})^e$
      and $\overline{T} \neq \textit{Err}$ then,
      $H; t\ \overline{T_0} \ReducesTo H'; \overline{T'}\ \overline{T_0}$ and
      $\overline{T'} = \PErr \lor \CastErr \lor \overline{T}$.
  \end{enumerate}
\end{enumerate}
\end{restatable}

The Dynamic Gradual Guarantee (\cref{th: dyn gradual}) uses a single step
reduction to guarantee that the capabilities are semantics preserving, modulo
permission and cast errors.  We extend the Dynamic Gradual Guarantee to account
for multi-step reductions, starting from an initial configuration until reaching a
terminal configuration, \ie{} $\epsilon; P \ReducesTo^* H; C$. To remove
non-determinism of program reductions, we define the trace of a program as a list
of pairs that contain the reduction step and the thread id on which the reduction happens.
We extend the reduction relation to account for the trace,
named the replay reduction relation, which is the standard reduction relation
except that it deterministically applies the expected reduction step on the
expected thread id (Appendix,
\cref{def:trace,def:tracereduction,def:tracereplay} and \cref{th: multistep dyn
  gradual}).
The basic idea is to reduce a safe program to a terminal configuration,
which produces a trace. We use this trace to replay the reductions on the capability
stripped (unsafe) program (and \textit{vice versa}), showing two programs reduce
to the same terminal configuration modulo cast errors and permission errors.
Since it obscures some cases where the identical step is taken, we show the single-step
theorem in the paper which highlights these cases.

\section{\LangNameImpl{}: A Prototype Implementation of \LangName{} in Grace}
\label{sec:implementation}

To be able to explore the applicability of our capability model, we have embedded it within a
preexisting general-purpose object-oriented language, Grace~\cite{grace}. We
extended an existing implementation to support run-time enforcement of
capabilities with the correct dynamic semantics. Programs in the formal syntax
have a straightforward translation into the Grace syntax, and the wider features
of the language are also usable with minimal limitations.

Our implementation, \LangNameImpl{}\footnote{\emph{Daddala} is a genus of moth in the family Erebid\ae.}, is fully embedded within the existing Grace
syntax, with no syntactic changes to the host language.
The embedded syntax will be familiar to the formal syntax above.

Objects in Grace are created with object literals
\ec{object \{ ... \}}, which instantiate new anonymous
objects with the given methods, fields, and inline initialisation code. To
specify the capability of the object we use the existing inheritance
system~\cite{GracesInheritance,ObjectInheritance} and write
\ec{object \{use isolate\}} (or immutable, local; unsafe
is the default).

\ec{let (e) in \{ x -> t \}},
\ec{spawn \{ x -> t \}}, and
\ec{consume(x)}
perform the corresponding roles.
Send and receive is \ec{c <- 5} and \ec{<- c}.
%
Casting is implemented straightforwardly as Grace
type annotations precede the capability name: \ec{var x : iso := y}.


\subsection{Extending Grace with \LangName{} Capabilities}

Our prototype extends Grace in two ways:
every assignment performs some check and
objects are tagged at creation with their capabilities. Objects are tagged unsafe
(default), immutable, local, free isolate, or bound isolate.

An isolate is \emph{free} if it is not currently stored in a variable
or field, otherwise \emph{bound}. These are the only two capability tags that are not fixed for the
object's lifetime (distinguishing this system from Grace's
brands~\cite{BrandObjects}, which are immutable). When a variable
or field containing a (bound) isolate is assigned to, the
assignment re-tags the object as a free isolate. When
\textit{any} storage location is assigned to, the assignment
checks the right-hand side and raises an exception if it is
a bound isolate (as this would otherwise violate alias-freedom),
or re-tags a free isolate as a bound isolate
before storing it. In essence, this pair behaves as a one-bit
reference count for isolate objects.

Assignments occur at assignment statements (\ec{x := y}), variable and field
declarations (\ec{var x := y}, \ec{def x = y}), binding of method parameters
(\ec{method foo(x) \{\}; foo(y)}), and binding of block (first-class
lambda) parameters (\ec{\{ x -> \}.apply(y)}). All of these
locations perform the free/bound isolate handling above,  while fields also
perform the necessary structural checks based on the capability of their
enclosing object and of the assigned object. These are the only places new kinds
of error have been introduced.


There is one special extension motivated by simplicity: the \ec{consume(x)}
operation is syntactically rewritten to \ec{(x := null)}, rather
than truly a method as the syntax suggests.

\subsection{Limitations Due to Embedding}
\label{sec:limit-due-embedd}
Some features of the Grace language run counter to the properties
of the formal model. While a program written according to the
model will execute correctly, it is possible to construct
pathological programs using other language features that violate
the imposed constraints in the present implementation. In
particular, Grace makes heavy use of lexical capture: all scopes
are closures, including object bodies~\cite{graceSigcse13}. A
program such as the one in \cref{fig:limit} will access the
isolate object that should no longer be reachable, because the
\ec{y} object retains the lexical scope of the surrounding object
and can implicitly call its method \ec{f}. Our prototype will not
raise an error at this point. This type of capture is unavoidable
without breaking fundamental aspects of the language (which uses
scoping to provide even builtins like
if-then-else~\cite{GraceDialects}), but the \textit{use} of
improperly-captured objects can be detected statically. While it
is therefore possible to eliminate these errors (and other
implementation strategies, such as proxies for isolates, can
similarly address the issue), our current prototype permits a
programmer who goes far outside the model to shoot themselves in
the foot.

\begin{figure}[t]
\begin{lstlisting}[style=minigrace]
var x := object {
    use isolate
    method f { }
    method g {
        object {
          method h { f }
} } }
def y = x.g
def c = spawn{v->(<-v)}
c <- consume(x)
y.h
\end{lstlisting}
  \caption{Limitations due to embedding: current unsoundness of our implementation. See \cref{sec:limit-due-embedd}.}
  \label{fig:limit}
\end{figure}

\subsection{Concluding Remarks}
Extending an existing system to support capabilities in \LangName{} style is
relatively straightforward; it is only the various kinds of variable assignment
that require meaningful intervention, plus a small object tag. We modified an
implementation that was previously designed with no concept of capabilities, so
that it now enforces them dynamically and supports a direct translation from the
formal model.

\section{Related Work}
\label{sec:rw}
\label{sec:related}



The \LangName{} capability model (immutables $<$ isolates $<$ thread-locals)
carefully selects a number of well-known concepts from the literature
\cite{DougLeaCPIJ2, Islands, MinimalOwnership, WrigstadPMZV09, RobustComposition,
  PonyTS}.  Dala's key contribution here is the careful combination: what we have
left out is at least as important as the features we have included.  Given that
an actor is essentially a thread plus thread-local storage \cite{Marr-taxonomy},
our distinctions are similar to many object-actor hybrid systems
\cite{Jcobox,AmbientTalk1,
  AmbientTalk2,OwnershipIsolation,marr-actor-domains1,marr-actor-domains2},
although, crucially, we follow Singularity \cite{Singularity} by incorporating
isolates for fast transfers.  Similarly, there are many more flexible models for
distinguishing between read-only and read-write objects: we adopt ``deep
immutability'' for its clear conceptual model Glacier \cite{Glacier}.  While
there are certainly more complicated models of permissions and capabilities for
data-race freedom (\eg{} \cite{MinimalOwnership, EncoreTS, PonyTS,
  DBLP:conf/tase/ZhaoB08} and many others), we consider our chosen set of
concepts a ``sweet spot'' in the balance between expressivity and complexity.


\subsection{Capabilities and Ownership}

Dala is also heavily influenced by static capability-based programming
languages~\cite{SFMEncore,EncoreTS,PonyLang,PonyTS, Gordon}.  Capability-based
programming languages require all programs to be fully annotated with
capabilities, and these annotations guarantee data-race freedom, with erasure
semantics. In contrast, our approach begins with a dynamic language
that allows data race and data race-free programs to interact, and we
maintain data race-freedom for programs with safe capabilities.










Sergey and Clarke~\cite{GradualOwnership} add gradual ownership to a
static language, introducing notions of parametric ownership and inserting
casts when needed;
they prove soundness and common
ownership invariants.
Our work has similarities in that isolated objects can be seen as
owners-as-dominators, and our local objects have threads as owners. \LangName{}
differs in that it is a dynamic, concurrent language, and we prove common invariants and
data race-freedom for safe~objects.



HJp~\cite{PermissionGradualTyping} enforces safe sharing of objects using a
permission-based gradual typing, which inserts run-time checks when necessary.
Objects are either in shared-read permission, which allows reading from multiple
threads but no mutation, or read-write permission, allows any mutation and
aliasing but no sharing. It also introduces storable permissions which allows a
permission to refer to a tree of objects. Our approach uses capabilities at
run-time; \immkw{} capabilities are similar to HJp
shared-read and read-write permissions are similar to \localkw{}. In addition,
\LangName{} also has the concept of \isokw{} which can move across threads
but do not allow any aliasing.

Roberts \textit{et al.}~\cite{TransientChecks} (and recently~\cite{moy2020corpse}) showed that run-time gradual type checks could have
minimal or no performance impact on a suitable virtual machine, despite what is
naively much extra checking for partially-typed programs. We chose to extend
their Moth virtual machine for our dynamically-checked implementation to take
advantage of their work.



\subsection{Capabilities In The Wild}

Castegren's \etal{} work~\cite{EncoreTS} seems
to be the first one where reference capabilities are orthogonal to the
concurrency model \ie{} the reference capabilities seem applicable to multiple concurrency models. 
Their capabilities have been formalised using fork-join style
but their implementation uses active objects~\cite{SFMEncore}.
%
The implementation of Gordon's \etal{} reference immutability work closely
follows the formal semantics; because their reference permissions apply
transitively to fields of the object, it is not clear whether the model is
general enough to be applicable to different concurrency models, \eg{} actor or tuplespace
model~\cite{ActorSurvey,Linda-TupleSpaces}.
Boyland et al's work~\cite{CapabilitiesForSharing} can encode 8
object capabilities to express different invariants and they argue that these can
be used in concurrent programs with no run-time cost, when programs are fully
typed. We believe the capability system is expressive enough to work
on different concurrency models, but it is not clear whether their capabilities enforce data
race freedom.
%
In languages such as E~\cite{E}, AmbientTalk~\cite{AmbientTalk2}, and
Newspeak~\cite{bracha2017newspeak} references (far and near) represent object
capabilities~\cite{ObjectCapability-DennisH66} and use a \textit{vat}-based concurrency model.
%
In contrast, our work is simple and uses 3 capabilities, allows
interoperation between safe and ``racy'' programs and (as far as we know) it is the
first one to use reference capabilities in a dynamic language where the
capabilities are orthogonal to the concurrency model.


\subsection{Race Detectors}
Although our capability checks guarantee data-race freedom,
they
are different to the checks that a data-race detector might employ \cite{Eraser}.
These checks are also in some sense ``eager'' or may cause false
positives. For example, a program that effectively transfers a
mutable object $o$ between two threads will execute without errors
if $o$ is isolated, but not if it is local and the non-owning thread dereferences
it. This is a somewhat pragmatic choice, but
guided by our desire to make our capabilities a tool for
programmers to capture their \emph{intent}. Thus, we expect that a
local object is explicitly demarcated local (at creation time) and
not isolated for a reason. Thus, \LangName{} helps programmers state
their intentions and check that the programs they write conform to
said intentions. This is different from a data race detector which
may only fire if a data race occurs (which may happen on
some runs but not others of the same program).

\section{Discussion and Future Work}
\label{sec:future}

Our claim of efficiency rests on absence of deep copying and turning
local accesses into asynchronous operations. That said, our capabilities
incur a cost on (most) accesses to objects -- \eg{} on writes to
fields, etc. To remove most of this cost will require
self-optimising run-times~\cite{Truffle} and techniques similar to
Grace's transient checks~\cite{TransientChecks} to reduce the
number of checks needed to satisfy the capability invariants.

Adding gradual capabilities at the type level will allow most
checks to be removed \cite{TransientChecks,moy2020corpse}, but more importantly help programmers
document the behaviour of code. Notably this addition will not need
escape hatches due to inflexible types as programmers can fall-back to
dynamic checks which are equally safe.

In this paper, the Dala capabilities only constrain heap
structures. Nothing prevents a stack variable in an immutable
object to point to an unsafe object. To reason about data-race
freedom of a method call on a safe object, we need to consider the
methods arguments' capabilities. Extending the structural
constraints to stack variables is an interesting point in the
design space: if immutable objects can only ``see'' other
immutable objects, method calls on immutables are guaranteed to be
side-effect free modulo allocation and GC. For isos, side-effects
are not possible, but preexisting objects may be updated in place.
Finally, local objects would only observe local objects belonging
to the same thread (which is probably very desirable), and permit
side-effects visible in the current thread only. Such a design can
reduce the number of checks (\eg{} all checks of thread-ownership
happen only when calling a local method in an unsafe context).

\section{Conclusion}
\label{sec:conclusion}

A data race is a fundamental, low-level aspect of a program which
is not tied to the intended semantics of a particular application.
While many race conditions stem from data-races, data-race freedom
does not mean absence of race conditions. Data-race freedom is
however still important: removing them removes many race
conditions and moreover makes a program's semantics independent on
the idiosyncrasies of a particular (weak) memory model. In the
case of programming languages like C and C++, data-races are
examples undefined behaviour. The Dala capabilities guarantee
absence of data races in safe objects by imposing restrictions on
all code that interacts with these objects. Safe and unsafe objects
can co-exist and the presence of the latter does not compromise
the guarantees of the former.

Dala helps programmers structure their programs with capabilities
including immutable, isolated, and thread-local. We support the \LangName{} design with a formal
model, clear and proven properties \wrt{} data-race freedom and
semantics preservation when capabilities are added to a program.

We provide \LangNameImpl{} (\cref{sec:implementation}), an early proof-of-concept prototype implementation
which is available as open source.\footnote{https://github.com/gracelang/moth-SOMns/tree/daddala}
Based on this last experience, we believe that our model can
provide opt-in data-race safety to programmers on top of existing
languages, with relatively little implementation difficulty and
overhead.



\bibliography{biblio}

\appendix

\newpage

\section{Expression Rules Producing Errors}
\label{sec: expression rules errors}

\cref{fig:errors extended} shows all the expression rules producing errors.

\begin{figure*}[!ht]
\[
\ntyperule{E-NoSuchField}
{~\\
  H(x.f)=\bot
}
{
H;  E[x.f]
\ReducesTo
H;  \err
}
\qquad
\ntyperule{E-NoSuchMethod}
{
  H(x) = \loc \quad
  H(\loc)=\Obj{\_}{\_}{\Ms} \\
  m \not\in \textit{names}(\Ms)
}
{
  H; E[x.m(v)]
  \ReducesTo
  H;  \err
}
\qquad
\ntyperule{E-NoSuchFieldAssign}
{ ~ \\
  H(x.f)=\bot
}
{
H;  E[x.f = v]
\ReducesTo
H;  \err
}
\]
\[
\ntyperule{E-SendBadTargetOrArgument}
{
  H(\loc)=\Obj{\_}{\_}{\_} \lor
  H(\loc')=\Chan{\_}
}
{
H;  E[\move{\loc}{\loc'}], T
\ReducesTo
H;  \err
}
\quad
\ntyperule{E-RecvBadTarget}
{ H(\loc)=\Obj{\_}{\_}{ \_}}
{H; E[\recv{\loc}] \ReducesTo H; \err}
\quad
\ntyperule{E-CastError}
{H(\loc) = \Obj{K'}{\_}{\_} \quad
  K' \neq K}
{H; E[\CC{K}{\loc}]
\ReducesTo
H; \CastErr
}
\]
\[
\ntyperule{E-AbsentVar}
{
H(x) = \Absent
}
{
H;  E[x]
\ReducesTo
H;  \CErr
}
\quad
\ntyperule{E-Consume}
{
H(x) = \Absent
}
{
H;  E[\consume{x}]
\ReducesTo
H;  \CErr
}
\quad
\ntyperule{E-AbsentTarget}
{
H(x) = \Absent
}
{
H; E[x.m(v)]
\ReducesTo
H; \CErr
}
\]
\[
\ntyperule{E-AbsentTargetAccess}
{H(x) = \Absent}
{H; E[x.f] \ReducesTo H; \CErr}
\qquad
\ntyperule{E-AbsentFieldAssign}
{H(x) = \Absent}
{H; E[x.f = v] \ReducesTo H; \CErr}
\qquad
\ntyperule{E-AbsentCopyTarget}
{H(x) = \Absent}
{H; E[\kopy{K}{x}] \ReducesTo H; \CErr}
\]
\[
\ntyperule{E-AliasIso}
{
  H(x)=\loc \\
  \Helper{isIso}{H, \loc}
}
{
H;  E[x]
\ReducesTo
H;  \PErr
}
\qquad
\ntyperule{E-IsoField}{
H(x) = \loc' \quad   H(x.f)=\loc \\
\Helper{localOwner}{H, i, \loc'} \quad
\Helper{isIso}{H, \loc} 
}{
  H;  E[x.f]^i
  \ReducesTo
  H;  \PErr
}
\qquad
\ntyperule{E-BadInstantiation}
{
  \exists v \in \overline{v}.\\
  \neg \OkRef{K}{v} \lor \Helper{notLocalOwner}{H, i, v}
}
{
  H;  E[\Obj{K}{\overline{f = v}}{\Ms}]^i
  \ReducesTo
  H;  \PErr
}
\]
\[
\ntyperule{E-BadFieldAssign}
{
  H(x) = \loc \quad H(\loc) = \Obj{K}{\_\ f = v'}{\Ms} \\
  \Helper{isImm}{H, \loc} \lor \neg \OkRef{K}{v} \lor
 \Helper{notLocalOwner}{H, i, \loc}
 \lor (\Helper{isLocal}{H, \loc} \land \Helper{isOwner}{H, i, \loc} \land
 \Helper{notLocalOwner}{H, i, v})
}
{
H;  E[x.f = v]^i
\ReducesTo
H;  \PErr
}
\]
\[
\ntyperule{E-CopyTarget}
{H(x) \neq \Absent \quad \Helper{notLocalOwner}{H, i, H(x)}}
{H; E[\kopy{K}{x}]^i \ReducesTo H; \PErr}
\quad
\ntyperule{E-LocalField}{
  H(x) = \loc \quad H(x.f) = v \qquad
\Helper{notLocalOwner}{H, i, \loc}
}{
  H;  E[x.f]^i
  \ReducesTo
  H;  \PErr
}
\]
\caption{Expression rules producing errors. To reduce clutter, we
  write $H(x.f)=v$ when
  $H(x) = \loc \land H(\iota)=\Obj{\_}{F}{\_}$ and $f=v\in F$, and
  $H(x.f)=\bot$ when $H(x) = \loc \land H(\loc)=\Obj{\_}{F}{\_}$
 and $f\not\in\dom{F}$, and $\Helper{notLocalOwner}{H, i, v} = \Helper{isLocal}{H, v} \land \neg \Helper{isOwner}{H, i, v}$.}
\label{fig:errors extended}
\end{figure*}

\section{Definition, Lemmas, and Proofs}
\label{sec: defs lems proofs}


\begin{restatable}[Initial Configuration]{definition}{INITIAL}\label{def:init}
An initial configuration is a closed term with empty heap, $\epsilon; t$.
\end{restatable}

\begin{restatable}[Terminal Configuration]{definition}{FINAL}\label{def:terminal}
  A well-formed configuration $\Gamma \vdash H;   \overline{T} $
  is terminal if
  it contains zero threads
  ($\overline{T} = \epsilon$),
  it is an error
  ($\overline{T} = \textit{Err}$),
  or if it is a deadlock configuration
  ($\Helper{Deadlock}{\Gamma \vdash H;  \overline{T}}$).
\end{restatable}

\begin{restatable}[Deadlock Configuration]{definition}{DEADLOCK}\label{def: deadlock}
  A deadlocked configuration is a well-formed configuration
  where all threads are blocked on sends and receives.
  \begin{align*}
  & \Helper{Deadlock}{\Gamma \vdash H;   \overline{T}} \iff \\
  &  \overline{T} \neq \epsilon \land \forall T' \in \overline{T}.\lor\left\{
    \begin{array}{l}
    T'=E[\recv{\loc}] \land H(\loc) = \Chan{\_, \varnothing} \\
    T'=E[\move{x}{\_}] \land H(\loc) = \Chan{\_, v} \\
    T'=E[\blacksquare_i\,\loc] \land H(\loc) = \Chan{i, \_}
    \end{array}\right.
  \end{align*}
\end{restatable}

\begin{restatable}[Data Race]{definition}{DATARACE}\label{def:data race}
  To \textup{formally} define a data race, we introduce the notion of a
  trace $R$, which is an ordered sequence of reductions $r_1$,
  $r_2$\ldots{} in the evaluation of a program $P$. We only concern
  ourselves with traces that do not evaluate to a permission or cast error.

  We define the precedes relation $<$ on reductions thus: If
  $R=r_1,R'$, then $r_1 < r_2$ for all $r_2\in R'$. Recall that
  threads have unique ids.

We are now ready to formally define a data race in Dalarna.
A program has a data race if it may give rise to at least one
(non-erring) trace with a data race. A trace $R$ has a data race if
it contains steps $r_1 < r_2$ such that
  \begin{enumerate}[a)]
    \item  $r_1 = H_1; E[x.f = \_]^i$, and $H_1(x) = l$ \Remark{(thread $i$ writes $l.f$)}
    \item  $r_2 = H_2; E[y.f = \_]^j$ or $r_2 = H_2; E[y.f]^j$, and $H_2(y) = l$ \Remark{(thread $j$ reads or writes $l.f$)}
    \item  $i\neq j$ \Remark{(the conflicting accesses take place in different threads}
  \end{enumerate}
  and there does not exist $r_3, r_4 \in R$ such that
  \begin{enumerate}[i]
  \item $r_1 < r_3 < r_4 < r_2$,
  \item $r_3 = H_3; E[\_ \leftarrow v_1]^i$ and $l \in ROG(H, v_1)$ \Remark{(thread $i$ sends $l$ somewhere)}
  \item $r_4 = H_4;[\leftarrow v_2]^j$ and $l \in ROG(H, v_2)$ \Remark{(thread $j$ receives $l$ from somewhere)}
  \end{enumerate}

\noindent
where $ROG(H, v)$ denotes the transitive closure of objects (the Reachable Object Graph) in $H$ rooted in $v$.
\end{restatable}

\begin{restatable}[Trace]{definition}{TRACE}\label{def:trace}
Define a trace $\mathcal{T}$ as a list of pairs containing the reduction rules and thread ids.
\end{restatable}

\begin{restatable}[Trace Of A Program]{definition}{TRACEPROGRAM}\label{def:tracereduction}
  Define the trace $\mathcal{T}$ of a program $P$ as the accumulation
  of pairs of reduction rules and thread ids on which the reduction rules take place,
  \eg{} $\epsilon; P \ReducesTo^* H'; \overline{T'}$ produces trace $\mathcal{T}$.
\end{restatable}

\begin{restatable}[Trace Reduction Replay]{definition}{TRACEREPLAY}\label{def:tracereplay}
\sloppy Define a trace replay reduction relation $(\mathcal{T:T'}); H;
\overline{T} \Replays{} \mathcal{T'}; H'; \overline{T'}$ as the reduction
relation that executes the first item of the trace $\mathcal{T:T'}$ at a time,
returning the trace list without the first item ($\mathcal{T}$), a new heap
($H'$), and a new thread configuration ($\overline{T'}$).
\end{restatable}

\cref{def: thread-local} states that if an object is reachable from threads $t$
and $t'$ then either $t=t'$ or the object is not local or the object can only be
dereference in a single thread (via ownership check). Thus, thread-local objects
are not reachable by multiple threads. The helper function $\Helper{ROG}{H,t}$
returns the set of objects in $H$ reachable from $t$ by traversing variables and
fields (a reflexive transitive closure of objects).

\begin{definition}[Thread-Local Objects]\label{def: thread-local}
  A configuration satisfies object ``thread-locality'' if no object with a local capability is reachable from more than one thread.
\begin{align*}
&\Helper{Local}{H, \overline{T}} \iff \forall t^i, t'^j\in\overline{T}. \\
& \qquad \qquad \loc \in (\Helper{ROG}{H, t^i}\cap\Helper{ROG}{H, t'^j}) \Rightarrow  \\
& \qquad \qquad (t=t' \land i = j) \lor (H(\loc) = \Obj{K}{\_}{\_} \land K \neq \localkw{}) \lor \\
& \qquad \qquad (t \neq t' \land i \neq j \land (\Helper{isLocal}{H, \loc} \\
& \qquad \qquad\Rightarrow \Helper{isOwner}{H, i, \loc} \land \neg \Helper{isOwner}{H, j, \loc})
\end{align*}
\end{definition}

Object isolation~(\cref{def: object iso}) states that all isolated objects have a single reference to them, from the heap or stack,
modulo borrowing. Borrowing allows calling methods on isolated objects without consuming
the target, leading to (temporal) aliases which are all on the stack of the same thread.
 We forbid consuming \self{}, thus
aliases of isolated objects (and \self{}) are only introduced through method calls. After a method
call, the alias is buried and non-accessible.

\begin{definition}[Object Isolation]\label{def: object iso}
  A configuration satisfies object isolation if no objects with an \isokw{} capability have more than one incoming pointer from the stack and heap, modulo borrowing.
    \begin{align*}
    &\Helper{Isolated}{H, \overline{T}} \iff  \\
    & \quad \forall\loc\in \Helper{Isos}{H} \,.\,
 |\Helper{Inc}{H, \overline{T}, \loc} \cup \Helper{Inc}{H, \loc}| > 1 \Rightarrow  \\
    & \quad \quad \exists t\in\overline{T} \,.\, \Helper{Inc}{H, \overline{T}, \loc}=\Helper{Inc}{H, t, \loc} \land \Helper{Inc}{H, \loc} = \emptyset
    \end{align*}
  where \textsf{Inc} collects the set of variables and fields aliasing a particular location:
  $\Helper{Inc}{H, v} = \{ \loc.f \,|\, \loc \in \dom{H} \land H(\loc.f) = v \} \cup \{ \loc \,|\, \loc \in \dom{H} \land H(\loc) = \Chan{i, v}\}$,
  $\Helper{Inc}{H, \overline{T}, v} = \bigcup_{t\in\overline{T}} \Helper{Inc}{H, t, v}$,
  $\Helper{Inc}{H, t, v} = \{ x \,|\, x \in \Helper{FreeVars}{t} \land H(x) = v \}
   \cup \{ v \mid v \in t \}$, and $\Helper{FreeVars}{t}$ is the set of all free variables in $t$.
\end{definition}

\begin{definition}[Object's Owner]
  Define the predicate $\Helper{isOwner}{H, i, v}$ for a heap ($H$), owner ($i$),
  and value ($v$) as:
\begin{align*}
\Helper{isOwner}{H, i, v} \iff H(\loc) = \Obj{K^i}{\_}{\_},\\
\qquad \textit{where}\ v = \loc \land \loc \in \dom{H}
\end{align*}
\end{definition}

\begin{definition}[Safe Erasure]\label{def:safe stripping}
  Define the function \textit{safe erasure}, $^e$, as the function that replaces
  safe capabilities to \unkw{} capability as follows:
\begin{align*}
&\erase{\Gamma, x : K} = \erase{\Gamma}, x : \unkw  \\
&\erase{\Gamma, \loc : K} = \erase{\Gamma}, \loc : \unkw \\
&\erase{H, x \mapsto \loc} = \erase{H}, x \mapsto \loc \\
&\erase{H, \loc \mapsto \Chan{i, v}} = \erase{H}, \loc \mapsto \Chan{i, v} \\
&\erase{H, \loc \mapsto \Obj{K}{\_}{\_}} = \erase{H}, \loc \mapsto \Obj{\unkw}{\_}{\_} \\
&\erase{t \ \overline{T}} = \erase{t}\ \erase{\overline{T}} \\
&\erase{t} = \begin{cases}
 \Obj{\unkw{}}{\overline{f=\erase{v}}}{\overline{\erase{M}}}, & t = \Obj{K}{\overline{f=v}}{\Ms} \\
\CC{\unkw}{\loc}, & t = \CC{K}{\loc} \\
\kopy{\unkw}{x}, & t = \kopy{K}{x} \\
\LetIn{x}{\erase{e}}{\erase{t}}, & t =  \LetIn{x}{e}{t}\\
\Spawn{x}{\erase{t}}, & t =  \Spawn{x}{t}\\
\Method{m}{x}{\erase{t}}, & t =  \Method{m}{x}{t}\\
t, & \textit{otherwise}
\end{cases}
\end{align*}
\end{definition}

\begin{definition}[Capability Extraction]
\[
\CK{H}{x} = \begin{cases}
K, & H(x) = \loc \land H(\loc) = \Obj{K}{\_}{\_} \\
\localkw{}, & H(x) = \loc \land H(\loc) = \Chan{\_, \_} \\
K, & H(x) = \Absent, \textit{ for some } K
\end{cases}
\]
\end{definition}

\begin{definition}[Thread's Variables]
The variables of a thread are all variables used in the thread.
\begin{align*}
& \Helper{Vars}{\LetIn{x}{e}{t}} = \{ x \} \cup \Helper{Vars}{e} \cup \Helper{Vars}{t} \\
& \Helper{Vars}{x} = \Helper{Vars}{\consume{x}}
                   = \Helper{Vars}{x.f} =  \Helper{Vars}{\kopy{K}{x}} =\{ x \}
\\
& \Helper{Vars}{x.f = e} = \Helper{Vars}{x.m(e)} = \{ x \} \cup \Helper{Vars}{e} \\
& \Helper{Vars}{\move{e}{e'}} = \Helper{Vars}{e} \cup \Helper{Vars}{e'} \\
& \Helper{Vars}{\CC{K}{e}} = \Helper{Vars}{\recv{e}} = \Helper{Vars}{e} \\
& \Helper{Vars}{\Obj{K}{\overline{f = w}}{\_}} = \Helper{Vars}{\overline{w}} \\
& \Helper{Vars}{\Spawn{x}{t}} = \Helper{Vars}{\_} = \epsilon
\end{align*}
\end{definition}

\begin{definition}[Reachable Object Graph]\label{app: rog}
Define the helper function $\Helper{ROG}{H, t}$ and $\Helper{ROG}{H, \loc}$ as
the function that returns the transitive closure of objects reachable from term
$t$ and location $\loc$.
\begin{align*}
\Helper{ROG}{H, t}  &= \bigcup_{x \in \Helper{DefVar}{H, t}} \Helper{ROG}{H, H(x)}
\cup \{ \Helper{ROG}{H, \loc} \mid \loc \in t \} \\
\Helper{ROG}{H, v'} &=
\begin{cases}
\bigcup_{f=v\in\Fs}\, \Helper{ROG}{H, v} \cup \{ \loc \},& v' = \loc \land
H(\loc) = \Obj{K}{\Fs}{\Ms}
\\
\emptyset, & \textit{otherwise}
\end{cases}
\\
\Helper{DefVar}{H, t} &= \{ x \mid x \in t \land x \in \dom{H} \}
\end{align*}
\end{definition}


\begin{lemma}[Well-formed Values Are Not Absent (Or Null)]
\label{lem:absent-value}
\[
\Gamma \vdash H;  E[v] \Rightarrow v \neq \varnothing \land v \neq \Absent
\]
\begin{proof}
By induction on the well-formedness of the configuration.
\end{proof}
\end{lemma}

\begin{lemma}[Fresh Variables]\label{lem:fresh-var}
  \begin{align*}
  &\Gamma \vdash H;  \LetIn{x}{e}{t} \Rightarrow x \notin \dom{\Gamma} \land x \notin \dom{H} \\
  &\Gamma \vdash H;  \Spawn{x}{t} \Rightarrow x \notin \dom{\Gamma} \land x \notin \dom{H}
  \end{align*}
  \textit{Explanation.} When a term or expression introduces a new variable, the variable is new to $\Gamma$ and $H$.
\begin{proof}
Immediate by definition of well-formed configuration.
\end{proof}
\end{lemma}

\begin{lemma}[Store Contains Top-Level Variables]\label{lem:store-var}
\begin{align*}
&\Gamma \vdash H;  \LetIn{x}{e}{t} \Rightarrow \Helper{FreeVars}{e} \cup \Helper{FreeLoc}{e} \subseteq \dom{H}
\end{align*}
\textit{Explanation.} Given a thread $T = \LetIn{x}{e}{t}$, the free variables
and locations of the expression $e$ are in the domain of the store.
\begin{proof}
Immediate from definition of well-formed configuration.
\end{proof}
\end{lemma}

\begin{lemma}[Weakened Well-formed Environment]\label{lem:weak-env-heap}
\[
\Gamma \vdash H \land x \notin \dom{\Gamma}  \Rightarrow  \Gamma, x : K \vdash H
\]
\textit{Explanation}. If the store is a well-formed \wrt{} an environment, then
extending the environment cannot contradict this fact.
\begin{proof}
Straightforward induction on the well-formedness rules.
\end{proof}
\end{lemma}

\begin{lemma}[Typing Permutation]\label{lem:permutation-env}
\[
\Gamma \vdash t \land \Delta \textit{ is a permutation of } \Gamma \Rightarrow \Delta \vdash t
\]
\begin{proof}
By induction on the typing derivation $\Gamma \vdash t$.
\end{proof}
\end{lemma}

\begin{lemma}[Weakening Typing Environment]\label{lem:weak-env}
\begin{align}
& \Gamma \vdash t \land x \notin \dom{\Gamma} \land x \notin \Helper{Vars}{t}
  \Rightarrow \Gamma, x : K \vdash t \label{weak1}  \\
& \label{weak2}
  \Gamma \vdash t \land \loc \notin \dom{\Gamma} \Rightarrow \Gamma, \loc : K \vdash t
\\
& \Gamma \vdash e \land x \notin \Gamma \Rightarrow \Gamma, x : K \vdash e
\label{weak3}
\end{align}
\begin{proof}
  By induction over the term well-formedness.
  There are multiple cases to handle.
  We show the case when $t = \LetIn{y}{e}{t'}$ which is the more interesting one.
  Case $t = \LetIn{y}{e}{t'}$. By the initial assumptions:
  \begin{align}
    &\Gamma \vdash \LetIn{y}{e}{t'} \label{wk:1} \\
    &x \notin \dom{\Gamma} \label{wk:2}\\
    &x \notin \Helper{Vars}{t} \label{wk:3}
  \end{align}
  \textbf{Need to show}
  \[
  \Gamma, x: K \vdash \LetIn{y}{e}{t'}
  \]
  We prove each of the components of rule \RN{WF-Let}.
  \begin{enumerate}
    \item By the initial assumption $\Gamma \vdash \LetIn{y}{e}{t'}$ (\ref{wk:1}) and \textsc{WF-Let}, $y \notin \dom{\Gamma}$. \label{wk:i1}
    \item By \ref{wk:i1} and \ref{wk:3}, we can conclude that $y \notin \dom{\Gamma, x : K}$.
    \item For any $e$, we must prove that $\Gamma, x: K \vdash e$ is well-formed.
      We apply the induction hypothesis (\cref{weak3}), and conclude that $\Gamma, x: K \vdash e$ is
      well-formed  \label{wk:i3}
    \item We need to show that $\Gamma, x : K, y : K' \vdash t'$. By the initial
      hypothesis $\Gamma \vdash \LetIn{y}{e}{t'}$ (\ref{wk:1}). As part of the assumptions
      $\Gamma, y : K' \vdash t'$ holds. By the induction hypothesis \cref{weak1},
      we conclude that $\Gamma, x : K, y : K' \vdash t'$. \label{wk:i4}
    \item From \ref{wk:i3} and \ref{wk:i4} and by typing rule \RN{WF-Let},
      we can conclude $\Gamma, x: K \vdash \LetIn{y}{e}{t'}$.
  \end{enumerate}

  The proof of \cref{weak2,weak3} follow the same proof strategy.
\end{proof}
\end{lemma}

\begin{lemma}[Weakening]\label{lem:weakening}
\begin{align*}
& \Gamma \vdash H;  t \land \Gamma' = \Gamma, \Gamma'' \land
(\dom{\Gamma''} \cap \dom{\Gamma} \cap \Helper{Vars}{t} = \epsilon)
 \land H' \supseteq H \\
 & \qquad \land  \Gamma' \vdash H' \Rightarrow \Gamma' \vdash H';  t
\end{align*}
\begin{proof}
  By induction on the shape of $t$. The most interesting case is $t = \LetIn{x}{e}{t'}$
  and the others follow the same proof technique. We need to prove:
\begin{align*}
& \Gamma \vdash H;  \LetIn{x}{e}{t'} \land \Gamma' = \Gamma \cup \Gamma' \land
\Gamma' \supseteq \Gamma \land H' \supseteq H \\
& \qquad \land \Gamma'' \cap \Gamma \cap \Helper{Vars}{\LetIn{x}{e}{t'}} = \epsilon \land  \Gamma' \vdash H' \\
& \qquad \Rightarrow \Gamma' \vdash H';  \LetIn{x}{e}{t'}
\end{align*}
  We prove that if all components are well-formed, then by rule \RN{WF-Term}
  $\Gamma' \vdash H';  \LetIn{x}{e}{t'}$.
  \begin{enumerate}
    \item From the assumptions, $\Gamma' \vdash H'$. \label{lem:weakening1}
    \item We want to prove that $\Gamma' \vdash \LetIn{x}{e}{t'}$.
      From the initial assumptions, $\Gamma \vdash \LetIn{x}{e}{t'}$ and
      $x \notin \dom{\Gamma}$ and $x \notin \Helper{Vars}{t}$ and $\Gamma' \supseteq \Gamma$.
      By multiple applications of \cref{lem:weak-env}, we conclude
      $\Gamma' \vdash \LetIn{x}{e}{t'}$.\label{lem:weakening2}
    \item From \ref{lem:weakening1} and \ref{lem:weakening2}, and by application
      of \RN{WF-Term}, conclude $\Gamma' \vdash H';  \LetIn{x}{e}{t'}$.
  \end{enumerate}
\end{proof}
\end{lemma}

\begin{lemma}[Weakening Maintains Thread Locality]\label{lem:local-values}
\begin{align*}
& \overline{T} \equiv t\ \overline{T'} \land \Helper{Local}{H, \overline{T}} \land \loc \in \Helper{ROG}{H, t}
  \\
  & \quad \land \Helper{isLocal}{H, \loc} \land x \notin \dom{H}
    \Rightarrow \Helper{Local}{(H, x \mapsto \loc), t \ \overline{T'}}
\end{align*}
\textit{Explanation.} Extending the heap with new variables does not affect thread-local
objects.
\begin{proof}
Trivially satisfied by induction on the definition of $\Helper{Local}{H, \overline{T}}$.
\end{proof}
\end{lemma}

\begin{lemma}[Binding Isolate Locations Maintain Isolatedness]\label{lem:linear-values}
\begin{align*}
\overline{T} \equiv t\ \overline{T'} \land \Helper{Isolated}{H, \overline{T}} \land
\loc \in t
\land \Helper{isIso}{H, \loc} \Rightarrow \Helper{Isolated}{(H, x \mapsto \loc), \overline{t\ T'}}
\end{align*}
\textit{Explanation.} Extending the heap with a stack variable over an isolated location
maintains isolation.
\begin{proof}
Trivially satisfied by induction on the definition of $\Helper{Isolated}{H, \overline{T}}$.
\end{proof}
\end{lemma}

\begin{lemma}[Thread-Local Term Substitution Maintains Locatity]\label{lem:thread-subs}
\begin{align*}
t' \subseteq t &\land \Helper{Local}{H, t\ \overline{T}} \land \loc \in \Helper{ROG}{H, t}
  \land \Helper{isLocal}{H, \loc} \\
&  \land x \notin \dom{H} \land \Helper{FreeVar}{t'} \subseteq \Helper{FreeVar}{t} \cup \{ x \} \\
&  \Rightarrow \Helper{Local}{(H, x \mapsto \loc),  t'\ \overline{T}}
\end{align*}
\textit{Explanation.} If an object $\loc$ is thread-local to term $t$, there is a subterm
$t'$ of $t$, and we extend the heap to map a stack variable to $\loc$, then
replacement of $t$ by $t'$ maintains thread-locality under the modified heap.
\begin{proof}
By induction on the shape of $t'$. $t'$ can only use free variables
available in $t$ and the other threads cannot point to the introduced variable
$x$ as $x \notin \dom{H}$. Hence, thread-locality is maintained for object in $\loc$.
\end{proof}
\end{lemma}

\begin{lemma}[Deterministic Evaluation]\label{lem:deterministic}
If $H; T\ \overline{T} \ReducesTo H'; \overline{T'}\ \overline{T}$ and $H; T\ \overline{T} \ReducesTo H''; \overline{T''}\ \overline{T}$, then $\overline{T'} \equiv \overline{T''}$
\begin{proof}
By induction on a derivation of $H; T\ \overline{T} \ReducesTo H'; \overline{T'}\ \overline{T}$.
\end{proof}
\end{lemma}

\begin{lemma}[Capability Stripping Preserves Terms and Expressions]\label{lem:term-rewritting}
$(t^i)^e = t^i[K/\unkw]$
  \textit{Explanation:} A capability stripped term is equivalent to a replacement
  of its capabilities $K$ by $\unkw{}$, \ie{} preserves the terms and expressions.
\begin{proof}
  By induction on the term structure.
\end{proof}
\end{lemma}

\THREADLOCAL*
\begin{proof}
  By induction on the thread structure.  Assume we are deadling with thread local
  objects.  The only way to create a thread local object with a reference to
  a thread local object with different owner is by rule
  \textsc{R-FieldUpdate}. But this rule does not satisfy the premises when the owner of the
  target and source objects have different owners. Instead, rule \textsc{E-BadFieldAssign} applies
  and throws a permission error. 
\end{proof}

\begin{restatable}[Unsafe Programs Do Not Raise Permission Errors]{lemma}{UnsafeErr}\label{lem:unsafe err}
If
$\Gamma^e \vdash (H;\overline{T})^e$
and
$(H; \overline{T})^e \ReducesTo H'; \overline{T'}$,
then
$\CastErr \not\equiv \overline{T'} \not\equiv \PErr{}$.
\end{restatable}
\begin{proof}
  By induction on a derivation of the reduction derivation and the thread
  configuration. Trivially, configurations that do not throw runtime errors
  satisfy the inductive hypothesis; configurations where $T' \equiv \err$ or $T' \equiv \CastErr$
  trivially satisfy the hypothesis; from the reduction rules that throw permission errors:
  \textsc{E-AliasIso}, \textsc{E-IsoField}, \textsc{E-BadFieldAssign}, \textsc{E-CopyTarget}, \textsc{E-BadInstantiation} and \textsc{E-LocalField}
  require safe capabilities to throw a permission error but by the initial assumptions,
  we have stripped the safe capabilities. Thus, these errors cannot happen.
  Reduction rule \textsc{E-BadInstantiation} cannot happen because all capabilities are
  \unkw{} and $\Helper{OkRef}{...}$ is trivially satisfied.
\end{proof}



\PROGRESS*
\begin{proof}
  \textbf{Assumptions $\Gamma \vdash H;  \overline{T} $.}

  We start by induction on the shape of $\overline{T}$.
  \begin{enumerate}
    \item $\overline{T} \equiv \textit{Err}$. Progress holds trivially.

    \item $\overline{T} \equiv \epsilon$. Progress holds trivially.

    \item $\overline{T} \equiv T \ \overline{T'}$. Induction hypothesis:
      \begin{description}
        \item[IH1] $T = \textit{Err} \lor \Helper{Deadlock}{\Gamma \vdash H; T} \lor T = \epsilon \lor H; T \ReducesTo H'; \overline{T''}$
        \item[IH2] $\overline{T'} = \textit{Err} \lor \Helper{Deadlock}{\Gamma \vdash H; \overline{T'}} \lor \overline{T'} = \epsilon \lor H; \overline{T'} \ReducesTo H'; \overline{T''}$
        \item[IH3] $\Helper{Deadlock}{\Gamma \vdash H; T\ \overline{T'}}$
      \end{description}
      We continue by case analysis over \textbf{IH1}.
      \begin{enumerate}
        \item $T \equiv \textit{Err}\ \overline{T'}$. By the equivalence rules,
          $T \equiv \textit{Err}$ and is a terminal configuration.
        \item $T \equiv \epsilon$, then we continue by \textbf{IH2}.
        \item $H; T \ReducesTo H'; \overline{T''}$ by \textsc{C-Eval}.
        \item $\Helper{Deadlock}{\Gamma \vdash H; T}$. We continue by case analysis over \textbf{IH2}:
          \begin{enumerate}
          \item $\overline{T'} \equiv \textit{Err}$. By the equivalence rules, $T\ \textit{Err} \equiv \textit{Err}$ and is a terminal configuration.
          \item $\overline{T'} = \epsilon$. Then $\Gamma \vdash H; T$ is a terminal configuration.
          \item $H; \overline{T'} \ReducesTo H'; \overline{T''}$, the configuration steps to
            $H; T\ \overline{T'} \ReducesTo H'; T\ \overline{T''}$ by rule \textsc{C-Eval}.
          \item $\Helper{Deadlock}{\Gamma \vdash H; \overline{T'}}$, then we proceed by \textbf{IH3},
          $\Helper{Deadlock}{\Gamma \vdash H;  T \ \overline{T'}}$, and progress holds trivially.
          \end{enumerate}
      \end{enumerate}

    \item $\overline{T} \equiv t$. By the induction hypothesis either $\Gamma \vdash H;
      t$ is a terminal configuration or it can make progress. \label{progress: t}

      \begin{description}
      \item[A1] $\Helper{Deadlock}{\Gamma \vdash H;  t} \lor H; t \ReducesTo H'; \overline{T'}$
      \end{description}
      We continue by case analysis over \textbf{A1}.
      \begin{itemize}
        \item $\Helper{Deadlock}{\Gamma \vdash H;  t}$. Progress holds trivially.
        \item $H; t \ReducesTo H'; \overline{T'} $. 

          From the initial assumption and \RN{WF-Configuration}:
          \begin{align}
          & \Gamma \vdash H \label{prg:a1} \\
          & \Gamma \vdash H;  t \label{prg:a2} \\
          & \forall t' \in \overline{T}.\ \Gamma \vdash H;  t' \label{prg:a3}
          \end{align}

          We proceed by induction over the shape of $t$, and assume that all
          subexpressions were reduced using the evaluation context $E$ rules.  We
          prove all the inductive cases where the evaluation context reduces an
          expression to a value.

\newcommand{\xIndomH}{$x \in \dom{H}$ by \cref{lem:store-var}.
}
\newcommand{\lIndomH}{$\loc \in \dom{H}$ by \cref{lem:store-var}
}

          \begin{enumerate}
          \item $t = \varnothing$. This cannot happen, follows from
            \cref{lem:absent-value} with \cref{prg:a2}.

          \item $t \equiv \LetIn{x}{v}{t'}$.
            \begin{enumerate}
            \item \label{itm:let1} $x \notin \dom{H}$ from \cref{lem:fresh-var}
            \item \ByWithWhere{R-Let}{itm:let1}{H, x \mapsto v; t'}{x \notin \dom{H}}
            \end{enumerate}
          \item $t \equiv v$. By the equivalence rules (\cref{fig: config
            concurrency}), $H; v \equiv H; \epsilon$ and we apply the induction
            hypothesis to this configuration.
          \item $t \equiv E[x]$. \label{progress: E[x]}
            \begin{enumerate}
              \item \label{itm:x1} \xIndomH{}
              \item \label{itm:x2} $x \mapsto v \in H$ by assumption \cref{prg:a1} with \cref{itm:x1}.
              \item \label{itm:x3} From \cref{itm:x2} we proceed by case analysis on the shape of $v$:
              \begin{itemize}
              \item $H(x) = \Absent$. \By{E-AbsentVar}{H; \CErr}
              \item $H(x) = \varnothing$.
                This cannot happen, follows from \cref{lem:absent-value}  with \cref{prg:a2}..
              \item $H(x) = \loc$. $\loc \in \dom{H}$ by assumption
                \cref{prg:a1}. There are two cases to consider, whether the
                location is to an isolated object or not.
                \begin{itemize}
                  \item $\neg \Helper{isIso}{H, \loc}$. \By{R-Var}{H; E[\loc]}
                  \item $\Helper{isIso}{H, \loc}$. \By{E-AliasIso}{H; \PErr}
                \end{itemize}
              \end{itemize}
            \end{enumerate}
          \item $t \equiv E[\consume{x}]$.
            \begin{enumerate}
              \item \label{itm:consume1} \xIndomH{}
              \item \label{itm:consume2} $x \mapsto v \in H$ by assumption \cref{prg:a1} with \cref{itm:consume1}.
              \item From \cref{itm:consume2} we proceed by case analysis on the
                shape of $v$:
              \begin{itemize}
              \item $v = \Absent$. \By{E-Consume}{H; \CErr}
              \item $v = \varnothing$.
                This cannot happen, follows from \cref{lem:absent-value}  with \cref{prg:a2}..
              \item $v = \loc$. $\loc \in \dom{H}$ by assumption
                \cref{prg:a1}, and $H; E[\consume{x}] \ReducesTo H[x \mapsto \Absent]; E[\loc]$
                by \textsc{R-Consume}.
              \end{itemize}
            \end{enumerate}
          \item $t \equiv E[\CC{K}{\loc}]$.
            \begin{enumerate}
              \item \label{itm:cc1} \lIndomH{}
              \item We proceed by case analysis
                on the object capability of the location $\loc$, checking whether
                the object capability matches the casted capability.
              \begin{itemize}
              \item $H(\loc) = \Obj{K}{\_}{\_}$. \By{R-CastLoc}{H; E[\loc]}
              \item $H(\loc) = \Obj{K'}{\_}{\_} \land K \neq K'$. \By{E-CastError}{H; \CErr}
              \end{itemize}
            \end{enumerate}
          \item $t \equiv E[x.f]$. There are multiple cases to handle, depending
            on the value of $x$ and its field $f$.
            \begin{enumerate}
              \item \label{itm:x.f1}\xIndomH{} \label{itm:x.f}
              \item \label{itm:x.f2} $x \mapsto v \in H$ by assumption \cref{prg:a1} with \cref{itm:x.f1}.
              \item $v = \Absent \lor \loc \lor \varnothing$.
                We proceed by case analysis over $v$:
                \begin{itemize}
                \item $v = \varnothing$.
                  This cannot happen. By assumption (\ref{prg:a1}),
                  $x \mapsto \varnothing \in H$ is not well-formed.
                \item $v = \Absent$. \By{E-AbsentTargetAccess}{H; \CErr}
                \item $v = \loc$. By the assumption \cref{prg:a1},
                    $\loc \mapsto \Obj{\_}{\_}{\_} \in H$. We proceed by case analysis on
                    the field.
                    \begin{itemize}
                    \item $H(\loc) = \Obj{\_}{\Fs}{\Ms}, f \notin \dom{\Fs}$. \By{E-NoSuchField}{H; \err}
                    \item $H(\loc) = \Obj{\_}{\_ f = v}{\Ms}$,\\
                      $\neg \Helper{isIso}{H, \loc}, \Helper{localOwner}{H, i, \loc}$. \By{R-Field}{H; v}
                    \item $H(\loc) = \Obj{\_}{\_ f = v}{\Ms}$,\\
                      $\Helper{isIso}{H, \loc}, \Helper{localOwner}{H, i, \loc}$. \By{E-IsoField}{H; \PErr}
                    \item $H(\loc) = \Obj{\_}{\_ f = v}{\Ms}$,\\ $\Helper{noLocalOwner}{H, i, \loc}$.
                      \By{E-LocalField}{H; \PErr}.
                    \end{itemize}
                \end{itemize}
            \end{enumerate}
          \item $t \equiv E[x.f = v]$ There are multiple cases to handle, depending
            on the value of $x$ and its field $f$.
            \begin{enumerate}
            \item \xIndomH{} \label{itm:x.f = v}
            \item $x \mapsto v' \in H$ by assumption \cref{prg:a1} with \cref{itm:x.f = v}.
            \item $v = \Absent \lor \loc \lor \varnothing$.
              We proceed by case analysis over $v'$ and $v$:
              \begin{itemize}
              \item $v' = \varnothing$. This cannot happen. By assumption (\ref{prg:a1}),
                $x \mapsto \varnothing \in H$ is not well-formed.
              \item $v' = \Absent$. \By{E-AbsentFieldAssign}{H; \CErr}.
              \item $v' = \loc'$. By the assumption \cref{prg:a1}, $\loc \mapsto \Obj{\_}{\_}{\_} \in H$.
                There are multiple cases to handle. We start by case analysis
                on the value of $v$, and then continue by case analysis on field $f$:
                \begin{itemize}
                \item $v = \varnothing$. This cannot happen, follows from \cref{lem:absent-value}  with \cref{prg:a2}..
                \item $v = \Absent$. This cannot happen, by \cref{lem:absent-value}  with \cref{prg:a2}..
                \item $v = \loc$. We proceed by case analysis over field $f$ below:
                \end{itemize}
                \begin{itemize}
                \item $H(\loc') = \Obj{\_}{\Fs}{\Ms}, f \notin \dom{\Fs}$.
                  \By{E-NoSuchFieldAssign}{H; \err}.

                \item $H(\loc') = \Obj{K}{\_ f = v}{\Ms} \land K = \immkw{}$.
                  \By{E-BadFieldAssign}{H; \PErr}
                \item $H(\loc) = \Obj{K}{\_ f = v''}{\Ms} \land K \neq \immkw{},
                  v'' = \loc''$.  If the object location $\loc''$ has a
                  capability $K'$, s.t. $\OkField{K, K'}$ and the source
                  and target are owned by the current thread, then we conclude by
                  direct application of \RN{R-FieldAssign}.  If the object location
                  $\loc''$ has a capability $K'$ s.t.  $K \nleq K'$
                  or if the current thread does not own the target or if
                  the current thread is the owner of the target but not of the
                  source, then we conclude by direct application of
                  \RN{E-BadFieldAssign}, which reduces to $H; \PErr$.
                \end{itemize}
              \end{itemize}
            \end{enumerate}

         \item $t \equiv E[\move{v}{v'}]$ Reminder. We are under the assumption
           that $\neg \Helper{Deadlock}{\Gamma \vdash H;  t\ \overline{T'}}$, so
           $t$ may be blocked but there is no deadlock in the global
           configuration ($\overline{T'}$). We proceed by case analysis on the shape
           of $v$ and $v'$.

           \begin{enumerate}

           \item $v = \varnothing \lor v' = \varnothing$.
             This cannot happen, follows from \cref{lem:absent-value}  with \cref{prg:a2}..

           \item $v = \Absent \lor v' = \Absent$.
             This cannot happen, follows from \cref{lem:absent-value}  with \cref{prg:a2}..

           \item $v = \loc \land v' = \loc'$.
             There are multiple cases to handle, depending on the shape
             of the objects pointed by $\loc$ and $\loc'$.
             \begin{itemize}
               \item $H(\loc) = \Obj{\_}{\_}{\_}$. \By{E-SendBadTargetOr\-Argument}{H;\err}
               \item $H(\loc') = \Chan{\_, \_}$. \By{E-SendBadTargetOr\-Argument}{H;\err}
               \item $H(\loc) = \Chan{\_, \varnothing}$, by \RN{R-SendBlock} the configuration steps.
             \end{itemize}
           \end{enumerate}

          \item $t \equiv E[\recv{\loc}]$. By \cref{lem:store-var}, $\loc \in \dom{H}$.
            There are two cases to handle, when $\loc$ maps to an object and to a channel.
            \begin{itemize}
              \item $\loc \mapsto \Obj{\_}{\_}{\_}$. \By{E-RecvBadTarget}{H;\err}
              \item $\loc \mapsto \Chan{\_, v}$. $v = \loc'$ by assumption
                (\ref{prg:a1}) and the assumption that it takes a reduction step
                (the deadlock was handled before) $\loc \mapsto \Chan{\_,
                \loc'}$. \By{R-Recv}{H[\loc \mapsto \Chan{\_, \varnothing}];
                E[\loc']}.
            \end{itemize}
          \item $t \equiv E[x.m(v)]$
            \begin{enumerate}
              \item \label{itm:mcall}$x \in \dom{H}$ by \cref{lem:store-var}.
              \item Given (\ref{itm:mcall}), we do case analysis on the shape of
                the value pointed by $x$ in the store.
                \begin{itemize}
                  \item $x \mapsto \Absent$. \By{E-AbsentTarget}{H;\CErr}
                  \item $x \mapsto \loc$. By the assumptions of \RN{WF-Configuration},
                    $\loc \in \dom{H}$ (with \RN{WF-H-Var}). We proceed by case analysis
                    on whether $m$ exists in $\Ms$:
                    \begin{itemize}
                    \item $m \notin \Helper{names}{\Ms}$.
                    we conclude with \RN{E-NoSuchMethod} and $H; \err$.
                    \item $m \in \Helper{names}{\Ms}$.  we conclude by direct
                      application of \RN{R-Call}, $H, x' \mapsto \loc, y' \mapsto
                      v; E[t[\self = x'][y = y']]$, where $x', y'$ are fresh
                      variables and $t$ is the method body of $m$.
                    \end{itemize}
                \end{itemize}
            \end{enumerate}
          \item $t \equiv E[\kopy{K}{x}]$.
            \begin{enumerate}
            \item $x \in \dom{H}$, by \cref{lem:store-var}.
            \item By assumption (\ref{prg:a1}), $x \mapsto v$ and $v = \Absent$
            or $v = \loc$ and $\loc \in \dom{H}$. We proceed by case analysis on $v$:
            \begin{itemize}
              \item $x \mapsto \Absent \in H$. \By{E-AbsentCopyTarget}{H; \CErr}.
              \item $x \mapsto \loc \in H \land \Helper{notLocalOwner}{H, i, \loc}$.
                \By{E-CopyTarget}{H; \PErr}.
              \item $x \mapsto \loc \in H \land \Helper{localOwner}{H, i, \loc}$
                By \textsc{WF-Copy}, $\isokw{} \neq K$.
                \By{R-Copy}{H';
                E[\loc]} where $\Helper{OkDup}{H, K, \loc} = (H', \loc')$ where
              $\loc'$ is the fresh location to the duplicated object graph and
              $H'$ is the new store containing the duplicated object graph.
            \end{itemize}
            \end{enumerate}

          \item $t \equiv E[\Spawn{x}{t'}]$. By well-formed configuration
            (\RN{WF-Configuration}) with \RN{WF-Spawn}, $x \notin \dom{\Gamma}$
            and $\Helper{FreeVar}{t'} \subseteq \{ x \}$. By \cref{lem:fresh-var},
            $x \notin \dom{H}$ either.  We conclude by direct application of
            \RN{R-Spawn}, $H, x \mapsto \loc, \loc \mapsto \Chan{i, \varnothing};
            E[\loc]\ t$ where $\loc, i$ are fresh, and $t$ is a new spawned
            thread.

          \item $t \equiv E[\blacksquare_i\,\loc]$. By \cref{lem:store-var},
            $\loc$ in $\dom{H}$.  By well-formed configuration
            (\RN{WF-Configuration}), we can further assume $\loc \mapsto
            \Chan{i', v}$ (\RN{WF-Term}). We proceed by case analysis
            on the index $i$ and value $v$.
            \begin{itemize}
              \item $i = i', v = \varnothing$, then by \RN{R-SendUnblock}
                we conclude with $H; E[\loc]$.
              \item $i = i', v \neq \varnothing$, then the configuration
                is blocked. By the assumption of well-formed configuration
                \RN{WF-Configuration} there is a $t' \in \overline{T}$
                that such that it is not deadlocked. We conclude by induction
                on the set of threads $\overline{T}$.
              \item $i \neq i', \textit{any}\ v$. \By{R-SendUnblock}{H; \loc}.
            \end{itemize}
          \item $t \equiv E[\Obj{K}{\overline{f = v}}{\Ms}]$. We proceed by case
            analysis on the compatibility of the capability $K$ and the object
            capability of each value, $\overline{v}$, and $\Helper{OkRef}{H, K, v}$.
            \begin{itemize}
              \item $\forall v \in \overline{v}.\ \Helper{OkRef}{H, K, v}$.
                We proceed by case analysis over $K$
                \begin{enumerate}
                \item $K = \localkw{}$. If all values are owned by the current thread,
                by \RN{R-New} the configuration steps to $H, \loc \mapsto
                \Obj{K}{\overline{f = v}}{\Ms}; E[\loc]$ where $\loc$ is fresh.
                If a value is not owned by the current thread, by \RN{E-BadInstantiation}
                the configuration reduces to $H; \PErr$.
                \item $K \neq \localkw{}$, by \RN{R-New} the configuration steps
                  to $H, \loc \mapsto \Obj{K}{\overline{f = v}}{\Ms}; E[\loc]$
                  where $\loc$ is fresh.
                \end{enumerate}
              \item $\exists v \in \overline{v}.\ \neg \Helper{OkRef}{H, K, v}$
                by \RN{E-BadInstantiation} the configuration reduces to $H; \PErr$.
            \end{itemize}
          \end{enumerate}
      \end{itemize}
  \end{enumerate}
\end{proof}


\PRESERVATION*
\begin{proof}

\textbf{Initial Assumptions:}
\begin{align*}
  & \Gamma \vdash H;  t\ \overline{T} & \RN{WF-Configuration}
\end{align*}

\textbf{Step:}
\begin{align*}
& H; t\ \overline{T}  \ReducesTo H';  \overline{T'}\ \overline{T} & \RN{C-Eval}
\end{align*}

\textbf{Need to show:}
\begin{align*}
  & \Gamma \vdash H;  \overline{T'}\ \overline{T}
\end{align*}

By induction on the shape of the term $t$.  If it holds for one thread, then it
will hold for all thread configurations. We use the rule \RN{C-Eval} to reduce
a single thread $t$ from the bunch of threads $t\ \overline{T}$.

\begin{enumerate}


\item $t = \LetIn{x}{v}{t'}$. By the induction hypothesis, with rule \RN{R-Let}
  \begin{align*}
    &H; \LetIn{x}{v}{t'}\ \ \overline{T}  \ReducesTo H';  t'\ \overline{T} \\
    & \qquad \Rightarrow \exists \Gamma'. \Gamma' \supseteq \Gamma, \Gamma';  H, x \mapsto v \vdash t'\ \overline{T}
  \end{align*}

  From initial assumption of well-formedness and \RN{WF-Configuration}:
  \begin{align}
      &\label{preserv:let0}\dom{\Gamma} = \dom{H} \\
      &\label{preserv:let2}\Gamma \vdash H \\
      &\label{preserv:let}\forall t'' \in \LetIn{x}{v}{t'}\ \overline{T}.\ \ \Gamma \vdash H;  t'' \\
      &\label{preserv:local}\Helper{Local}{H, \LetIn{x}{v}{t'}\ \overline{T}} \\
      &\label{preserv:iso}\Helper{Isolated}{H, \LetIn{x}{v}{t'}\ \overline{T}}
  \end{align}

  \textbf{Need to show}
  \[\exists  \Gamma'. \Gamma' \supseteq \Gamma \land \Gamma' \vdash H';
    t'\ \overline{T}, \textit{ where } H' = H, x \mapsto v\]

  The strategy of the proof is to show that all components are well-formed.
  By rule \RN{WF-Configuration}, we need to show:
  \begin{align}
    & \Gamma' \supseteq \Gamma \label{p-let-gamma} \\
    & \dom{\Gamma'} = \dom{H'} \label{p-let-gamma-heap} \\
    & \Gamma' \vdash H, x \mapsto v \label{p-let-wf}\\
    & \forall t'' \in t' \ \overline{T}. \Gamma'; H, x \mapsto v \vdash t'' \label{p-let-t} \\
    & \Helper{Local}{H', t'\ \overline{T}} \label{p-let-local}\\
    & \Helper{Isolated}{H', t'\ \overline{T}} \label{p-let-iso}
  \end{align}

  We pick $\Gamma' = \Gamma, x: \CK{H}{x}$, and $H' = H, x \mapsto v$.  We now
  proof each of the components:
  \begin{itemize}
  \item We picked $\Gamma' = \Gamma, x : \CK{H}{x}$, this trivially satisfies (\ref{p-let-gamma}), $\Gamma' \supseteq \Gamma$.

    \item Show that $\dom{\Gamma'} = \dom{H'}$. We know that $\Gamma' = \Gamma, x : \CK{H}{x}$
      and $H' = H, x \mapsto v$. To show $\dom{\Gamma, x : \CK{H}{x}} = \dom{H, x \mapsto v}$:
      \begin{enumerate}
        \item $\dom{\Gamma} = \dom{H}$ by assumption (\ref{preserv:let0}). \label{p-let-dom1}
        \item Trivially, $x \in \dom{\Gamma, x : \CK{H}{x}}$ and $x \in \dom{H, x \mapsto v}$.\label{p-let-dom2}
      \end{enumerate}
      From (\ref{p-let-dom1}) and (\ref{p-let-dom2}), we conclude that $\dom{\Gamma'} = \dom{H'}$.

    \item Show $\Gamma, x: \CK{H}{x} \vdash H, x \mapsto v$ (\ref{p-let-wf}).
      We proceed by case analysis on the shape of $v$.
      \begin{itemize}
      \item $v = \loc$ (and by our pick of $\Gamma'$ we have $x : \CK{H}{x}$).
        We proceed by checking that the environment $\Gamma'$ is well-formed
        \wrt{} $H'$.
        \begin{enumerate}
          \item \label{p-let-ck1} From the assumption of well-formed threads (\cref{preserv:let}) and
            typing rule \RN{WF-Let}: $x \notin \dom{\Gamma}$

          \item We prove that $\loc \in \dom{H'}$
            \begin{enumerate}
            \item From thread well-formedness assumption \cref{preserv:let}, $\Gamma \vdash H;  t''$ for
            all threads $t'' \in t \ \overline{T}$. From this fact and by rule \RN{WF-Term},
            $\Gamma \vdash \LetIn{x}{\loc}{t'}$. \label{p-let-ck12}
            \item By \cref{lem:store-var} with \cref{p-let-ck12},
            $\loc \in \dom{\Gamma}$. \label{p-let-ck13}
            \item From the initial assumption, $\dom{\Gamma} = \dom{H}$. \label{p-let-ck14}
            \item From \cref{p-let-ck13} and \cref{preserv:let0} we conclude that
              $\loc \in \dom{H}$.\label{p-let-ck15}
            \item From $\Gamma' = \Gamma, x : \_ \supseteq \Gamma$ and the proved
              step $\dom{\Gamma'} = \dom{H'}$, we can conclude that $\loc \in
              \dom{H'}$ as $H' \supseteq H$ and $\loc \in H$ by \ref{p-let-ck15}.
            \end{enumerate}

          \item From the assumption $\Gamma \vdash H$, by rule \RN{WF-H-Object}
            any $\loc \in \dom{H}$ with capability $K$ must have had $\Gamma(\loc) = K$.
            By our pick of $\Gamma'$ where $x : \CK{H}{\loc}$ and
            by definition of $\CK{H}{\loc}$, it follows that $\Gamma(x) = \Gamma(\loc) = K$.
          \item From assumption (\ref{preserv:let2}), $\Gamma \vdash H$. \label{p-let-ck2}
        \end{enumerate}
        From steps (\ref{p-let-ck1}) -- (\ref{p-let-ck2}) and by rule \RN{WF-H-Var},
        we conclude that $\Gamma, x: \CK{H}{x} \vdash H, x \mapsto v$.

      \item $v = \Absent$. We need to show that $x \in \dom{\Gamma}$ and $\Gamma' \vdash H$.
        \begin{enumerate}
          \item By our pick of $\Gamma'$, trivially $x \in \dom{\Gamma'}$, \label{p-let-wf-absent}
          \item By our assumption that $\Gamma \vdash H$ (\ref{preserv:let2}) and $\Gamma' \supseteq \Gamma$
            and direct application of \cref{lem:weak-env-heap}, $\Gamma' \vdash H$ \label{p-let-wf-absent2}
         \item From \ref{p-let-wf-absent} and \ref{p-let-wf-absent2} and by application of
        rule \RN{WF-H-Absent}, we conclude that $\Gamma' \vdash H'$.
        \end{enumerate}
      \end{itemize}

    \item Show $\forall t'' \in t' \ \overline{T}. \Gamma'; H, x \mapsto v \vdash
      t''$.  There are multiple cases to consider. For all $t'' \in \overline{T}$
      and by \cref{lem:weakening}, extending the environment and heap do not
      affect the well-formedness of a thread.  For the case where $t'' = t'$, we
      proceed by case analysis on the shape of $t'$.
      \begin{itemize}
        \item $t' = x$.
          \begin{enumerate}
          \item We just proved $\Gamma, x: \CK{H}{x} \vdash H, x \mapsto v$ (\ref{p-let-wf}).\label{p-let-wf-t}
          \item Since $x \in \dom{\Gamma, x: \CK{H}{x}}$ and by our assumptions
            of well-formedness $\vdash \Gamma$ and $x \notin \Gamma$, by direct
            application of \RN{WF-Env-Var}, $\vdash \Gamma, x :
            \CK{H}{x}$. \label{p-let-wf-t2}
          \item From \ref{p-let-wf-t2} and direct application of
            \RN{WF-Var}, $\Gamma, x : \CK{H}{x} \vdash x$.  \label{p-let-wf-t3}
          \item From \ref{p-let-wf-t} and \ref{p-let-wf-t3} and by
            application of \RN{WF-Term}, we conclude $\Gamma'; H, x \mapsto v \vdash x$
            which is what we had to show.
          \end{enumerate}

        \item $t' = \LetIn{x}{e}{t'''}$.
          Follows the same steps as the prove for $t' = x$.
        \item $t' = \consume{x}$.
          Follows the same steps as the prove for $t' = x$.
        \item $t' = \CC{K}{w}$.
          Follows the same steps as the prove for $t' = x$.
      \end{itemize}

    \item Show $\Helper{Local}{H', t'\ \overline{T}}$ (\ref{p-let-local}).
      There are multiple cases to consider, depending on whether
      there is a local variable and if this one belongs to $t'$ or $\overline{T}$
      \begin{enumerate}
        \item No local variables. Then the induction hypothesis is vacuously satisfied, since there is
          no location to which we can apply it.
        \item $\loc \in \Helper{ROG}{H', t'} \land \Helper{isLocal}{H, \loc}$.
          \begin{enumerate}
          \item \label{p-let-local-00}From the initial assumptions (\ref{preserv:local}) and the
            $\Helper{Local}{H, t\ \overline{T}}$ and $x \notin \dom{H}$ and $\loc$ is local to $t$
            and $H' = H, x \mapsto \loc$ and by direct application of
            \cref{lem:local-values}, extending the heap with a fresh variable
            does not affect other threads. Thus, $\Helper{Local}{H', \overline{T}}$.
            With this, we know that $x$ and $\loc$ is still not reachable from other
            threads.
          \item Now, we need to show that $t'$ maintains the thread-locality of $t$
            and extending the heap does not allow to reach to local variables from other threads.

          We need to show that $\Helper{Local}{H', t' \ \overline{T}}$.
          Clearly $t' \subseteq t$, $t'$ only has access to the free variables available
          in $t$, by the assumptions of the case, \ref{p-let-local-00} and by direct application of
          \cref{lem:thread-subs}, we conclude $\Helper{Local}{H', t'\ \overline{T}}$.
          \end{enumerate}
        \item $\loc \notin \Helper{ROG}{H', t'} \land \Helper{isLocal}{H, \loc}$.
          It must be the case that $\loc$ is local to some other thread. By application
          of \cref{lem:local-values}, we can conclude that $\Helper{Local}{H', t'\ \overline{T}}$.
      \end{enumerate}
      Thus, we can conclude that $\Helper{Local}{H', t'\ \overline{T}}$.

    \item Show $\Helper{Isolated}{H', t'\ \overline{T}}$, (\ref{p-let-iso}).
      From the initial assumption $\Helper{Isolated}{H,
        \LetIn{x}{v}{t}\ \overline{T}}$ (\ref{preserv:iso}), an isolate location
      can either be in a field or multiple times in the stack accesible to a single thread.
      If there are multiple references to it, then it must be on the stack.
      We consider the case where there are multiple references, as the predicate is vacuously
      isolated otherwise.
      \begin{enumerate}
        \item Assume $\Helper{isIso}{H, v}$. Since there are multiple references
          we need to further assume its owner. Assume $v$ is only accesible from
          $\Helper{Isolated}{H, \LetIn{x}{v}{t}}$. Then, extending the environment
            with a stack variable cannot break isolation, \cref{lem:linear-values}.
      \end{enumerate}
  \end{itemize}


\item $t = E[x]$. By the induction hypothesis,
  \[H; E[x]\ \overline{T} \ReducesTo H; \overline{T'}\ \overline{T} \Rightarrow \exists
  \Gamma'. \Gamma' \supseteq \Gamma \land \Gamma' \vdash H';  \overline{T'}\ \overline{T}\]

  We proceed by case analysis over the \textbf{step}.

  \begin{description}
  \item[R-Var], $H; E[x] \ReducesTo H; E[v]$.

    From initial assumption and \RN{WF-Configuration}:
    \begin{align}
      &  \Gamma \vdash H;  E[x] \label{pre-var-a1} \\
      &  \Gamma \vdash H \label{pre-var-a2} \\
      &\label{pre-var-a3}
        \forall t' \in E[x]\ \overline{T}.\ \ \Gamma \vdash H;  t' \\
      &\label{pre-var-a4}
        \Helper{Local}{H, E[x]\ \overline{T}} \\
      &\label{pre-var-a5}
        \Helper{Isolated}{H, E[x]\ \overline{T}}
    \end{align}

      \textbf{Need to show}
      \[\exists  \Gamma'. \Gamma' \supseteq \Gamma \land \Gamma' \vdash H';
        E[v]\ \overline{T}, \textit{ where } H' = H\]

  The strategy of the proof is to show that all components are well-formed.
  By rule \RN{WF-Configuration} and IH, we need to show:
  \begin{align}
    & \Gamma' \supseteq \Gamma \label{p-var-gamma} \\
    & \Gamma' \vdash H';  E[v] \label{p-var-wf-let} \\
    & \Gamma' \vdash H \label{p-var-wf}\\
    & \forall t'' \in E[v] \ \overline{T}. \Gamma' \vdash H;  t'' \label{p-var-t} \\
    & \Helper{Local}{H', E[v]\ \overline{T}} \label{p-var-local}\\
    & \Helper{Isolated}{H', E[v]\ \overline{T}} \label{p-var-iso}
  \end{align}

  We pick $\Gamma' = \Gamma$ and $H' = H$ by \RN{R-Var}.
  We now proof each of the components of the configuration:
  \begin{itemize}
    \item We picked $\Gamma' = \Gamma, x : K$, this trivially satisfies (\ref{p-var-gamma}).

    \item Show that $\Gamma' \vdash H';  E[v]$ (\ref{p-var-wf-let}).
      \begin{enumerate}
        \item \label{pre-var-a1-a1} By (\ref{pre-var-a1}), with \RN{WF-Let} and \RN{WF-Loc},
          we conclude that $\Gamma' \vdash E[v]$ is well-formed
        \item By \RN{WF-Term-Let} with (\ref{pre-var-a1-a1}), we conclude that $\Gamma' \vdash H';  E[v]$.
      \end{enumerate}

    \item Show $\Gamma' \vdash H$ (\ref{p-var-wf}).
      Trivially satisfied from (\ref{pre-var-a2}).

    \item Show $\forall t'' \in E[v] \ \overline{T}. \Gamma' \vdash H;  t''$.
      The case for $\overline{T}$ is trivially from (\ref{pre-var-a3}).
      The case for $E[v]$ was shown in above (\ref{p-var-wf-let}).
    \item Show $\Helper{Local}{H', t'\ \overline{T}}$ (\ref{p-var-local}).
      This is trivially satisfied from (\ref{pre-var-a4}).
    \item Show $\Helper{Isolated}{H', t'\ \overline{T}}$, (\ref{p-var-iso}).
      This is trivially satisfied from (\ref{pre-var-a5}).
  \end{itemize}

  By \RN{WF-Configuration}, we conclude that $\Gamma \vdash H;   E[v]\ \overline{T}$

  \item[E-AbsentVar] $H; E[x] \ReducesTo H; \CErr$.
    From initial assumption and \RN{WF-Configuration}:
    \begin{align}
      &  \Gamma \vdash H;  E[x] \label{pre-absentvar-a1} \\
      &  \Gamma \vdash H \label{pre-absentvar-a2}
      \end{align}

    By the equivalence rules, we immediately reduce $\CErr\ \overline{T} \equiv \CErr$.

    \textbf{Need to show}
    \[\exists  \Gamma'. \Gamma' \supseteq \Gamma \land \Gamma' \vdash H';
    \CErr, \textit{ where } H' = H\]

  The strategy of the proof is to show that all components are well-formed.
  By rule \RN{WF-Configuration} and IH, we need to show:
  \begin{align}
    & \Gamma' \supseteq \Gamma \label{p-absentvar-gamma} \\
    & \Gamma' \vdash H' \label{p-absentvar-wf} \\
  \end{align}

  We pick $\Gamma' = \Gamma$ and $H' = H$ by \RN{R-AbsentVar}.
  We now proof each of the components of the configuration:
  \begin{itemize}
    \item We picked $\Gamma' = \Gamma$, this is trivially satisfied by (\ref{p-absentvar-gamma}).
    \item Show that $\Gamma' \vdash H';  \CErr$. Given that $\Gamma' = \Gamma$ and $H = H$,
      this is trivially satisfied from (\ref{pre-absentvar-a2}).
  \end{itemize}

  We conclude by direct application of \RN{WF-Configuration} that
  $\Gamma \vdash H;  \CErr$ is well-formed.

  \item[E-AliasIso]$H; E[x] \ReducesTo H; \PErr$.
    From initial assumption and \RN{WF-Configuration}:
    \begin{align}
      &  \Gamma \vdash H \label{pre-aliasvar-a1}
      \end{align}

    By the equivalence rules, we immediately reduce $\PErr\ \overline{T} \equiv \PErr$.

    \textbf{Need to show}
    \[\exists  \Gamma'. \Gamma' \supseteq \Gamma \land \Gamma' \vdash H';
    \PErr, \textit{ where } H' = H\]

  The strategy of the proof is to show that all components are well-formed.
  By rule \RN{WF-Configuration} and IH, we need to show:
  \begin{align}
    & \Gamma' \supseteq \Gamma \label{p-aliasvar-gamma} \\
    & \Gamma' \vdash H' \label{p-aliasvar-wf}
  \end{align}

  We pick $\Gamma' = \Gamma$ and $H' = H$ by \RN{R-AliasIso}.
  We now proof each of the components of the configuration:
  \begin{itemize}
    \item We picked $\Gamma' = \Gamma$, this is trivially satisfied from (\ref{p-aliasvar-gamma}).
    \item Show that $\Gamma' \vdash H';  \CErr$. Given that $\Gamma' = \Gamma$ and $H = H$,
      this is trivially satisfied from (\ref{pre-aliasvar-a1}).
  \end{itemize}

  We conclude by direct application of \RN{WF-Configuration} that
  $\Gamma \vdash H;  \PErr$ is well-formed.
  \end{description}


\item $t = E[\consume{x}]$. By the induction hypothesis,
  \begin{align*}
    &H; E[x]\ \overline{T} \ReducesTo H; E[v]\ \overline{T}
    & \qquad \Rightarrow \exists
  \Gamma'. \Gamma' \supseteq \Gamma \land \Gamma' \vdash H';  E[v]\ \overline{T}
  \end{align*}

  There are two rules, which depend on whether $x$ is $\Absent$ or not.

  \begin{description}
  \item[R-Consume], $H; E[\consume{x}] \ReducesTo H; E[v]$.
    From initial assumption and \RN{WF-Configuration}:
    \begin{align}
      &  \Gamma \vdash H;  E[\consume{x}] \label{pre-cons-a1} \\
      &  \Gamma \vdash H \label{pre-cons-a2} \\
      &\label{pre-cons-a3}
        \forall t' \in E[\consume{x}]\ \overline{T}.\ \ \Gamma \vdash H;  t' \\
      &\label{pre-cons-a4}
        \Helper{Local}{H, E[\consume{x}]\ \overline{T}} \\
      &\label{pre-cons-a5}
        \Helper{Isolated}{H, E[\consume{x}]\ \overline{T}}
    \end{align}

    \textbf{Need to show:}
    \[\exists
    \Gamma'. \Gamma' \supseteq \Gamma \land \Gamma' \vdash H';  E[v]\ \overline{T}\]

    The strategy of the proof is to show that all components are well-formed.
  By rule \RN{WF-Configuration} and IH, we need to show:
  \begin{align}
    & \Gamma' \supseteq \Gamma \label{p-cons-gamma} \\
    & \Gamma' \vdash H';  E[v] \label{p-cons-wf-let} \\
    & \Gamma' \vdash H \label{p-cons-wf}\\
    & \forall t'' \in E[v] \ \overline{T}. \Gamma' \vdash H;  t'' \label{p-cons-t} \\
    & \Helper{Local}{H', E[v]\ \overline{T}} \label{p-cons-local}\\
    & \Helper{Isolated}{H', E[v]\ \overline{T}} \label{p-cons-iso}
  \end{align}

  We pick $\Gamma' = \Gamma$ and $H' = H$ by \RN{R-Consume}.
  We now proof each of the components of the configuration:
  \begin{itemize}
    \item We picked $\Gamma' = \Gamma$, this trivially satisfies (\ref{p-cons-gamma}).

    \item Show that $\Gamma' \vdash H';  E[v]$ (\ref{p-cons-wf-let}).
      \begin{enumerate}
        \item \label{pre-cons-a1-a1} By \RN{WF-Let} and \RN{WF-Loc} with (\ref{pre-cons-a1})
          we conclude that $\Gamma' \vdash E[v]$ is well-formed.
        \item \label{pre-cons-a1-a2} We need to show that $v = \loc \Rightarrow v
          \in \dom{H}$.  This was part of the initial assumptions in
          (\ref{pre-cons-a1}), where $x \mapsto \loc \in \dom{H}$.
        \item By \RN{WF-Term-Let} with (\ref{pre-cons-a1-a1}) and
          (\ref{pre-cons-a1-a2}),
          we can conclude we conclude that $\Gamma' \vdash H';  E[v]$ is well-formed.
      \end{enumerate}

    \item Show $\Gamma' \vdash H$ (\ref{p-cons-wf}).
      Trivially satisfied by (\ref{pre-cons-a2}).

    \item Show $\forall t'' \in E[v] \ \overline{T}. \Gamma' \vdash H;  t''$.
      The case for $\overline{T}$ is trivially satisfied by (\ref{pre-cons-a3}).
      The case for $E[v]$ was shown in above (\ref{p-cons-wf-let}).

    \item Show $\Helper{Local}{H', t'\ \overline{T}}$ (\ref{p-cons-local}).
      This is trivially satisfied by (\ref{pre-cons-a4}).

    \item Show $\Helper{Isolated}{H', t'\ \overline{T}}$, (\ref{p-cons-iso}).
      This is trivially satisfied by (\ref{pre-cons-a5}).
  \end{itemize}

  By \RN{WF-Configuration}, we conclude that $\Gamma \vdash H;   E[v]\ \overline{T}$

  \item[E-Consume] $H; E[\consume{x}] \ReducesTo H; \CErr$
    The proof follows the same steps as \RN{E-AliasIso}.
  \end{description}


\item $t = E[x.f]$. By the induction hypothesis,
  \begin{align*}
    &H; E[x.f]\ \overline{T} \ReducesTo H; \overline{T'}\ \overline{T} \\
    & \qquad \Rightarrow \exists
  \Gamma'. \Gamma' \supseteq \Gamma \land \Gamma' \vdash H';  \overline{T'}\ \overline{T}
  \end{align*}
  There are multiple reductions rules based on whether $x$ maps to $\Absent$
  value (\RN{E-AbsentTarget}), field $f$ does not exist in the object pointed by
  $x$ (\RN{E-NoSuchField}), when the field access is to an isolate
  \RN{E-IsoField} and when the field access is not to an isolate (\RN{R-Field}).
  We proceed by case analysis on the \textbf{step}.

  \begin{description}
  \item[R-Field], $H; E[x.f] \ReducesTo H; E[v]$.
    From initial assumption and \RN{WF-Configuration}:
    \begin{align}
      & x \mapsto \loc \in H \label{pre-fac-a0}\\
      & H(\loc) = \Obj{\_}{\_ \ f = v}{\Ms} \label{pre-fac-a00}\\
      & \neg \Helper{isIso}{H, v} \label{pre-fac-a000} \\
      &  \Gamma \vdash H;  E[x.f] \label{pre-fac-a1} \\
      &  \Gamma \vdash H \label{pre-fac-a2} \\
      &\label{pre-fac-a3}
        \forall t' \in E[x.f]\ \overline{T}.\ \ \Gamma \vdash H;  t' \\
      &\label{pre-fac-a4}
        \Helper{Local}{H, E[x.f]\ \overline{T}} \\
      &\label{pre-fac-a5}
        \Helper{Isolated}{H, E[x.f]\ \overline{T}}
    \end{align}

    \textbf{Need to show:}
    \[\exists
    \Gamma'. \Gamma' \supseteq \Gamma \land \Gamma' \vdash H';  E[v]\ \overline{T}\]

    The strategy of the proof is to show that all components are well-formed.
    By rule \RN{WF-Configuration} and IH, we need to show:
  \begin{align}
    & \Gamma' \supseteq \Gamma \label{p-fac-gamma} \\
    & \Gamma' \vdash H';  E[v] \label{p-fac-wf-let} \\
    & \Gamma' \vdash H \label{p-fac-wf}\\
    & \forall t'' \in E[v] \ \overline{T}. \Gamma' \vdash H;  t'' \label{p-fac-t} \\
    & \Helper{Local}{H', E[v]\ \overline{T}} \label{p-fac-local}\\
    & \Helper{Isolated}{H', E[v]\ \overline{T}} \label{p-fac-iso}
  \end{align}

  We pick $\Gamma' = \Gamma$ and $H' = H$ by \RN{R-Field}.
  We now proof each of the components of the configuration:
  \begin{itemize}
    \item We picked $\Gamma' = \Gamma$, this trivially satisfies (\ref{p-fac-gamma}).

    \item Show that $\Gamma' \vdash H';  E[v]$ (\ref{p-fac-wf-let}).
      \begin{enumerate}
        \item \label{pre-fac-a1-a1} By \RN{WF-Let} and \RN{WF-Loc} with (\ref{pre-fac-a1}),
          we conclude that $\Gamma' \vdash E[v]$ is well-formed.
        \item \label{pre-fac-a1-a2} By initial assumptions (\ref{pre-fac-a0}), (\ref{pre-fac-a00}), (\ref{pre-fac-a000}), we have that $x \mapsto \loc \in H$ and $\loc \in \dom{H}$ and $H(x.f) = v$
          and $v$ is not isolated.
        \item By \RN{WF-Term-Let} with (\ref{pre-fac-a1-a1}) and
          (\ref{pre-fac-a1-a2}),
          we can conclude we conclude that $\Gamma' \vdash H';  E[v]$ is well-formed.
      \end{enumerate}

    \item Show $\Gamma' \vdash H$ (\ref{p-fac-wf}).
      Trivially satisfied by (\ref{pre-fac-a2}).

    \item Show $\forall t'' \in E[v] \ \overline{T}. \Gamma' \vdash H;  t''$.
      The case for $\overline{T}$ is trivially saitisfied by (\ref{pre-fac-a3}).
      The case for $E[v]$ was shown in above (\ref{p-fac-wf-let}).

    \item Show $\Helper{Local}{H', E[v]\ \overline{T}}$ (\ref{p-fac-local}).
      This is trivially satisfied by  (\ref{pre-fac-a4}).

    \item Show $\Helper{Isolated}{H', E[v]\ \overline{T}}$, (\ref{p-fac-iso}).
      This is vacuously satisfied by  (\ref{pre-fac-a5}) and
      (\ref{pre-fac-a000}), since $H(x.f) = v$ and $v$ is not an isolate.
  \end{itemize}

  By \RN{WF-Configuration}, we conclude that $\Gamma \vdash H;   E[v]\ \overline{T}$ is well-formed.

  \item[E-NoSuchField] $H; E[x.f] \ReducesTo H; \err$

    From initial assumption and \RN{WF-Configuration}:
    \begin{align}
      & f \textit{ does not exist in the object} \\
      &  \Gamma \vdash H \label{pre-fac-nofield-a1}
      \end{align}

    By the equivalence rules, we immediately reduce $\err\ \overline{T} \equiv \err$.
    \textbf{Need to show}
    \[\exists  \Gamma'. \Gamma' \supseteq \Gamma \land \Gamma' \vdash H';
    \err, \textit{ where } H' = H\]

  The strategy of the proof is to show that all components are well-formed.
  By rule \RN{WF-Configuration} and IH, we need to show:
  \begin{align}
    & \Gamma' \supseteq \Gamma \label{p-fac-nofield-gamma} \\
    & \Gamma' \vdash H' \label{p-fac-nofield-wf} \\
  \end{align}

  We pick $\Gamma' = \Gamma$ and $H' = H$ by \RN{E-NoSuchField}.
  We now proof each of the components of the configuration:
  \begin{itemize}
    \item We picked $\Gamma' = \Gamma$, this is trivially satisfied by (\ref{p-fac-nofield-gamma}).
    \item Show that $\Gamma' \vdash H';  \err$. Given that $\Gamma' = \Gamma$ and $H = H$,
      this is trivially satisfied from initial assumption (\ref{pre-fac-nofield-a1}).
  \end{itemize}

  We conclude by direct application of \RN{WF-Configuration} that
  $\Gamma \vdash H;  \err$ is well-formed.

  \item[E-AbsentTargetAccess], $H; E[x.f] \ReducesTo H; \CErr$.
    From initial assumption and \RN{WF-Configuration}:
    \begin{align}
      &  H(x) = \Absent \\
      &  \Gamma \vdash H \label{pre-fac-absent-a1}
      \end{align}

    By the equivalence rules, we immediately reduce $\CErr\ \overline{T} \equiv \CErr$.
    \textbf{Need to show}
    \[\exists  \Gamma'. \Gamma' \supseteq \Gamma \land \Gamma' \vdash H';
    \CErr, \textit{ where } H' = H\]

  The strategy of the proof is to show that all components are well-formed.
  By rule \RN{WF-Configuration} and IH, we need to show:
  \begin{align}
    & \Gamma' \supseteq \Gamma \label{p-fac-absent-gamma} \\
    & \Gamma' \vdash H' \label{p-fac-absent-wf}
  \end{align}

  We pick $\Gamma' = \Gamma$ and $H' = H$ by \RN{E-AbsentTargetAccess}.
  We now proof each of the components of the configuration:
  \begin{itemize}
    \item We picked $\Gamma' = \Gamma$, this trivially satisfies (\ref{p-fac-absent-gamma}).
    \item Show that $\Gamma' \vdash H';  \CErr$. Given that $\Gamma' = \Gamma$ and $H = H$,
      this is trivially satisfied with (\ref{pre-fac-absent-a1}).
  \end{itemize}

  We conclude by direct application of \RN{WF-Configuration} that
  $\Gamma \vdash H;  \CErr$ is well-formed.

  \item[E-LocalField], $H; E[x.f]^i \ReducesTo H; \PErr$.
    From initial assumption and \RN{WF-Configuration}:
    \begin{align}
      &  H(x.f) = \loc \\
      & \Helper{notLocalOwner}{H, i, \loc} \\
      &  \Gamma \vdash H \label{pre-fac-iso-a111}
      \end{align}

    By the equivalence rules, we immediately reduce $\PErr\ \overline{T} \equiv \PErr$.
    \textbf{Need to show}
    \[\exists  \Gamma'. \Gamma' \supseteq \Gamma \land \Gamma' \vdash H';
    \PErr, \textit{ where } H' = H\]

  The strategy of the proof is to show that all components are well-formed.
  By rule \RN{WF-Configuration} and IH, we need to show:
  \begin{align}
    & \Gamma' \supseteq \Gamma \label{p-fac-iso-gammax} \\
    & \Gamma' \vdash H' \label{p-fac-iso-wfx}
  \end{align}

  We pick $\Gamma' = \Gamma$ and $H' = H$ by \RN{E-LocalField}.
  We now proof each of the components of the configuration:
  \begin{itemize}
    \item We picked $\Gamma' = \Gamma$, this is trivially satisfied by (\ref{p-fac-iso-gammax}).
    \item Show that $\Gamma' \vdash H';  \PErr$. Given that $\Gamma' = \Gamma$ and $H = H$,
      this is trivially satisfied by (\ref{pre-fac-iso-a111}).
  \end{itemize}

  We conclude by direct application of \RN{WF-Configuration} that
  $\Gamma \vdash H;  \PErr$ is well-formed.

  \item[E-IsoField], $H; E[x.f] \ReducesTo H; \PErr$.
    From initial assumption and \RN{WF-Configuration}:
    \begin{align}
      &  H(x.f) = \loc \\
      & \Helper{isIso}{H, \loc} \\
      &  \Gamma \vdash H \label{pre-fac-iso-a1}
      \end{align}

    By the equivalence rules, we immediately reduce $\PErr\ \overline{T} \equiv \PErr$.
    \textbf{Need to show}
    \[\exists  \Gamma'. \Gamma' \supseteq \Gamma \land \Gamma' \vdash H';
    \PErr, \textit{ where } H' = H\]

  The strategy of the proof is to show that all components are well-formed.
  By rule \RN{WF-Configuration} and IH, we need to show:
  \begin{align}
    & \Gamma' \supseteq \Gamma \label{p-fac-iso-gamma} \\
    & \Gamma' \vdash H' \label{p-fac-iso-wf}
  \end{align}

  We pick $\Gamma' = \Gamma$ and $H' = H$ by \RN{E-IsoField}.
  We now proof each of the components of the configuration:
  \begin{itemize}
    \item We picked $\Gamma' = \Gamma$, this is trivially satisfied by (\ref{p-fac-iso-gamma}).
    \item Show that $\Gamma' \vdash H';  \PErr$. Given that $\Gamma' = \Gamma$ and $H = H$,
      this is trivially satisfied by (\ref{pre-fac-iso-a1}).
  \end{itemize}

  We conclude by direct application of \RN{WF-Configuration} that
  $\Gamma \vdash H;  \PErr$ is well-formed.

  \end{description}


\item $t = E[x.f = v]$. By the induction hypothesis,
  \begin{align*}
    &H; E[x.f = v]\ \overline{T} \ReducesTo H; \overline{T'}\ \overline{T} \\
    & \qquad \Rightarrow \exists
  \Gamma'. \Gamma' \supseteq \Gamma \land \Gamma' \vdash H';  \overline{T'}\ \overline{T}
  \end{align*}

  There are multiple cases to handle, depending on whether the variable $x$ is
  $\Absent$ (\RN{E-AbsentFieldAssign}), whether field $f$ exists
  (\RN{E-NoSuchFieldAssign}), whether $x$ is immutable or the capability of $x$
  is compatible with the capability of the value $v$ (\RN{E-BadFieldAssign}), and
  finally the case where $x$ is not immutable and the capability of $v$ is
  compatible with the capability of $x$ (\RN{R-FieldAssign}).
  We proceed by case analysis over the \textbf{step}.

  \begin{description}
  \item[E-AbsentFieldAssign], $H; E[x.f = v] \ReducesTo H; \CErr$.

    From initial assumption and \RN{WF-Configuration}:
    \begin{align}
      &  H(x) = \Absent \\
      &  \Gamma \vdash H \label{pre-fass-absent-a1}
      \end{align}

    By the equivalence rules, we immediately reduce $\CErr\ \overline{T} \equiv \CErr$.
    \textbf{Need to show}
    \[\exists  \Gamma'. \Gamma' \supseteq \Gamma \land \Gamma' \vdash H';
    \CErr, \textit{ where } H' = H\]

  The strategy of the proof is to show that all components are well-formed.
  By rule \RN{WF-Configuration} and IH, we need to show:
  \begin{align}
    & \Gamma' \supseteq \Gamma \label{p-fass-absent-gamma} \\
    & \Gamma' \vdash H' \label{p-fass-absent-wf}
  \end{align}

  We pick $\Gamma' = \Gamma$ and $H' = H$ by \RN{E-AbsentFieldAssign}.
  We now proof each of the components of the configuration:
  \begin{itemize}
    \item We picked $\Gamma' = \Gamma$, which trivially satisfies (\ref{p-fass-absent-gamma}).
    \item Show that $\Gamma' \vdash H';  \CErr$. Given that $\Gamma' = \Gamma$ and $H = H$,
      this is trivially satisfied with (\ref{pre-fass-absent-a1}).
  \end{itemize}

  We conclude by direct application of \RN{WF-Configuration} that
  $\Gamma \vdash H;  \CErr$ is well-formed.

  \item[E-BadFieldAssign], $H; E[x.f = v] \ReducesTo H; \PErr$.

    \textbf{Assumptions:}
    From initial assumption and \RN{WF-Configuration}:
    \begin{align}
      &  H(x) = \loc \\
      &  H(\loc) = \Obj{K}{\_ f=v}{\Ms} \\
      & \Helper{isImm}{H, \loc} \lor \neg \Helper{OkRef}{H, K, v} \label{pre-fass-bfield-a0}\\
      &  \Gamma \vdash H \label{pre-fass-bfield-a1}
      \end{align}

    By the equivalence rules, we immediately reduce $\PErr\ \overline{T} \equiv \PErr$.
    \textbf{Need to show}
    \[\exists  \Gamma'. \Gamma' \supseteq \Gamma \land \Gamma' \vdash H';
    \PErr, \textit{ where } H' = H\]

  The strategy of the proof is to show that all components are well-formed.
  By rule \RN{WF-Configuration} and IH, we need to show:
  \begin{align}
    & \Gamma' \supseteq \Gamma \label{p-fass-bfield-gamma} \\
    & \Gamma' \vdash H' \label{p-fass-bfield-wf}
  \end{align}

  We pick $\Gamma' = \Gamma$ and $H' = H$ by \RN{E-BadFieldAssign}.
  We now proof each of the components of the configuration:
  \begin{itemize}
    \item We picked $\Gamma' = \Gamma$, this trivially satisfies (\ref{p-fass-bfield-gamma}).
    \item Show that $\Gamma' \vdash H';  \PErr$. Given that $\Gamma' = \Gamma$ and $H = H$,
      this is trivially satisfied by the assumption (\ref{pre-fass-bfield-a1}).
  \end{itemize}

  We conclude by direct application of \RN{WF-Configuration} that
  $\Gamma \vdash H;  \PErr$ is well-formed.

  \item[E-NoSuchFieldAssign], $H; E[x.f = v] \ReducesTo H; \err$.

    From initial assumption and \RN{WF-Configuration}:
    \begin{align}
      & f \textit{ does not exist in the object} \\
      &  \Gamma \vdash H \label{pre-fass-nofield-a1}
      \end{align}

    By the equivalence rules, we immediately reduce $\err\ \overline{T} \equiv \err$.
    \textbf{Need to show}
    \[\exists  \Gamma'. \Gamma' \supseteq \Gamma \land \Gamma' \vdash H';
    \err, \textit{ where } H' = H\]

  The strategy of the proof is to show that all components are well-formed.
  By rule \RN{WF-Configuration} and IH, we need to show:
  \begin{align}
    & \Gamma' \supseteq \Gamma \label{p-fass-nofield-gamma} \\
    & \Gamma' \vdash H' \label{p-fass-nofield-wf}
  \end{align}

  We pick $\Gamma' = \Gamma$ and $H' = H$ by \RN{E-NoSuchFieldAssign}.
  We now proof each of the components of the configuration:
  \begin{itemize}
    \item We picked $\Gamma' = \Gamma$, this trivially satisfies (\ref{p-fass-nofield-gamma}).
    \item Show that $\Gamma' \vdash H';  \err$. Given that $\Gamma' = \Gamma$ and $H = H$,
      this is trivially satisfied by the assumption (\ref{pre-fass-nofield-a1}).
  \end{itemize}

  We conclude by direct application of \RN{WF-Configuration} that
  $\Gamma \vdash H;  \err$ is well-formed.

  \item[R-FieldAssign], $H; E[x.f = v] \ReducesTo H; E[v']$.

    From initial assumption and \RN{WF-Configuration}:
    \begin{align}
      & x \mapsto \loc \in H \label{pre-fass-a0}\\
      & H(\loc) = \Obj{K}{\_ \ f = v'}{\Ms} \label{pre-fass-a00}\\
      & \neg \Helper{isImm}{H, \loc} \label{pre-fass-a000} \\
      & \Helper{OkRef}{H, K, v} \label{pre-fass-a0000} \\
      &  \Gamma \vdash H;  E[x.f = v] \label{pre-fass-a1} \\
      &  \Gamma \vdash H \label{pre-fass-a2} \\
      &\label{pre-fass-a3}
        \forall t' \in E[x.f = v]\ \overline{T}.\ \ \Gamma \vdash H;  t' \\
      &\label{pre-fass-a4}
        \Helper{Local}{H, E[x.f = v]\ \overline{T}} \\
      &\label{pre-fass-a5}
        \Helper{Isolated}{H, E[x.f = v]\ \overline{T}}
    \end{align}

    \textbf{Need to show:}
    \[\exists
    \Gamma'. \Gamma' \supseteq \Gamma \land \Gamma' \vdash H';  E[v']\ \overline{T}\]

    The strategy of the proof is to show that all components are well-formed.
    By rule \RN{WF-Configuration} and IH, we need to show:
  \begin{align}
    & \Gamma' \supseteq \Gamma \label{p-fass-gamma} \\
    & \Gamma' \vdash H \label{p-fass-wf}\\
    & \Gamma' \vdash H';  E[v'] \label{p-fass-wf-let} \\
    & \forall t'' \in E[v'] \ \overline{T}. \Gamma' \vdash H;  t'' \label{p-fass-t} \\
    & \Helper{Local}{H', E[v']\ \overline{T}} \label{p-fass-local}\\
    & \Helper{Isolated}{H', E[v']\ \overline{T}} \label{p-fass-iso}
  \end{align}

  We pick $\Gamma' = \Gamma$ and $H' = H[\loc \mapsto \Obj{K}{\_ f = v'}{\Ms}]$ by \RN{R-FieldAssign}.
  We now proof each of the components of the configuration:
  \begin{itemize}
    \item We picked $\Gamma' = \Gamma$, this trivially satisfies (\ref{p-fass-gamma}).

    \item Show $\Gamma' \vdash H'$ (\ref{p-fass-wf}).
      To do this, it amounts to showing that the object is still well-formed
      after field substitution.
      \begin{enumerate}
        \item \label{pre-fass-wf-00}Show that substituting $\Obj{K}{\_\ f=v'}{\Ms}$ by
          $\Obj{K}{\_\ f=v}{\Ms}$ maintains well-formedness.  To do this, we
          first show that all fields in the object satisfy
          $\Helper{OkRefEnv}{\Gamma', K, v}$. For all fields where $v \neq v'$,
          this is satisfied by the assumption (\ref{pre-fass-a1}). When $f = v$,
          with (\ref{pre-fass-a2}) guarantees that $\Gamma(l) = K$,
          the assumptions of the reduction rule (\ref{pre-fass-a0000}) guarantees
          that $\Helper{OkRef}{H, K, v}$, and by $\RN{WF-H-Object}$ with $\Helper{OkRefEnv}{\Gamma, K, v}$,
          we can conclude that $\Gamma \vdash H, \loc \mapsto \Obj{K}{\_ f=v}{\Ms}$
          is well-formed.
        \item By the well-formed configuration rules with (\ref{pre-fass-wf-00}),
          we conclude that $\Gamma' \vdash H'$.
      \end{enumerate}

    \item Show that $\Gamma' \vdash H';  E[v]$ (\ref{p-fass-wf-let}).
      \begin{enumerate}
        \item \label{pre-fass-a1-a1} By the assumption \RN{WF-Let} and \RN{WF-Loc} with (\ref{pre-fass-a1}),
          we conclude that $\Gamma' \vdash E[v]$ is
          well-formed.
        \item \label{pre-fass-a1-a2} By the initial assumptions (\ref{pre-fass-a0}), (\ref{pre-fass-a00}), (\ref{pre-fass-a000}), then $x \mapsto \loc \in H$ and $\loc \in \dom{H}$ and $H(x.f) = v'$
          and $K$ is not immutable.
        \item By \RN{WF-Term-Let} with (\ref{pre-fass-a1-a1}) and
          (\ref{pre-fass-a1-a2}),
          we can conclude we conclude that $\Gamma' \vdash H';  E[v']$ is well-formed.
      \end{enumerate}

    \item Show $\forall t'' \in E[v'] \ \overline{T}. \Gamma' \vdash H;  t''$.
      The case for $\overline{T}$ is trivially saitisfied by the assumptions, (\ref{pre-fass-a3}).
      The case for $E[v]$ was shown above (\ref{p-fass-wf-let}).

    \item Show $\Helper{Local}{H', E[v']\ \overline{T}}$ (\ref{p-fass-local}).
      This is trivially satisfied by the assumption (\ref{pre-fass-a4}).

    \item Show $\Helper{Isolated}{H', E[v']\ \overline{T}}$, (\ref{p-fass-iso}).
      This is vacuously satisfied by the assumption (\ref{pre-fass-a5}).
  \end{itemize}

  By \RN{WF-Configuration}, we conclude that $\Gamma \vdash H;   E[v']\ \overline{T}$ is well-formed.
  \end{description}


\item $t = E[\CC{K}{\loc}]$. By the induction hypothesis,
  \begin{align*}
    &H; E[\CC{K}{\loc}]\ \overline{T} \ReducesTo H; \overline{T'}\ \overline{T} \\
    & \qquad \Rightarrow \exists
  \Gamma'. \Gamma' \supseteq \Gamma \land \Gamma' \vdash H';  \overline{T'}\ \overline{T}
  \end{align*}

  There are multiple cases to handle, depending on whether the location matches
  the casted capability or not.
  We proceed by case analysis over the \textbf{step}.

  \begin{description}
  \item[R-CastLoc], $H; E[\CC{K}{\loc}] \ReducesTo H; E[\loc]$.

    From initial assumption and \RN{WF-Configuration}:
    \begin{align}
      & H(\loc) = \Obj{K}{\_}{\_} \label{pre-cast-a00}\\
      &  \Gamma \vdash H;  E[\CC{K}{\loc}] \label{pre-cast-a1} \\
      &\label{pre-cast-a3}
        \forall t' \in E[\CC{K}{\loc}]\ \overline{T}.\ \ \Gamma \vdash H;  t' \\
      &\label{pre-cast-a4}
        \Helper{Local}{H, E[\CC{K}{\loc}]\ \overline{T}} \\
      &\label{pre-cast-a5}
        \Helper{Isolated}{H, E[\CC{K}{\loc}]\ \overline{T}}
    \end{align}

    \textbf{Need to show:}
    \[\exists
    \Gamma'. \Gamma' \supseteq \Gamma \land \Gamma' \vdash H';  \overline{T'}\ \overline{T}\]

    The strategy of the proof is to show that all components are well-formed.
    By rule \RN{WF-Configuration} and IH, we need to show:
  \begin{align}
    & \Gamma' \supseteq \Gamma \label{p-cast-gamma} \\
    & \Gamma' \vdash H \label{p-cast-wf}\\
    & \Gamma' \vdash H';  E[v'] \label{p-cast-wf-let} \\
    & \forall t'' \in E[v'] \ \overline{T}. \Gamma' \vdash H;  t'' \label{p-cast-t} \\
    & \Helper{Local}{H', E[v']\ \overline{T}} \label{p-cast-local}\\
    & \Helper{Isolated}{H', E[v']\ \overline{T}} \label{p-cast-iso}
  \end{align}

  We pick $\Gamma' = \Gamma$, and $H' = H$ and $\overline{T'} = E[\loc]$ by \RN{R-CastLoc}.
  We now proof each of the components of the configuration:
  \begin{itemize}
    \item We picked $\Gamma' = \Gamma$, this trivially satisfies (\ref{p-cast-gamma}).

    \item Show $\Gamma' \vdash H'$ (\ref{p-cast-wf}).
      This is trivially satisfied from (\ref{pre-cast-a1}).

    \item Show that $\Gamma' \vdash H';  E[\loc]$ (\ref{p-cast-wf-let}).
      This is trivially satisfied from assumption (\ref{pre-cast-a1}).

    \item Show $\forall t'' \in E[\loc] \ \overline{T}. \Gamma' \vdash H;  t''$.
      The case for $\overline{T}$ is trivially satisfied by the assumptions, (\ref{pre-cast-a3}).
      The case for $E[\loc]$ was shown above (\ref{p-cast-wf-let}).

    \item Show $\Helper{Local}{H', E[\loc]\ \overline{T}}$ (\ref{p-cast-local}).
      This is trivially satisfied from (\ref{pre-cast-a4}).

    \item Show $\Helper{Isolated}{H', E[\loc]\ \overline{T}}$, (\ref{p-cast-iso}).
      This is vacuously satisfied from (\ref{pre-cast-a5}).
  \end{itemize}

  By \RN{WF-Configuration}, we conclude that $\Gamma \vdash H;   E[\loc]\ \overline{T}$ is well-formed.

  \item[E-CastError], $H; E[\CC{K}{\loc}] \ReducesTo H; \CastErr$.

    From initial assumption and \RN{WF-Configuration}:
    \begin{align}
      &  H(\loc) = \Obj{K'}{\_}{\_}\ \quad \textit{where } K \neq K'  \\
      &  \Gamma \vdash H \label{pre-casterr-a1}
      \end{align}

    By the equivalence rules, we immediately reduce $\CastErr\ \overline{T} \equiv \CastErr$.
    \textbf{Need to show}
    \[\exists  \Gamma'. \Gamma' \supseteq \Gamma \land \Gamma' \vdash H';
    \err, \textit{ where } H' = H\]

  The strategy of the proof is to show that all components are well-formed.
  By rule \RN{WF-Configuration} and IH, we need to show:
  \begin{align}
    & \Gamma' \supseteq \Gamma \label{p-casterr-gamma} \\
    & \Gamma' \vdash H' \label{p-casterr-wf}
  \end{align}

  We pick $\Gamma' = \Gamma$ and $H' = H$ by \RN{E-CastError}.
  We now proof each of the components of the configuration:
  \begin{itemize}
    \item We picked $\Gamma' = \Gamma$, this trivially satisfies (\ref{p-casterr-gamma}).
    \item Show that $\Gamma' \vdash H';  \CastErr$. Given that $\Gamma' = \Gamma$ and $H = H$,
      this is trivially satisfied by the assumption (\ref{pre-casterr-a1}).
  \end{itemize}

  We conclude by direct application of \RN{WF-Configuration} that
  $\Gamma \vdash H;  \CastErr$ is well-formed.
  \end{description}


\item $t = E[\Obj{K}{\overline{f = v}}{\Ms}]$. By the induction hypothesis,
  \begin{align*}
    &H; E[\Obj{K}{\overline{f = v}}{\Ms}]\ \overline{T} \ReducesTo H; \overline{T'}\ \overline{T} \\
    & \qquad \Rightarrow \exists
  \Gamma'. \Gamma' \supseteq \Gamma \land \Gamma' \vdash H';  \overline{T'}\ \overline{T}
  \end{align*}

  There are multiple cases to handle, depending on whether object fields are
  compatible (\RN{R-New}) or not (\RN{E-BadInstantiation}). We proceed by case
  analysis over the \textbf{step}.

  \begin{description}
  \item[R-New], $H; E[\Obj{K}{\overline{f = v}}{\Ms}] \ReducesTo H'; E[\loc]$.

    From initial assumption and \RN{WF-Configuration}:
    \begin{align}
      &  \Gamma \vdash H;  E[\Obj{K}{\overline{f = v}}{\Ms}] \label{pre-new-a1} \\
      &\label{pre-new-a2}
      \forall v \in \overline{v}.\ \Helper{OkRef}{H, K, v} \\
      &\label{pre-new-a3}
        \forall t' \in E[\Obj{K}{\overline{f = v}}{\Ms}]\ \overline{T}.\ \ \Gamma \vdash H;  t' \\
      &\label{pre-new-a4}
        \Helper{Local}{H, E[\Obj{K}{\overline{f = v}}{\Ms}]\ \overline{T}} \\
      &\label{pre-new-a5}
        \Helper{Isolated}{H, E[\Obj{K}{\overline{f = v}}{\Ms}]\ \overline{T}}
    \end{align}

    \textbf{Need to show:}
    \[\exists
    \Gamma'. \Gamma' \supseteq \Gamma \land \Gamma' \vdash H';  \overline{T'}\ \overline{T}\]

    The strategy of the proof is to show that all components are well-formed.
    By rule \RN{WF-Configuration} and IH, we need to show:
  \begin{align}
    & \Gamma' \supseteq \Gamma \label{p-new-gamma} \\
    & \Gamma' \vdash H \label{p-new-wf}\\
    & \Gamma' \vdash H';  E[\loc] \label{p-new-wf-let} \\
    & \forall t'' \in E[\loc] \ \overline{T}. \Gamma' \vdash H;  t'' \label{p-new-t} \\
    & \Helper{Local}{H', E[\loc]\ \overline{T}} \label{p-new-local}\\
    & \Helper{Isolated}{H', E[\loc]\ \overline{T}} \label{p-new-iso}
  \end{align}

  We pick $\Gamma' = \Gamma, \loc: K$, and $H' = H, \loc \mapsto
  \Obj{K}{\overline{f=v}}{\Ms}$ and $\overline{T'} = E[\loc]$ by \RN{R-New}.  We
  now proof each of the components of the configuration:
  \begin{itemize}
    \item We picked $\Gamma' = \Gamma, \loc : K$, this trivially satisfies (\ref{p-new-gamma}).

    \item Show $\Gamma' \vdash H'$ (\ref{p-new-wf}).
      To show that $\Gamma, \loc : K \vdash H, \loc \mapsto
      \Obj{K}{\overline{f=v}}{\Ms}$, we proceed by proving well-formedness
      from each component in rule \RN{WF-H-Object}.
      \begin{enumerate}
      \item \label{new1} By assumption \ref{pre-new-a2} values are compatible with
      the object capability.
      \item \label{new2}By assumption \ref{pre-new-a1}, all values are in $\Gamma$.
      \item From our choosing of $\Gamma'$,  $\Gamma'(\loc) = K$.
      \item By \RN{WF-H-Object} and \RN{Helper-OkRefEnv} with (\ref{new1}) and (\ref{new2}),
        we conclude that $\Gamma' \vdash H'$.
      \end{enumerate}

    \item Show that $\Gamma' \vdash H';  E[\loc]$ (\ref{p-new-wf-let}).
      This is trivially satisfied from assumption (\ref{pre-new-a1}) with (\ref{p-new-wf}).

    \item Show $\forall t'' \in E[\loc] \ \overline{T}. \Gamma' \vdash H;  t''$.
      The case for $\overline{T}$ is trivially satisfied by the assumptions, (\ref{pre-new-a3}).
      The case for $E[\loc]$ was shown above (\ref{p-new-wf-let}).

    \item Show $\Helper{Local}{H', E[\loc]\ \overline{T}}$ (\ref{p-new-local}).
      This is trivially satisfied by the assumptions (\ref{pre-new-a4}).

    \item Show $\Helper{Isolated}{H', E[\loc]\ \overline{T}}$, (\ref{p-new-iso}).
      This is satisfied by the assumptions, (\ref{pre-new-a5}).
  \end{itemize}

  By \RN{WF-Configuration}, we conclude that $\Gamma \vdash H;   E[\loc]\ \overline{T}$ is well-formed.

  \item[E-BadInstantiation], $H; E[\Obj{K}{\overline{f = v}}{\Ms}] \ReducesTo H; \PErr$.

    From initial assumption and \RN{WF-Configuration}:
    \begin{align}
      &  \Gamma \vdash H \label{pre-new-bad-a1}
      \end{align}

    By the equivalence rules, we immediately reduce $\PErr\ \overline{T} \equiv \PErr$.
    \textbf{Need to show}
    \[\exists  \Gamma'. \Gamma' \supseteq \Gamma \land \Gamma' \vdash H';
    \PErr, \textit{ where } H' = H\]

  The strategy of the proof is to show that all components are well-formed.
  By rule \RN{WF-Configuration} and IH, we need to show:
  \begin{align}
    & \Gamma' \supseteq \Gamma \label{p-new-bad-gamma} \\
    & \Gamma' \vdash H' \label{p-new-bad-wf}
  \end{align}

  We pick $\Gamma' = \Gamma$ and $H' = H$ by \RN{E-BadInstantiation}.
  We now proof each of the components of the configuration:
  \begin{itemize}
    \item We picked $\Gamma' = \Gamma$, this is trivially satisfied by (\ref{p-new-bad-gamma}).
    \item Show that $\Gamma' \vdash H';  \PErr$. Given that $\Gamma' = \Gamma$ and $H = H$,
      this is trivially satisfied by the assumption (\ref{pre-new-bad-a1}).
  \end{itemize}

  We conclude by direct application of \RN{WF-Configuration} that
  $\Gamma \vdash H;  \PErr$ is well-formed.
  \end{description}


\item $t = E[x.m(v)]$. By the induction hypothesis,
  \begin{align*}
    &H; E[x.m(v)]\ \overline{T} \ReducesTo H; \overline{T'}\ \overline{T} \\
    & \qquad \Rightarrow \exists
  \Gamma'. \Gamma' \supseteq \Gamma \land \Gamma' \vdash H';  \overline{T'}\ \overline{T}
  \end{align*}

  There are multiple cases to handle, depending on whether object exists
  (\RN{R-Call}) or not (\RN{E-AbsentTarget}) and whether $m$ is a valid method name
  We proceed by case analysis over the \textbf{step}.

  \begin{description}
  \item[R-Call], $H; E[x.m(v)] \ReducesTo H; E[t[\self{}=x'][y=y']]$.
    From initial assumption and \RN{WF-Configuration}:
    \begin{align}
      & H(x) = \loc \land H(\loc) = \Obj{K}{\_}{\Ms} \land m \in \Ms \\
      &  \Gamma \vdash H;  E[x.m(v)] \label{pre-mcall-a1} \\
      &\label{pre-mcall-a3}
        \forall t' \in E[x.m(v)]\ \overline{T}.\ \ \Gamma \vdash H;  t' \\
      &\label{pre-mcall-a4}
        \Helper{Local}{H, E[x.m(v)]\ \overline{T}} \\
      &\label{pre-mcall-a5}
        \Helper{Isolated}{H, E[x.m(v)]\ \overline{T}}
    \end{align}

    \textbf{Need to show:}
    \[\exists
    \Gamma'. \Gamma' \supseteq \Gamma \land \Gamma' \vdash H';  \overline{T'}\ \overline{T}\]

    The strategy of the proof is to show that all components are well-formed.
    By rule \RN{WF-Configuration} and IH, we need to show:
  \begin{align}
    & \Gamma' \supseteq \Gamma \label{p-mcall-gamma} \\
    & \Gamma' \vdash H \label{p-mcall-wf}\\
    & \Gamma' \vdash H';  E[t[\self{}=x'][y=y']] \label{p-mcall-wf-let} \\
    & \forall t'' \in E[t[\self{}=x'][y=y']] \ \overline{T}. \Gamma' \vdash H;  t'' \label{p-mcall-t} \\
    & \Helper{Local}{H', E[t[\self{}=x'][y=y']]\ \overline{T}} \label{p-mcall-local}\\
    & \Helper{Isolated}{H', E[t[\self{}=x'][y=y']]\ \overline{T}} \label{p-mcall-iso}
  \end{align}

  We pick $\Gamma' = \Gamma, x' : K, y' : \Gamma(v)$, and $H' = H, x' \mapsto
  \loc, y' \mapsto v$ and $\overline{T'} = E[t[\self{}=x'][y=y']]$ by
  \RN{R-Call}.  We now proof each of the components of the configuration:
  \begin{itemize}
    \item We picked $\Gamma' = \Gamma, , x' : K, y' : \Gamma(v)$, this trivially
      satisfies (\ref{p-mcall-gamma}).

    \item Show $\Gamma' \vdash H'$ (\ref{p-mcall-wf}).
      To show that $\Gamma, x' : K, y' : \Gamma(v) \vdash H, x' \mapsto \loc
      y' \mapsto v$, it suffices to show that their capabilities match. By the assumptions
      (\ref{pre-mcall-a1}), $x$ and $\loc$ have capability $K$, and from \RN{WF-H-Var},
      we conclude that $\Gamma' \vdash H, x' \mapsto \loc$. Repeating these steps for
      $y$, lets us conclude $\Gamma' \vdash H'$.

    \item Show that $\Gamma' \vdash H';  E[t[\self{}=x'][y=y']]$ (\ref{p-mcall-wf-let}).
      This is trivially satisfied from assumption (\ref{pre-mcall-a1}) with (\ref{p-mcall-wf}).

    \item Show $\forall t'' \in E[t[\self{}=x'][y=y']] \ \overline{T}. \Gamma' \vdash H;  t''$.
      The case for $\overline{T}$ is trivially satisfied by the assumptions, (\ref{pre-mcall-a3}).
      The case for $E[t[\self{}=x'][y=y']]$ was shown above (\ref{p-mcall-wf-let}).

    \item Show $\Helper{Local}{H', E[t[\self{}=x'][y=y']]\ \overline{T}}$ (\ref{p-mcall-local}).
      This is trivially satisfied by the assumptions (\ref{pre-mcall-a4}).

    \item Show $\Helper{Isolated}{H', E[t[\self{}=x'][y=y']]\ \overline{T}}$, (\ref{p-mcall-iso}).
      This is satisfied by the assumptions, (\ref{pre-mcall-a5}). Lets proceed by
      case analysis over the isolated assumptions and show that it maintains isolateness.
      \begin{itemize}
        \item $\neg \Helper{isIso}{H, x}$. Thus, by the assumptions
          (\ref{pre-mcall-a5}), this is trivially satisfied.

        \item $\Helper{isIso}{H, x} \land x \mapsto \loc \in H'$.
          Lets proceed by case analysis over the incoming references being greater than 1, or not.
          \begin{itemize}
            \item $|\Helper{Inc}{H, E[t[\self{}=x'][y=y']]\ \overline{T}, \loc} \cup |\Helper{Inc}{H, \loc}|  \leq 1$. Cannot happen, as $x \mapsto \loc \in H$ already guarantees that the incoming references
              $|\Helper{Inc}{H, E[t[\self{}=x'][y=y']]\ \overline{T}, \loc}| \geq 1$

            \item $|\Helper{Inc}{H, E[t[\self{}=x'][y=y']]\ \overline{T}, \loc} \cup |\Helper{Inc}{H, \loc}|  \geq 1$. Proceed by case analysis over the possible incoming references. We know that
              $|\Helper{Inc}{H, E[t[\self{}=x'][y=y']]\ \overline{T}, \loc}| \geq 1$ since
              $x \mapsto \loc \in H$ and $\loc$ is isolated and we assume (\ref{pre-mcall-a5}).
              \begin{itemize}
                \item $|\Helper{Inc}{H, \loc}|  \geq 1$. Cannot happen, by assumption (\ref{pre-mcall-a5})
                  that violates $\Helper{Inc}{H, \loc} = \emptyset$.
                \item $|\Helper{Inc}{H, E[t[\self{}=x'][y=y']]\ \overline{T},
                  \loc}| \geq 1$. We pick
                  the thread $E[t[\self{}=x'][y=y']]$ and show that it trivially
                  satisfies that $\Helper{Inc}{H, E[t[\self{}=x'][y=y']]\ \overline{T}, \loc} =
                    \Helper{Inc}{H, E[t[\self{}=x'][y=y']], \loc}$, by the assumption (\ref{pre-mcall-a5}),
                      and maintains that $\Helper{Inc}(H', \loc) = \emptyset$.
                      Thus, $\Helper{Isolated}{H', E[t[\self{}=x'][y=y']]\ \overline{T}}$.
              \end{itemize}
          \end{itemize}
      \end{itemize}
  \end{itemize}

  By \RN{WF-Configuration}, we conclude that $\Gamma \vdash H;   E[t[\self{}=x'][y=y']]\ \overline{T}$ is well-formed.

  \item[E-NoSuchMethod], $H; E[x.m(v)] \ReducesTo H; \err$.
  This proceeds in the same fashion as \RN{E-NoSuchField}

  \item[E-AbsentTarget], $H; E[x.m(v)] \ReducesTo H; \CErr$.
  This proceeds in the same fashion as \RN{E-AbsentVar}
  \end{description}


\item $t = E[\Spawn{x}{t'}]$. By the induction hypothesis,
  \begin{align*}
    &H; E[\Spawn{x}{t'}]\ \overline{T} \ReducesTo H; \overline{T'}\ \overline{T} \\
    & \qquad \Rightarrow \exists
  \Gamma'. \Gamma' \supseteq \Gamma \land \Gamma' \vdash H';  \overline{T'}\ \overline{T}
  \end{align*}

  There is only a single reduction rule.
  \begin{description}
  \item[R-Spawn], $H; E[\Spawn{x}{t'}] \ReducesTo H; E[\loc]\ t'$.
    From initial assumption and \RN{WF-Configuration}:
    \begin{align}
      &  \Gamma \vdash H;  E[\Spawn{x}{t'}] \overline{T} \label{pre-spawn-a1} \\
      &\label{pre-spawn-a3}
        \forall t' \in E[\Spawn{x}{t'}]\ \overline{T}.\ \ \Gamma \vdash H;  t' \\
      &\label{pre-spawn-a4}
        \Helper{Local}{H, E[\Spawn{x}{t'}]\ \overline{T}} \\
      &\label{pre-spawn-a5}
        \Helper{Isolated}{H, E[\Spawn{x}{t'}]\ \overline{T}}
    \end{align}

    \textbf{Need to show:}
    \[\exists
    \Gamma'. \Gamma' \supseteq \Gamma \land \Gamma' \vdash H';  \overline{T'}\ \overline{T}\]

    The strategy of the proof is to show that all components are well-formed.
    By rule \RN{WF-Configuration} and IH, we need to show:
  \begin{align}
    & \Gamma' \supseteq \Gamma \label{p-spawn-gamma} \\
    & \Gamma' \vdash H' \label{p-spawn-wf}\\
    & \Gamma' \vdash H';  E[\loc]\ t' \label{p-spawn-wf-let} \\
    & \forall t'' \in E[\loc]\ t' \ \overline{T}. \Gamma' \vdash H;  t'' \label{p-spawn-t} \\
    & \Helper{Local}{H', E[\loc]\ t'\ \overline{T}} \label{p-spawn-local}\\
    & \Helper{Isolated}{H', E[\loc]\ t'\ \overline{T}} \label{p-spawn-iso}
  \end{align}

  We pick $\Gamma' = \Gamma, \loc : \localkw{}, x : \localkw{}$, and $H' = H, x' \mapsto
  \loc, \loc \mapsto \Chan{i, \varnothing}$ and $\overline{T'} = E[\loc]\ t'$ by
  \RN{R-Spawn}.  We now proof each of the components of the configuration:
  \begin{itemize}
    \item We picked $\Gamma' = \Gamma, \loc : \localkw{}, x : \localkw{}$, this trivially
      satisfies (\ref{p-spawn-gamma}).

    \item Show $\Gamma' \vdash H'$ (\ref{p-spawn-wf}).
      By the assumption (\ref{pre-spawn-a1}), we have $\Gamma \vdash H$.
      We proceed by proving that each component is well-formed.
      \begin{enumerate}
      \item \label{p-spawn-i1} $Gamma(x) = \Gamma(\loc) = \localkw{}$
      \item \label{p-spawn-i2} By \RN{WF-H-Var} with (\ref{p-spawn-i1}), we conclude that $\Gamma' \vdash H, x \mapsto \loc$
      \item By \RN{WF-H-Chain} with (\ref{p-spawn-i1}) and (\ref{p-spawn-i2}), we conclude
        that $\Gamma' \vdash H'$.
      \end{enumerate}

    \item Show that $\Gamma' \vdash H';  E[\loc] \ t'$ (\ref{p-spawn-wf-let}).
      This is trivially satisfied from assumption (\ref{pre-spawn-a1}) with (\ref{p-spawn-wf}),
      given that it is part of the assumptions that t' is well-formed from \RN{WF-Term-Let}.

    \item Show $\forall t'' \in E[\loc] \ t' \ \overline{T}. \Gamma' \vdash H;  t''$.
      The case for $\overline{T}$ is trivially satisfied by the assumptions, (\ref{pre-spawn-a3}).
      The case for $E[\loc]$ was shown above (\ref{p-spawn-wf-let}).

    \item Show $\Helper{Local}{H', E[\loc]\ t' \ \overline{T}}$ (\ref{p-spawn-local}).
      This is vacuously satisfied by the assumption (\ref{pre-spawn-a4}).

    \item Show $\Helper{Isolated}{H', E[\loc]\ t' \overline{T}}$, (\ref{p-spawn-iso}).
      This is satisfied by the assumption (\ref{pre-spawn-a5}), since spawning a thread
      does not affect the \textbf{Isolated} predicate.
  \end{itemize}

  By \RN{WF-Configuration}, we conclude that $\Gamma \vdash H;   E[\loc]\ t'\ \overline{T}$ is well-formed.
  \end{description}


\item $t = E[\recv{\loc}]$. By the induction hypothesis,
  \begin{align*}
    &H; E[\recv{\loc}]\ \overline{T} \ReducesTo H; \overline{T'}\ \overline{T} \\
    & \qquad \Rightarrow \exists
  \Gamma'. \Gamma' \supseteq \Gamma \land \Gamma' \vdash H';  \overline{T'}\ \overline{T}
  \end{align*}

  There are multiple rules, depending on whether the location maps to
  an object \RN{E-RecvBadTarget} or to a channel (\RN{R-Recv}).

  \begin{description}
  \item[R-RecvBadTarget], $H; E[\recv{\loc}] \ReducesTo H; \err$.
    This proof is similar to \RN{E-AbsentVar}, except that $H(\loc)$ maps
    to an object, instead of to $\Absent$.

  \item[R-Recv], $H; E[\recv{\loc}] \ReducesTo H; E[\loc']$.
    From initial assumption and \RN{WF-Configuration}:
    \begin{align}
      &  \Gamma \vdash H;  E[\recv{\loc}] \overline{T} \label{pre-recv-a1} \\
      &\label{pre-recv-a3}
        \forall t' \in E[\recv{\loc}]\ \overline{T}.\ \ \Gamma \vdash H;  t' \\
      &\label{pre-recv-a4}
        \Helper{Local}{H, E[\recv{\loc}]\ \overline{T}} \\
      &\label{pre-recv-a5}
        \Helper{Isolated}{H, E[\recv{\loc}]\ \overline{T}}
    \end{align}

    \textbf{Need to show:}
    \[\exists
    \Gamma'. \Gamma' \supseteq \Gamma \land \Gamma' \vdash H';  \overline{T'}\ \overline{T}\]

    The strategy of the proof is to show that all components are well-formed.
    By rule \RN{WF-Configuration} and IH, we need to show:
  \begin{align}
    & \Gamma' \supseteq \Gamma \label{p-recv-gamma} \\
    & \Gamma' \vdash H' \label{p-recv-wf}\\
    & \Gamma' \vdash H';  E[\loc'] \label{p-recv-wf-let} \\
    & \forall t'' \in E[\loc']\ \overline{T}. \Gamma' \vdash H;  t'' \label{p-recv-t} \\
    & \Helper{Local}{H', E[\loc']\ \overline{T}} \label{p-recv-local}\\
    & \Helper{Isolated}{H', E[\loc']\ \overline{T}} \label{p-recv-iso}
  \end{align}

  We pick $\Gamma' = \Gamma$, and $H' = H[\loc \mapsto \Chan{i, \varnothing}]$
  and $\overline{T'} = E[\loc']$ by \RN{R-Recv}.  We now proof each of the
  components of the configuration:
  \begin{itemize}
    \item We picked $\Gamma' = \Gamma$, this trivially satisfies
      (\ref{p-recv-gamma}).

    \item Show $\Gamma' \vdash H'$ (\ref{p-recv-wf}).
      By (\ref{pre-recv-a1}), all the locations and variables
      match the expected capability and are defined in $\Gamma$ and $H$.
      By \RN{WF-H-Class} with initial assumptions, we conclude
      $\Gamma' \vdash H'$ is well-formed.

    \item Show that $\Gamma' \vdash H';  E[\loc']$ (\ref{p-recv-wf-let}).  By
      \RN{WF-Term-Let} with (\ref{pre-recv-a1}) and (\ref{p-recv-wf}) where
      $\loc' \in \dom{H}$, we conclude $\Gamma' \vdash H';  E[\loc']$.

    \item Show $\forall t'' \in E[\loc'] \ \overline{T}. \Gamma' \vdash H;  t''$.
      The case for $\overline{T}$ is trivially satisfied from (\ref{pre-recv-a3}).
      The case for $E[\loc]$ was shown above (\ref{p-recv-wf-let}).

    \item Show $\Helper{Local}{H', E[\loc'] \ \overline{T}}$ (\ref{p-recv-local}).
      This is trivially satisfied from (\ref{pre-recv-a4}). Explanation: Before the reduction
      step, by definition of \textit{ROG}, locations leading to channels are not considered
      part of a \textit{ROG}. Thus, $\loc'$ was unreachable by any thread. By initial assumption
      of well-formedness
      $\Gamma \vdash H$ and $\Gamma(\loc') \neq \localkw{}$. Thus, $\loc'$ is in a single
      thread, and we can conclude that it satisfies $\Helper{Local}{H', E[\loc'] \ \overline{T}}$.

    \item Show $\Helper{Isolated}{H', E[\loc']\ \overline{T}}$, (\ref{p-recv-iso}).
      This is satisfied from initial assumption (\ref{pre-recv-a5}). Explanation:
      From initial assumption and the helper function \textit{Isolated},
      if $\loc'$ is an isolate, then it can only be in a channel. By taking a step and removing
      $\loc'$ from the channel, we place it in the stack and the incoming number of references
      is 1, so it vacuously satisfies $\Helper{Isolated}{H', E[\loc']\ \overline{T}}$.
  \end{itemize}

  By \RN{WF-Configuration}, we conclude that $\Gamma \vdash H;   E[\loc']\  \overline{T}$ is well-formed.
  \end{description}


\item $t = E[\move{\loc}{v}]$. By the induction hypothesis,
  \begin{align*}
    &H; E[\move{\loc}{v}]\ \overline{T} \ReducesTo H; \overline{T'}\ \overline{T} \\
    & \qquad \Rightarrow \exists
  \Gamma'. \Gamma' \supseteq \Gamma \land \Gamma' \vdash H';  \overline{T'}\ \overline{T}
  \end{align*}

  There are multiple rules, depending on whether the element to send contains
  local elements (\RN{E-SendingLocal}), whether $\loc$ is a channel or not (\RN{R-SendBlock} and \RN{E-SendBadTargetOrArgument})
  and whether $v = \loc'$ and maps to a channel (\RN{E-SendBadTargetOrArgument})).

  \begin{description}
  \item[R-SendBlock], $H; E[\move{\loc}{v}] \ReducesTo H; E[\Blocked{i}{\loc}]$.
    From initial assumption and \RN{WF-Configuration}:
    \begin{align}
      &  \Gamma \vdash H;  E[\move{\loc}{v}] \overline{T} \label{pre-send-a1} \\
      &\label{pre-send-a3}
        \forall t' \in E[\move{\loc}{v}]\ \overline{T}.\ \ \Gamma \vdash H;  t' \\
      &\label{pre-send-a4}
        \Helper{Local}{H, E[\move{\loc}{v}]\ \overline{T}} \\
      &\label{pre-send-a5}
        \Helper{Isolated}{H, E[\move{\loc}{v}]\ \overline{T}}
    \end{align}

    \textbf{Need to show:}
    \[\exists
    \Gamma'. \Gamma' \supseteq \Gamma \land \Gamma' \vdash H';  \overline{T'}\ \overline{T}\]

    The strategy of the proof is to show that all components are well-formed.
    By rule \RN{WF-Configuration} and IH, we need to show:
  \begin{align}
    & \Gamma' \supseteq \Gamma \label{p-send-gamma} \\
    & \Gamma' \vdash H' \label{p-send-wf}\\
    & \Gamma' \vdash H';  E[\loc] \label{p-send-wf-let} \\
    & \forall t'' \in E[\loc]\ \overline{T}. \Gamma' \vdash H;  t'' \label{p-send-t} \\
    & \Helper{Local}{H', E[\loc]\ \overline{T}} \label{p-send-local}\\
    & \Helper{Isolated}{H', E[\loc]\ \overline{T}} \label{p-send-iso}
  \end{align}

  We pick $\Gamma' = \Gamma$, and $H' = H[\loc \mapsto \Chan{i, v}]$
  and $\overline{T'} = E[\loc]$ by \RN{R-SendBlock}.  We now proof each of the
  components of the configuration:
  \begin{itemize}
    \item We picked $\Gamma' = \Gamma$, this trivially satisfies
      (\ref{p-send-gamma}).

    \item Show $\Gamma' \vdash H'$ (\ref{p-send-wf}).
      From (\ref{pre-send-a1}), all the locations and variables
      match the expected capability and are defined in $\Gamma$ and $H$.
      By \RN{WF-H-Chan} with the initial assumptions, we conclude
      $\Gamma' \vdash H'$ is well-formed.

    \item Show that $\Gamma' \vdash H';  E[\loc]$ (\ref{p-send-wf-let}).  By
      \RN{WF-Term-Let} with (\ref{pre-send-a1}) and (\ref{p-send-wf}) where
      $\loc \in \dom{H}$, we conclude $\Gamma' \vdash H';  E[\loc]$.

    \item Show $\forall t'' \in E[\loc] \ \overline{T}. \Gamma' \vdash H;  t''$.
      The case for $\overline{T}$ is trivially satisfied from (\ref{pre-send-a3}).
      The case for $E[\loc]$ was shown above (\ref{p-send-wf-let}).

    \item Show $\Helper{Local}{H', E[\loc] \ \overline{T}}$ (\ref{p-send-local}).
      This is trivially satisfied from (\ref{pre-send-a4}).

    \item Show $\Helper{Isolated}{H', E[\loc]\ \overline{T}}$, (\ref{p-send-iso}).
      This is satisfied from initial assumption (\ref{pre-send-a5}). Explanation:
      From initial assumption and the helper function \textit{Isolated},
      if $v$ is an isolate, then it can only be in the stack and is unique. By taking a step and removing
      $v$ from the stack, we place it in channel. The number of stack incoming references for $v$ is 0,
      and the number of heap incoming references is 1. Thus, the configuration vacuously satisfies
      $\Helper{Isolated}{H', E[\loc]\ \overline{T}}$, as the antecedent of the implication is false.
  \end{itemize}

  By \RN{WF-Configuration}, we conclude that $\Gamma \vdash H;   E[\loc]\  \overline{T}$ is well-formed.

  \item[E-SendBadTargetOrArgument], $H; E[\move{\loc}{\loc'}] \ReducesTo H; \err$.
    This proof follows the same structure of \RN{E-RecvBadTarget}

  \item[E-SendingLocal], $H; E[\move{v}{\loc}] \ReducesTo H; \PErr$.
    This proof follows the same structure of \RN{E-BadInstantiation}
  \end{description}


\item $t = E[\Blocked{i}{\loc}]$. By the induction hypothesis,
  \begin{align*}
    &H; E[\Blocked{i}{\loc}]\ \overline{T} \ReducesTo H; \overline{T'}\ \overline{T} \\
    & \qquad \Rightarrow \exists
  \Gamma'. \Gamma' \supseteq \Gamma \land \Gamma' \vdash H';  \overline{T'}\ \overline{T}
  \end{align*}

  There is a single reduction rule that can make progress, \RN{R-SendUnblock}.

  \begin{description}
  \item[R-SendUnblock], $H; E[\Blocked{i}{\loc}] \ReducesTo H'; E[\loc]$.
    From initial assumption and \RN{WF-Configuration}:
    \begin{align}
      &  \Gamma \vdash H;  E[\Blocked{i}{\loc}] \overline{T} \label{pre-block-a1} \\
      &\label{pre-block-a3}
        \forall t' \in E[\Blocked{i}{\loc}]\ \overline{T}.\ \ \Gamma \vdash H;  t' \\
      &\label{pre-block-a4}
        \Helper{Local}{H, E[\Blocked{i}{\loc}]\ \overline{T}} \\
      &\label{pre-block-a5}
        \Helper{Isolated}{H, E[\Blocked{i}{\loc}]\ \overline{T}}
    \end{align}

    \textbf{Need to show:}
    \[\exists
    \Gamma'. \Gamma' \supseteq \Gamma \land \Gamma' \vdash H';  \overline{T'}\ \overline{T}\]

    The strategy of the proof is to show that all components are well-formed.
    By rule \RN{WF-Configuration}, we need to show:
  \begin{align}
    & \Gamma' \supseteq \Gamma \label{p-block-gamma} \\
    & \Gamma' \vdash H' \label{p-block-wf}\\
    & \Gamma' \vdash H';  E[\loc] \label{p-block-wf-let} \\
    & \forall t'' \in E[\loc]\ \overline{T}. \Gamma' \vdash H;  t'' \label{p-block-t} \\
    & \Helper{Local}{H', E[\loc]\ \overline{T}} \label{p-block-local}\\
    & \Helper{Isolated}{H', E[\loc]\ \overline{T}} \label{p-block-iso}
  \end{align}

  We pick $\Gamma' = \Gamma$, and $H' = H$
  and $\overline{T'} = E[\loc]$ by \RN{R-SendUnblock}.  We now proof each of the
  components of the configuration:
  \begin{itemize}
    \item We picked $\Gamma' = \Gamma$, this trivially satisfies
      (\ref{p-block-gamma}).

    \item Show $\Gamma' \vdash H'$ (\ref{p-block-wf}).
      This is part of the initial assumptions.

    \item Show that $\Gamma' \vdash H';  E[\loc]$ (\ref{p-block-wf-let}).  This was
      part of the initial assumptions (\ref{pre-block-a1}).

    \item Show $\forall t'' \in E[\loc] \ \overline{T}. \Gamma' \vdash H;  t''$.
      The case for $\overline{T}$ is trivially satisfied from (\ref{pre-block-a3}).
      The case for $E[\loc]$ was shown above (\ref{p-block-wf-let}).

    \item Show $\Helper{Local}{H', E[\loc] \ \overline{T}}$ (\ref{p-block-local}).
      This is trivially satisfied from the initial assumption (\ref{pre-block-a4}).

    \item Show $\Helper{Isolated}{H', E[\loc]\ \overline{T}}$, (\ref{p-block-iso}).
      This is satisfied from initial assumption (\ref{pre-block-a5}),
      as $\Gamma(\loc) = \localkw{}$.
  \end{itemize}

  By \RN{WF-Configuration}, we conclude that $\Gamma \vdash H;   E[\loc]\  \overline{T}$ is well-formed.

  \end{description}


\item $t = E[\kopy{K}{x}]$. By the induction hypothesis,
  \begin{align*}
    &H; E[\kopy{K}{x}]\ \overline{T} \ReducesTo H; \overline{T'}\ \overline{T} \\
    & \qquad \Rightarrow \exists
  \Gamma'. \Gamma' \supseteq \Gamma \land \Gamma' \vdash H';  \overline{T'}\ \overline{T}
  \end{align*}

  There are two reduction rules, which depend on the value in $x$.

  \begin{description}
  \item[R-Copy], $H; E[\kopy{K}{x}] \ReducesTo H'; E[\loc']$.

    From initial assumption and \RN{WF-Configuration}:
    \begin{align}
      &  \Gamma \vdash H;  E[\kopy{K}{x}] \overline{T} \label{pre-kopy-a1} \\
      &\label{pre-kopy-a3}
        \forall t' \in E[\kopy{K}{x}]\ \overline{T}.\ \ \Gamma \vdash H;  t' \\
      &\label{pre-kopy-a4}
        \Helper{Local}{H, E[\kopy{K}{x}]\ \overline{T}} \\
      &\label{pre-kopy-a5}
        \Helper{Isolated}{H, E[\kopy{K}{x}]\ \overline{T}}
    \end{align}

    \textbf{Need to show:}
    \[\exists
    \Gamma'. \Gamma' \supseteq \Gamma \land \Gamma' \vdash H';  \overline{T'}\ \overline{T}\]

    The strategy of the proof is to show that all components are well-formed.
    By rule \RN{WF-Configuration} and IH, we need to show:
  \begin{align}
    & \Gamma' \supseteq \Gamma \label{p-kopy-gamma} \\
    & \Gamma' \vdash H' \label{p-kopy-wf}\\
    & \Gamma' \vdash H';  E[\loc'] \label{p-kopy-wf-let} \\
    & \forall t'' \in E[\loc']\ \overline{T}. \Gamma' \vdash H; t'' \label{p-kopy-t} \\
    & \Helper{Local}{H', E[\loc]\ \overline{T}} \label{p-kopy-local}\\
    & \Helper{Isolated}{H', E[\loc]\ \overline{T}} \label{p-kopy-iso}
  \end{align}

  We pick $\Gamma'$ to be:
  \begin{align*}
  \Gamma' =& \{ x : K \mid x \in \dom{H} \land H(x) = \loc \land H(\loc) = \Obj{K}{\_}{\_} \} \\
  & \cup \{ \loc : K \mid \loc \in \dom{H} \land H(\loc) = \Obj{K}{\_}{\_} \} \\
  & \cup \{ \loc : \localkw{} \mid \loc \in \dom{H} \land H(\loc) = \Chan{i, v} \}
  \end{align*}

  and $H' = H, H''$ where $H''$ is a copy of the \textit{ROG} with fresh location names
  and where field values in the \textit{ROG} refer to the updated fresh locations.
  $\overline{T'} = E[\loc']$ by \RN{R-Copy}.  We now proof each of the
  components of the configuration:
  \begin{itemize}
    \item Clearly, $\Gamma' \supseteq \Gamma$ and satisfies
      (\ref{p-kopy-gamma}).

    \item Show $\Gamma' \vdash H'$ (\ref{p-kopy-wf}).  Trivial by direct application of
      \RN{WF-H-Object}, \RN{WF-H-Var}, and \RN{WF-H-Chan}.

    \item Show that $\Gamma' \vdash H';  E[\loc']$ (\ref{p-kopy-wf-let}).  By
      \RN{WF-Term-Let} with (\ref{pre-kopy-a1}) and (\ref{p-kopy-wf}) where
      $\Helper{ROG}{H', \loc'} \subseteq \dom{H}$, we conclude $\Gamma' \vdash H';  E[\loc]$.

    \item Show $\forall t'' \in E[\loc'] \ \overline{T}. \Gamma' \vdash H'; t''$.
      The case for $\overline{T}$ is trivially satisfied from (\ref{pre-kopy-a3}).
      The case for $E[\loc']$ was shown above (\ref{p-kopy-wf-let}).

    \item Show $\Helper{Local}{H', E[\loc'] \ \overline{T}}$ (\ref{p-kopy-local}).
      This is trivially satisfied since it returns a new copy of the objects.

    \item Show $\Helper{Isolated}{H', E[\loc']\ \overline{T}}$,
      (\ref{p-kopy-iso}).  This is trivially satisfied by the initial
      assumptions, since it only extends the store with a new copy of the objects.
  \end{itemize}

  By \RN{WF-Configuration}, we conclude that $\Gamma \vdash H;   E[\loc']\  \overline{T}$ is well-formed.

  \item[E-AbsentCopyTarget], $H; E[\kopy{K}{x}] \ReducesTo H; \CErr$.
    This proof follows the same structure as \RN{E-AbsentVar}.
  \end{description}
\end{enumerate}
\end{proof}


\DRF*
  \begin{proof}
    From \cref{th: preserv dyn} (Preservation), and
    \RN{WF-Config\-uration}, every configuration satisfies the
    \PN{Local} and \PN{Isolated} predicates. This means that at no
    point will a local object be reachable from two threads, which
    excludes these from taking part in data-races. By
    \RN{R-FieldAssign} and \RN{E-BadFieldAssign}, immutable objects
    cannot be modified, which excludes these from taking part in
    data-races. This leaves isolated objects. For a data-race to
    happen on an isolated object $\loc$, the following must be
    possible: at time $t_0$, a thread reads/writes $\loc$, at time
    $t_1 > t_0$ (later reduction step), another thread
    reads/writes $\loc$, and nowhere between $t_0$ and $t_1$ is
    $\loc$ moved from the first thread to the second. Because of
    \PN{Isolated}, if there are more than one
    reachable\footnote{Note that as we do not discard old stack
      variables, there may be variables in the store which are
      unreachable to the program.} pointers to $\loc$, all those
    pointers are on the stack of a single thread. Thus, for two
    threads to be able to reach the same isolated object without
    moving it, the isolated object must be stored in a field $f$
    in an object $o$ reachable from both threads (WLOG, assume
    directly reachable from stack as we require unravelling paths
    field-by-field). From before, we know $o$ cannot be local or
    isolated. Immutable objects cannot contain isolated objects
    (\RN{R-New} and \PN{OkRef}). Thus, this is only possible when
    $o$ is unsafe. Thus, we conclude that \LangFormal{} is data-race
    free modulo unsafe objects.
  \end{proof}


\DynamicGradualGuarantee*
\begin{proof}
\begin{enumerate}
\item $\Gamma\vdash H; t\ \overline{T_0} \Rightarrow \Gamma^e \vdash (H; t\ \overline{T_0})^e$.

  This is trivially satisfied from the definition of capability stripping (\cref{def:safe stripping}).
  Any configuration with safe capabilities can replace the safe capabilities by \unkw{} and
  will still be well-formed. \qed

\item
\begin{enumerate}
\item  If $H; t\ \overline{T_0}  \ReducesTo H'; \overline{T_1}\ \overline{T_0}$
    and $\overline{T_1} \neq \textit{Err}$ then,
    $(H; t\ \overline{T_0})^e  \ReducesTo (H'; \overline{T_1}\ \overline{T_0})^e$.

  By induction on a derivation of $H; t\ \overline{T_0} \ReducesTo H';
  \overline{T_1}\ \overline{T_0}$.
  \textbf{We need to show}:
  \[(H; t\ \overline{T_0})^e  \ReducesTo (H'; \overline{T_1}\ \overline{T_0})^e\]

  From the induction hypothesis, we assume that the reduction rule does not produce an error,
  \ie{} $T_1 \neq \textit{Err}$, and that the term to reduce is fixed, \ie{} $t$.

  \begin{enumerate}
  \item \RN{R-Let}. $H; \LetIn{x}{v}{t}\ \overline{T_0}  \ReducesTo H, x \mapsto v; t\ \overline{T_0}$\\
    \begin{enumerate}
    \item \label{gg21} $\erase{H; \LetIn{x}{v}{t}} = \erase{H}; \LetIn{x}{v}{t}$
    by \cref{lem:term-rewritting} with $t = \LetIn{x}{v}{t}$.
    \item $\erase{H; \LetIn{x}{v}{t}\ \overline{T_0}} \ReducesTo \erase{H, x \mapsto v; t\ \overline{T_0}}$ by
      application \RN{C-Eval}, \RN{R-Let} with \cref{gg21}.
    \end{enumerate}
  \item \RN{R-Var}. $H; E[x]\ \overline{T_0}  \ReducesTo H; E[v]\ \overline{T_0}$\\
    \begin{enumerate}
      \item $\neg \Helper{isIso}{...}$ by the assumptions.
      \item \label{gg22} $\erase{H; E[x]} = \erase{H}; E'[x]$ where $E'[\bullet] = \erase{E'}[\bullet]$
        by \cref{lem:term-rewritting} with $t = E[x]$.
      \item $\erase{H; E[x]\ \overline{T_0}} \ReducesTo \erase{H;
        E[v]\ \overline{T_0}}$ by application \RN{C-Eval}, \RN{R-Var} with
        \cref{gg22}.
    \end{enumerate}
  \item \RN{R-Consume}. $H; E[\consume{x}]\ \overline{T_0}  \ReducesTo H, x\mapsto \Absent; E[v]\ \overline{T_0}$\\
    \begin{enumerate}
      \item $H(x) = \loc$ by the assumptions.
      \item \label{gg23} $\erase{H; E[\consume{x}]} = \erase{H}; E'[\consume{x}]$
        where $E'[\bullet] = \erase{E'}[\bullet]$ by \cref{lem:term-rewritting}
        with $t = E[\consume{x}]$.
      \item $\erase{H; E[\consume{x}]\ \overline{T_0}} \ReducesTo \erase{H, x\mapsto \Absent;
        E[v]\ \overline{T_0}}$ by application \RN{C-Eval}, \RN{R-Consume} with
        \cref{gg23}, as required.
    \end{enumerate}

  \item \RN{R-Field}. $H; E[x.f]^i\ \overline{T_0}  \ReducesTo H; E[v]^i\ \overline{T_0}$\\
    \begin{enumerate}
      \item \label{gg241} $H(x) = \loc, f=v \in \Fs, \neg \Helper{isIso}{H, v},
        \Helper{localOwner}{H, i, \loc}$ by the assumptions.
      \item \label{gg242} $\erase{H; E[x.f]} = \erase{H}; E'[x.f]$
        where $E'[\bullet] = \erase{E'}[\bullet]$ by \cref{lem:term-rewritting}
        with $t = E[x.f]$.
      \item \label{gg243} $\erase{H(x)} = \loc, f=v \in \Fs, \neg
        \Helper{isIso}{\erase{H}, \erase{v}}, \Helper{localOwner}{\erase{H}, i,
          \loc}$ by \cref{def:safe stripping} with \cref{gg241} and
        \cref{gg242}.
      \item $\erase{H; E[x.f]^i\ \overline{T_0}} \ReducesTo \erase{H;
        E[v]^i\ \overline{T_0}}$ by application \RN{C-Eval}, \RN{R-Field} with
        \cref{gg243}, as required.
    \end{enumerate}
  \item \RN{R-FieldAssign}. $H; E[x.f = v]^i\ \overline{T_0}  \ReducesTo H'; E[v']^i\ \overline{T_0}$\\
    \begin{enumerate}
      \item \label{gg251} $H(x) = \loc, f=v' \in \Fs, \neg \Helper{isImm}{H,
        \loc},$\\ $\Helper{OkRef}{H, K, v}, \Helper{isLocal}{H, \loc} \Rightarrow
        (\Helper{isOwner}{H, i, \loc} \land \Helper{localOwner}{H, i, v})$ by the assumptions.

      \item \label{gg252} $\erase{H; E[x.f = v]} = \erase{H}; E'[x.f = v]$
        where $E'[\bullet] = \erase{E'}[\bullet]$ by \cref{lem:term-rewritting}
        with $t = E[x.f = v]$.
      \item \label{gg253} $\erase{H(x)} = \loc, f=v' \in \Fs, \neg \Helper{isImm}{H,
        \loc}$,\\ $\Helper{OkRef}{H, \unkw, v}$ by \cref{def:safe stripping} with
        \cref{gg251,gg252}.
      \item $\erase{H; E[x.f = v]^i\ \overline{T_0}} \ReducesTo \erase{H';
        E[v']^i\ \overline{T_0}}$ by application \RN{C-Eval}, \RN{R-FieldAssign} with
        \cref{gg253}, as required.
    \end{enumerate}

  \item \RN{R-New}. $H; E[\Obj{K}{\overline{f=v}}{\Ms}]^i\ \overline{T_0}
    \ReducesTo H, \loc \mapsto \Obj{K^i}{\overline{f=v}}{\Ms};
    E[v']^i\ \overline{T_0}$\\
    \begin{enumerate}
      \item \label{gg261} $ \forall f = v \in \overline{f = v}.\ \OkRef{K}{v}
        \land (K = \localkw{} \land \Helper{isLocal}{H, v}) \Rightarrow
        \Helper{isOwner}{H, i, v}, \loc\ \textit{fresh}$ by the assumptions.
      \item \label{gg262} $\erase{H; E[\Obj{K}{\overline{f=v}}{\Ms}]} =$ \\
        $\erase{H}; E'[\Obj{\unkw}{\overline{f=v}}{\Ms}]$ where $E'[\bullet] =
        \erase{E'}[\bullet]$ by \cref{lem:term-rewritting} with $t = E[\Obj{K}{\overline{f=v}}{\Ms}]$.
      \item \label{gg263} $\forall f = v \in \overline{f = v}.\ \OkRef{\unkw}{v}$
         by application of \cref{def:safe stripping} with \cref{gg261,gg262}.
      \item $\erase{H; E[\Obj{\unkw}{\overline{f=v}}{\Ms}]^i\ \overline{T_0}}
        \ReducesTo H, \loc \mapsto \Obj{\unkw^i}{\overline{f=v}}{\Ms};
        E[v']^i\ \overline{T_0}^e$ by application \RN{C-Eval}, \RN{R-New} with
        \cref{gg263}, as required.
    \end{enumerate}

  \item \RN{R-Call}. $H; E[x.m(v)]\ \overline{T_0} \ReducesTo H, x' \mapsto \loc,
    y' \mapsto v; E[t[\self{} = x'][y = y']]^i\ \overline{T_0}$\\
    \begin{enumerate}
      \item \label{gg271} $x \mapsto \loc, \loc \mapsto
        \Obj{\_}{\_}{\Ms\ \Method{m}{y}{t}}$ by the assumptions.
      \item \label{gg272} $\erase{H; E[x.m(v)]} = \erase{H}; E'[x.m(v)]$ where
        $E'[\bullet] = \erase{E'}[\bullet]$ by \cref{lem:term-rewritting} with $t
        = E[x.m(v)]$.
      \item \label{gg273} $x', y' \ \textit{fresh}$ by assumption $\erase{\Gamma}
        \vdash \erase{H; E[x.m(v)}]$
      \item $\erase{H; E[x.m(v)]\ \overline{T_0}} \ReducesTo \erase{H}, x'
        \mapsto \loc, y' \mapsto v; \erase{E}[\erase{t[\self{} = x'][y =
              y']}]\ \overline{T_0}^e$ by application \RN{C-Eval}, \RN{R-Call} with
        \cref{gg271,gg272,gg273}, as required.
    \end{enumerate}

  \item \RN{R-CastLoc}. $H; E[\CC{K}{\loc}]\ \overline{T_0} \ReducesTo H ; E[\loc]\ \overline{T_0}$\\
    \begin{enumerate}
      \item \label{gg281} $\loc \in \dom{H}$ by the assumptions.
      \item \label{gg282} $\erase{H; E[\CC{K}{\loc}]} = \erase{H}; E'[\CC{\unkw}{\loc}]$ where
        $E'[\bullet] = \erase{E'}[\bullet]$ by \cref{lem:term-rewritting} with $t
        = E[\CC{K}{\loc}]$.
      \item \label{gg283} $\loc \in \dom{\erase{H}}$ by assumption
        $\erase{\Gamma} \vdash \erase{H}; E[\CC{K}{\loc}]$
      \item $\erase{H; E[\CC{K}{\loc}]\ \overline{T_0}} \ReducesTo \erase{H};
        \erase{E}[\loc]\ \erase{\overline{T_0}}$ by application \RN{C-Eval},
        \RN{R-CastLoc} with \cref{gg281,gg282,gg283}, as required.
    \end{enumerate}

  \item \RN{R-Copy}. $H; E[\kopy{K}{x}]^i\ \overline{T_0} \ReducesTo H' ; E[\loc]\ \overline{T_0}$\\
    \begin{enumerate}
      \item \label{gg291} $x \in \dom{H}$ by the assumptions.
      \item \label{gg292} $\erase{H; E[\kopy{K}{x}]} = \erase{H}; E'[\kopy{\unkw}{x}]$ where
        $E'[\bullet] = \erase{E'}[\bullet]$ by \cref{lem:term-rewritting} with $t
        = E[\kopy{K}{x}]$.
      \item \label{gg293} $x \mapsto \loc' \in \erase{H}$ by assumption
        $\erase{\Gamma} \vdash \erase{H; E[\kopy{K}{x}]}$
      \item \label{gg294} $\Helper{localOwner}{H, i, \loc'}$ is vacuously
        satisfied.
      \item $\erase{H; E[\kopy{K}{x}]\ \overline{T_0}} \ReducesTo \erase{H''};
        \erase{E}[\loc]\ \erase{\overline{T_0}}$ by application \RN{C-Eval},
        \RN{R-Copy} with \cref{gg291,gg292,gg293,gg294}, where $\erase{H'} = H''$, as required.
    \end{enumerate}

  \item \RN{R-Spawn}, \RN{R-Recv}, \RN{R-SendBlock}, and \RN{R-SendUnblock}
    are all straightforward.

  \end{enumerate}

\item If $H; t\ \overline{T_0}  \ReducesTo H'; \overline{T_1}\ \overline{T_0}$ and
$\textit{T_1} = \CErr \lor \err$ then,
$(H; t\ \overline{T})^e  \ReducesTo \erase{H'';  \overline{T_1}\ \overline{T_0}}$

By
induction on a derivation of $H; t\ \overline{T_0}  \ReducesTo H';  \overline{T_1}\ \overline{T_0}$ and by case analysis of $t$. Reminder: assumption $\overline{T_1} = \CErr \lor \err$.
\textbf{Need to show:}
\[(H; t\ \overline{T})^e  \ReducesTo H'';  \textit{Err}\]

    We proceed by induction on a derivation $H; t \ReducesTo H'; \overline{T_1}$.
    \begin{itemize}
    \item By all terms that use the derivation \RN{E-NoSuchField},
      \RN{E-NoSuchMethod}, \RN{E-NoSuchFieldAssign},
      \RN{E-SendBadTargetOrArgument}, and \RN{E-RecvBadTarget}, a missing field
      or method in the unsafe configuration, $(H; t\ \overline{T_0})^e \ReducesTo
      H''; \overline{T_1}\ \overline{T_0}$ reduces using the same reduction rule
      producing the same error.
    \item \sloppy By all terms that use the derivation \RN{E-AbsentVar},
      \RN{E-Consume}, \RN{E-AbsentTarget}, \RN{E-AbsentTargetAccess},
      \RN{E-AbsentFieldAssign}, \RN{E-AbsentCopyTarget}, a missing target in the
      unsafe configuration, $(H; t\ \overline{T})^e \ReducesTo H''; \textit{Err}$
      reduces using the same reduction rule producing the same error.
    \end{itemize}\qed
\end{enumerate}

\item
\begin{enumerate}
\item
      If $(H; t\ \overline{T_0})^e  \ReducesTo (H';  \textit{Err})^e$
      then, $H; t\ \overline{T_0}  \ReducesTo H'; \textit{Err}$

By induction on a derivation of $(H; t\ \overline{T_0})^e$.
Need to show: \[H; t\ \overline{T_0}  \ReducesTo H'; \textit{Err}\]
\begin{enumerate}
\item \sloppy \RN{E-NoSuchField}, \RN{E-NoSuchMethod}, \RN{E-NoSuchFieldAssign}, \RN{E-SendBadTargetOrArgument}, and \RN{E-RecvBadTarget}. Any configuration $(H; t\ \overline{T_0})^e$ that reduces to  $(H';  \textit{Err})^e$.
By \cref{lem:deterministic} with $H; t\ \overline{T_0}$ deterministically takes a single reduction step.
The fact that $H; t\ \overline{T_0}$ may have safe capabilities does not stop normal errors to occur.

\item \RN{E-AbsentVar}, \RN{E-Consume}, \RN{E-AbsentTarget},
      \RN{E-AbsentTargetAccess}, \RN{E-AbsentFieldAssign},
      \RN{E-AbsentCopyTarget}. From the induction hypothesis and the rules of the case,
      regardless of safe capabilities, the configuration will throw a \CErr{}.
\end{enumerate}\qed

\item If $(H; t\ \overline{T_0})^e \ReducesTo (H';
\overline{T}\ \overline{T_0})^e$ and $\overline{T} \neq \textit{Err}$ then, $H;
t\ \overline{T_0} \ReducesTo H'; \overline{T'}\ \overline{T_0}$ and
$\overline{T'} = \PErr \lor \CastErr \lor \overline{T}$.

By induction on a derivation of $(H; t\ \overline{T_0})^e$.
Need to show:
\[
H; t\ \overline{T_0} \ReducesTo H'; \overline{T'}\ \overline{T_0} \land
\overline{T'} = \PErr \lor \CastErr \lor \overline{T}
\]

\begin{enumerate}
\item \RN{R-Let}. If $(H; \LetIn{x}{v}{t'}\ \overline{T_0})^e \ReducesTo (H, x
  \mapsto v; t'\ \overline{T_0})^e$, then $H; t\ \overline{T_0} \ReducesTo H';
  \overline{T'}\ \overline{T_0}$ and $\overline{T'} = \PErr \lor \CastErr \lor
  \overline{T}$. $x \notin \dom{H}$ by assumption $\Gamma \vdash H;
  t\ \overline{T_0}$, and by application of \RN{R-Let}, $H;
  \LetIn{x}{v}{t'}\ \overline{T_0} \ReducesTo H, x \mapsto v; t'\ \overline{T_0}$ as required.
\item \RN{R-Var}, $(H; E[x]\ \overline{T_0})^e \ReducesTo (H; E[\loc]\ \overline{T_0})^e$ under assumption $\neg \Helper{isIso}{H, \loc}$. We proceed by cases analysis over the capability of $\loc$ (omit the case were $H(x) = \Absent$, as that would have thrown an error in the hypothesis):
  \begin{itemize}
    \item $H(\loc) = \Obj{\isokw}{\_}{\_}$. By application of \RN{E-AliasIso},
      $H; E[x]\ \overline{T_0} \ReducesTo (H; \PErr\ \overline{T_0}$, as required.
    \item $H(\loc) = \Obj{K}{\_}{\_} \land K \neq \isokw{}$. By application of \RN{R-Var},
      $H; E[x]\ \overline{T_0} \ReducesTo H; E[\loc]\ \overline{T_0}$ as required.
  \end{itemize}
\item \RN{R-Consume}, $(H; E[\consume{x}]\ \overline{T_0})^e \ReducesTo (H[x \mapsto \Absent]; E[\loc]\ \overline{T_0})^e$. From the assumption of the derivation, $H(x) = \loc$. By application of
  \RN{R-Consume}, $H; E[\consume{x}]\ \overline{T_0} \ReducesTo H, x \mapsto \Absent; E[\loc]\ \overline{T_0}$ as required.
\item \RN{R-Field}, $(H; E[x.f]^i\ \overline{T_0})^e \ReducesTo (H; E[v]^i\ \overline{T_0})^e$
under assumptions $H(x) = \loc$, $f = v \in \Fs$, $\neg \Helper{isIso}{H, v}$, and
$\Helper{localOwner}{H, i, v}$. We proceed by case analysis over the capability of $\loc$.
\begin{itemize}
\item $\Helper{isLocal}{H, \loc}$. We proceed by case analysis over its ownership capability $K$:
  \begin{itemize}
    \item $K^i$. We proceed by case analysis over $f$:
      \begin{itemize}
        \item $\Helper{isIso}{H, v}$. By \RN{E-IsoField}, it reduces to $H; \PErr$.
        \item $\neg \Helper{isIso}{H, v}$. By \RN{R-Field},
          $H; E[x.f]^i\ \overline{T_0} \ReducesTo H; E[v]^i\ \overline{T_0}$ as required.
        \item $H(v) = \Obj{K}{\_}{\_} \land K \neq \isokw{}$.
          By \RN{R-Field},
          $H; E[x.f]^i\ \overline{T_0} \ReducesTo H; E[v]^i\ \overline{T_0}$ as required.
      \end{itemize}
    \item $K^j \land i \neq j$. By \RN{E-LocalField}, it reduces to $H; \PErr$.
  \end{itemize}
\item $\Helper{isIso}{H, v}$. By application of \RN{E-IsoField}, the configuration steps
  to $H; \PErr$ as required.
\item $\neg \Helper{isIso}{H, v}$. By cases analysis similar to the above case.
\item Remaining cases. By \RN{R-Field}, $H; E[x.f]^i\ \overline{T_0} \ReducesTo
  H; E[v]^i\ \overline{T_0}$ as required.
\end{itemize}

\item \RN{R-FieldAssign}, $(H; E[x.f = v]^i\ \overline{T_0})^e \ReducesTo (H'; E[v']^i\ \overline{T_0})^e$.
  For all cases where $H(x) = \loc$ and $\neg \Helper{OkRef}{H, K, \loc}$  by \RN{E-BadFieldAssign},
      $H; E[x.f = v]^i\ \overline{T_0} \ReducesTo H; \PErr$ as required. For all cases
  where $\Helper{OkRef}{H, K, \loc}$ we proceed by case analysis on the capability of $\loc$:
  \begin{itemize}
    \item $\Helper{isImm}{H, \loc}$. By \RN{E-BadFieldAssign},
      $H; E[x.f = v]^i\ \overline{T_0} \ReducesTo H; \PErr$ as required.
    \item $\neg \Helper{isImm}{H, \loc}$
    \item $\Helper{notLocalOwner}{H, i, \loc}$. By \RN{E-BadFieldAssign},
      $H; E[x.f = v]^i\ \overline{T_0} \ReducesTo H; \PErr$ as required.
    \item $\Helper{isLocal}{H, \loc} \land \Helper{isOwner}{H, i, \loc}$.
      By case analysis on the capability of $f = v \in \Fs$.
      \begin{itemize}
        \item $\Helper{notLocalOwner}{H, i, v}$. By \RN{E-BadFieldAssign},
      $H; E[x.f = v]^i\ \overline{T_0} \ReducesTo H; \PErr$ as required.
        \item $\Helper{localOwner}{H, i, v}$. By \RN{R-FieldAssign},
      $H; E[x.f = v]^i\ \overline{T_0} \ReducesTo H'; E[v']\ \overline{T_0}$ as required.
        \item Remaining cases. By \RN{R-FieldAssign},
      $H; E[x.f = v]^i\ \overline{T_0} \ReducesTo H'; E[v']\ \overline{T_0}$ as required.
      \end{itemize}
    \item Remaining cases. By \RN{R-FieldAssign},
      $H; E[x.f = v]^i\ \overline{T_0} \ReducesTo H'; E[v']\ \overline{T_0}$ as required.
  \end{itemize}
\item \RN{R-New}, $(H; E[\Obj{K}{\overline{f=v}}{\Ms}]^i\ \overline{T_0})^e \ReducesTo (H'; E[\loc]^i\ \overline{T_0})^e$. For all cases where $\neg \Helper{OkRef}{H, K, v} \forall f=v \in \overline{f=v}$,
by application of \RN{E-BadInstantiation}, the configuration reduces to $H; \PErr$.
Under the assumption that $\Helper{OkRef}{H, K, v} \forall f=v \in \overline{f=v}$ we proceed
by case analysis over $K$:
  \begin{itemize}
  \item $K = \localkw{}$. If $\exists f=v \in \overline{f=v}$, s.t. $\Helper{notLocalOwner}{H, i, v}$
  then by direct application of \RN{E-BadInstantiation} the configuration reduces to $H; \PErr$.
  Otherwise, by application of $R-New$, $H; E[\Obj{K}{\overline{f=v}}{\Ms}]^i\ \overline{T_0} \ReducesTo H';   E[\loc]^i\ \overline{T_0}$ as required.
  \item $K \neq \localkw{}$. Under the assumption $\Helper{OkRef}{H, K, v} \forall
  f=v \in \overline{f=v}$, by application of $R-New$, $H;
  E[\Obj{K}{\overline{f=v}}{\Ms}]^i\ \overline{T_0} \ReducesTo H';
  E[\loc]^i\ \overline{T_0}$ as required.
  \end{itemize}
\item The remaining rules follow the same pattern and structure.
\end{enumerate}
\end{enumerate}
\end{enumerate}

\end{proof}

\begin{restatable}[Multistep Dynamic Gradual Guarantee]{theorem}{MultistepDynamicGradualGuarantee}\label{th: multistep dyn gradual}
\begin{itemize}
\item
A program $P$ reduces to a terminal configuration $C$ and generates a trace $\mathcal{T}$ from
$\epsilon; P \ReducesTo^* H; C$.
$C$ may or may not be an error configuration.
Let $P'$ be $P^e$, and let $n = |\mathcal{T}|$:
\begin{enumerate}
\item If $\CastErr \neq C \neq \PErr$, then $\mathcal{T}; \epsilon; P' \Replays^n \mathcal{T'}; H; C$
\item If $C \equiv \CastErr \lor \PErr$,
      then $\mathcal{T}; \epsilon; P' \Replays^n \mathcal{T'}; H'; \overline{T}$, and $\overline{T} \neq C$
\end{enumerate}

\item
A program $P^e$ reduces to a terminal
configuration $C$ and generates a trace $\mathcal{T}$ from  $\epsilon; P^e \ReducesTo^* H^e; C$.
$C$ may or may not be an error configuration.
Let $P'$ be $P$, and let $n = |\mathcal{T}|$. If $P'$ executes $\mathcal{T}$, then either:
\begin{enumerate}
\item if $C \not\equiv \textit{Err}$,
      then $\mathcal{T}; \epsilon; P' \Replays^m \mathcal{T'}; H'; C'$ and $C' \equiv C \lor \PErr \lor \CastErr$ and $m \leq n$.
\item if $C \equiv \textit{Err}$, then $\mathcal{T}; \epsilon; P' \Replays^n \mathcal{T'}; H; C$
\end{enumerate}
\end{itemize}
\end{restatable}

The proof follows directly from \cref{th: dyn gradual}.
\qed

\end{document}